\documentclass[11pt]{article}
\usepackage{graphicx,amsmath,bm, amsthm,mathrsfs,amssymb, braket, verbatim}
\usepackage{caption}
\usepackage{subcaption}
\usepackage[usenames]{color}
\usepackage{ulem,mathtools}
\usepackage{pdfpages}
\usepackage{lscape}
\usepackage{authblk}
\usepackage{cite}
\usepackage{tcolorbox}
\usepackage[section]{placeins}
\usepackage{placeins}

\newcommand{\ee}{\end{equation}}

\newcommand{\reff}[1]{(\ref{#1})}
\newcommand{\beq}{\begin{equation}}
\newcommand{\eeq}[1]{\label{#1}\end{equation}}
\newcommand{\beqa}{\begin{eqnarray}}
\newcommand{\eea}{\end{eqnarray}}
\newcommand{\eeqa}[1]{\label{#1}\end{eqnarray}}
\newcommand{\beg}{\begin{equation*}}
\newcommand{\eeg}{\end{equation*}}

\newcommand{\p}{\!+\!}
\newcommand{\m}{\!-\!}

\newcommand{\bsplit}{\begin{split}}
\newcommand{\esplit}{\end{split}}

\usepackage{circuitikz} 
\usepackage[capposition=bottom]{floatrow} 
\usepackage{epigraph} 
\usepackage{appendix}
\usepackage{rotating}
\allowdisplaybreaks

\title{Two types of series expansions valid at strong coupling}
\author[]{Ariel Edery\thanks{aedery@ubishops.ca}}
\affil[]{Department of Physics and Astronomy, Bishop's University, 2600 College Street, Sherbrooke, Qu\'{e}bec, Canada, J1M 1Z7.\vspace{1em}}
\begin{document}
\date{}
\maketitle
\begin{abstract}
\noindent 
It is known that perturbative expansions in powers of the coupling in quantum mechanics (QM) and quantum field theory (QFT) are asymptotic series. This can be useful at weak coupling but fails at strong coupling. In this work, we present two types of series expansions valid at strong coupling. We apply the series to a basic integral as well as a QM path integral containing a quadratic and quartic term with coupling constant $\lambda$. The first series is the usual asymptotic one, where the quartic interaction is expanded in powers of $\lambda$. The second series is an expansion of the quadratic part where the interaction is left alone. This yields an absolutely convergent series in inverse powers of $\lambda$ valid at strong coupling. For the basic integral, we revisit the first series and identify what makes it diverge even though the original integral is finite. We fix the problem and obtain, remarkably, a series in powers of the coupling which is absolutely convergent and valid at strong coupling. We explain how this series avoids Dyson's argument on convergence. We then consider the QM path integral (discretized with time interval divided into $N$ equal segments). As before, the second series is absolutely convergent and we obtain analytical expressions in inverse powers of $\lambda$ for the $n$th order terms by taking functional derivatives of generalized hypergeometric functions. The expressions are functions of $N$ and we work them out explicitly up to third order. The general procedure has been implemented in a Mathematica program that generates the expressions at any order $n$. We present numerical results at strong coupling for different values of $N$ starting at $N=2$. The series matches the exact numerical value for a given $N$ (up to a certain accuracy). The continuum is formally reached when $N\to \infty$ but in practice this can be reached at small $N$. 
\end{abstract}

\setcounter{page}{1}
\newpage
\section{Introduction}\label{Intro}
The path integral (or generating functional) in quantum mechanics (QM) and quantum field theory (QFT) is usually expanded in a perturbative series of the interaction term in powers of of the coupling (whereas the kinetic term, which is quadratic in the fields, is left alone). This expansion in powers of the coupling is well-known to yield an asymptotic series \cite{Strocchi, Marino1,QCD}. Though the asymptotic series ultimately diverges, if the coupling is sufficiently small (weak coupling), the series can plateau to a value at lower orders before departing from this value at higher orders. Therefore, standard model calculations at weak coupling based on this series -- typically carried out using Feynman diagrams -- can be (and are) very accurate after summing the first few lowest orders. However, asymptotic series fail completely at large coupling (strong coupling); they do not plateau to a given value before diverging. This is noticeable early on, after just a few orders. This is not due to a breakdown of the physical theory at strong coupling. For example, lattice QCD simulations confirm that QCD is perfectly valid at strong coupling, as it leads to computational results in agreement with observations e.g. the masses of light hadrons \cite{LQCD}. Yet, its perturbative series expansion is an asymptotic series which is not reliable at strong coupling. A goal of this paper is to develop series expansions that are absolutely convergent and valid at strong coupling.      

We study two different types of series expansions of a basic integral and a QM path integral containing quadratic terms plus a quartic interaction with coupling $\lambda$ (the QM system is commonly referred to as the anharmonic oscillator). The first type of expansion is the usual perturbative series of the interaction term in powers of the coupling $\lambda$ (where the quadratic part is left alone). In the second case, we perform a series expansion of the \textit{quadratic part} and leave the quartic interaction containing the coupling alone. This yields an expansion in \textit{inverse powers} of the coupling $\lambda$. This second series is useful at strong coupling and is less familiar than the first. 

We begin with the basic integral in section 2; this is a one-dimensional integral over a variable $x$. This reveals important properties of the two types of series expansions without the more complicated mathematics of the QM case. The integrand is given by  $e^{-\lambda \,x^4 -a\,x^2}$  where $\lambda$ and $a$ are positive constants. The advantage of starting with this basic integral is that we know the exact analytical result. We can therefore easily compare the two series expansions to the exact result. The first expansion in powers of $\lambda$ yields an asymptotic series; in fact, the basic integral is the prototypical example used by authors studying non-perturbative QFT (e.g. \cite{Strocchi, Marino1}) to demonstrate how a series expansion about zero coupling yields an asymptotic series. We plot it as a function of the order $n$ for different values of the coupling $\lambda$. At small $\lambda$, it plateaus to the exact value of the original integral (to within eight digits) for a certain range of the order $n$ before ultimately diverging. At large $\lambda$, the series fails completely; it departs from the correct value right from the start (at low orders) and then diverges. The plots are useful in providing intuition as to how asymptotic series behave with coupling and order. The second series expansion in inverse powers of the coupling yields an absolutely convergent series and is valid at both large and small $\lambda$. We plot it as a function of the order $n$ for different values of the coupling $\lambda$. For large $\lambda$ the convergence is rapid (only a few orders $n$ are required) whereas at small $\lambda$, the convergence is slow (very large $n$ are required). So this second series is perfectly suited to the strong coupling regime. In contrast to the first expansion, it is an absolutely convergent series and is an exact representation of the original integral. 

We revisit the first series and ask the following question: why do we obtain an asymptotic series when it originates from an integral which is finite and whose integrand is an exponential with an exact series representation? We identify precisely where the problem originates, not only for this basic integral but also for QM and QFT path integrals. Moreover, when this problem is fixed, we obtain, remarkably, a series in powers of the coupling $\lambda$ which is absolutely convergent and valid for both small and large $\lambda$ i.e. valid at weak and strong coupling. Using a series expansion of the incomplete gamma function, we show that it is an exact representation of the original integral. The series converges quickly at small $\lambda$ but slowly at large $\lambda$. So it is more convenient to use at weak coupling whereas the second series is more convenient to use at strong coupling.  The two series are different expansions but yield the same result. We therefore have a strong-weak coupling duality. This duality is not a conjecture; we prove that one can be transformed into the other. 

We then consider in section 3 the QM path integral for the anharmonic oscillator whose potential contains a quadratic and quartic part. This can be viewed as the multi(infinite)-dimensional version of the previous basic one-dimensional integral. The amplitude for a particle to go from point $x_a$ at time $t_a$ to point $x_b$ at time $t_b$ is a path integral denoted as $K(x_a,t_a; x_b,t_b)$ and sometimes referred to as the Kernel. The way that modern authors \cite{Larkoski,Marino1,Marino2} as well as Feynman and Hibbs in their classic text \cite{Feynman} solve the path integral is that they first determine the classical path $x_c(t)$ and then consider fluctuations $y(t)$ about it. This is convenient since the spatial end points $x_a$ and $x_b$ appear only in the classical action $S_c$ which can be extracted out of the path integral as an exponential factor. We will however be using a different mathematical approach and solve for the Kernel directly, without extracting the classical action. This will have the advantage that once the amplitude $K_H$ of the harmonic oscillator is obtained, it is easy to incorporate a force term $J(t)\,x$ to obtain the generating functional $K_{FH}[J]$, which is the amplitude of the forced harmonic oscillator as a function of the end points. We introduce in section 4 the Euclidean version of the Kernel, $K_E(x_a,\tau_a; x_b,\tau_b)$ where $\tau$ is the Euclidean time; this can readily be obtained from the ordinary amplitude $K$ via a Wick rotation ($t \to -i\,\tau$).  In section 5, the quartic part $\lambda\, x^4$ is treated perturbatively and expanded in powers of the coupling $\lambda$ with coefficients obtained from taking functional derivatives with respect to the source $J(\tau)$ of $K_{{FH}_E}[J]$. We obtain novel analytical expressions for the coefficients as \textit{functions of the end points}. In contrast, the end points do not appear in the coefficients of an energy expansion or in the expansion of the thermal partition function with source $J$ (see \cite{Marino1} for a clear exposition). We evaluate the series at different values of $\lambda$. For small $\lambda$, it plateaus quickly at low orders to a value which is the correct result (to within a certain accuracy) but at large $\lambda$ it never plateaus to any given value and within a few orders departs significantly from the correct result. So the QM path integral for the first expansion yields an asymptotic series just like our basic integral. The generating functional $K_{{FH}_E}[J]$ is a Gaussian that can be viewed as the fundamental ``building block'' for the first series in the sense that all terms in the expansion are derived from it via functional derivatives. 

In section 6 we study the second series for the QM path integral where we expand the \textit{quadratic part} and leave the quartic part containing the coupling $\lambda$ as well as the source term $J(\tau) \,x$ alone. The path integral for the quartic part plus source term, which we call the generating functional $Z[J]$, can be evaluated exactly and yields products of generalized hypergeometric functions. So the fundamental ``building block'' for the second series are hypergeometric functions in contrast to Gaussians for the first series. The terms in this expansion are obtained by taking functional derivatives with respect to $J(\tau)$ of  $Z[J]$ and then setting $J$ to zero. The time interval is divided into $N$ equal parts and the path integral is an $N-1$-dimensional integral. We derive analytical expressions for the $n$th order terms in the series as functions of $N$. This is mathematically complicated so we begin by first working out explicitly by hand the expressions for the first three orders. This provides the reader with a solid and intuitive grasp of the procedure before treating the case of a general order $n$. The general procedure has been implemented in a Mathematica program that generates the analytical expressions for the terms at any order $n$ (in Appendix A we write out the analytical expression generated by the program for $n=5$). The series is absolutely convergent for any given $N$ and any coupling $\lambda$. We present numerical values at strong coupling for the series and the exact numerical integration for various values of $N$ and the two results match (to within a certain level of accuracy depending on $N$. Calculating the series at higher $N$ requires more numerical precision and hence more computational resources). To reach the continuum, one formally requires to take the limit as $N$ tends to infinity. However, in practice, the exact numerical integration will often reach the continuum value at a relatively low value of $N$ for a desired level of accuracy.  In our case, for the parameters chosen, the exact numerically integrated value could be obtained to three decimal places without going beyond $N=9$.       

Freeman Dyson provided an insightful argument as to why our usual perturbative expansions in powers of the coupling yields an asymptotic series. In his classic paper \cite{Dyson}, Dyson focused on quantum electrodynamics (QED) but his main points are quite general. The argument is that if the power series expansion is analytic about zero coupling (hence absolutely convergent), that would also hold for a negative coupling of sufficiently small magnitude. However, the theory turns out to be qualitatively different at negative coupling as it has an unstable vacuum (Dyson refers to quantum tunneling from an ordinary state to a ``pathological'' state that is unbounded from below). Such a physical situation cannot be governed by a power series that is analytic about zero and this in turn implies that the series with positive coupling must ultimately diverge and hence be an asymptotic series. We discuss how the absolutely convergent series in powers of the coupling circumvents Dyson's argument. 
     
One of the best known analytical methods used to probe the strong coupling regime of quantum field theories is the $1/N$ expansion where $N$ is the number of fields, which is typically taken to be very large (for recent examples see \cite{Beca1,Beca2, Beca3, Giombi}). Though $N$ is typically much larger than the number of fields of the original/realistic field theory, it still provides physical insights into a regime which is not accessible via regular perturbation theory. The $1/N$ expansion in QCD was introduced by 't Hooft \cite{tHooft} and this triggered a vast literature on this topic (for a review see \cite{Manohar}). In the $N\to \infty$ limit, the main aspects of QCD can be captured by considering the planar sector of the theory. For example, at small 't Hooft parameter, the leading contribution to the free energy function can be evaluated by summing planar diagrams for the first few orders in perturbation theory. However, to calculate correlation functions beyond perturbation theory, one needs to resum all planar diagrams, which is beyond our present analytical capabilities. In contrast to two-dimensional models, we do not have a solid analytical grasp of the strong coupling regime of four-dimensional QCD even in the $N \to \infty$ limit (let alone the realistic case of $N=3$). The AdS/CFT correspondence \cite{Malda} has of course also been extremely useful for probing the strong coupling regime of quantum field theories (see reviews in \cite{Aharony,Hubeny}). In general, when the quantum field theory is strongly interacting, the dual gravitational side will be weakly interacting which is easier to handle mathematically. This has found applications in many areas of physics, including nuclear and condensed matter physics. An important result was the calculation of the ratio $\eta/s$ of a quark–gluon plasma using five-dimensional black holes where $\eta$ is the shear viscosity and $s$ is the volume density of entropy \cite{Kovtun}. Despite the numerous triumphs of the AdS/CFT correspondence, the QFTs are limited to conformal field theories, which does not apply to many QFTs, including real four-dimensional QCD. The two types of series expansions we considered in this paper can be applied directly to the original quantum system in question. In going from the basic integral to the QM path integral, the terms in the series expansion became more complicated but this did not affect whether the series converged or not. In particular, the second series for the QM path integral converged at strong coupling just like the basic integral. In going from a QM to a QFT path integral, one expects the same thing to happen: the analytical terms for the second series expansion will be more complicated in the QFT case but it is expected that the series should be absolutely convergent and valid at strong coupling\footnote{By this we mean that the series expansion of correlation functions, which embody the physics, will be absolutely convergent and valid at strong coupling.}. This is discussed further in the conclusion.  

\section{One-dimensional integral containing quadratic plus a quartic term}
In QM or QFT, the Lagrangian is usually composed of a quadratic and interaction part. In the path integral formulation of QM or QFT, we perform a series expansion of the interaction part (in powers of the coupling constant) but leave the quadratic part as is and evaluate it as a Gaussian integral. The elementary (one-dimensional) Gaussian integral is given by
\beq 
\int_{-\infty}^{\infty} dx \,e^{-\frac{1}{2} a \,x^2 + J\,x}= \sqrt{\frac{2\,\pi}{a}}\,e^{\frac{J^2}{2\,a}}
\eeq{Gauss}
where $a$ is a positive real constant and $J$ is a real constant (but the formula works also if one replaces $a$  by $i \,a$ and J by $i \,J$ so that the result extends naturally to the complex numbers). We can generalize the above to an $N$-dimensional Gaussian integral given by
\beq
\int_{-\infty}^{\infty} d\vec{x}\,e^{-\frac{1}{2}\vec{x} \text{A} \,\vec{x} +\vec{J}\vec{x}}=\sqrt{\frac{(2 \pi)^N}{\text{det A}}}\,e^{\frac{1}{2} \vec{J} \,A^{-1} \vec{J}}
\eeq{NGauss}
where $\vec{x}$ is an $N$-dimensional vector and A is a symmetric $N \times N$ matrix. Gaussian integrals in QM and QFT are based in large part on the continuum limit of the above integral \cite{Schwartz}.

Since most of the path integrals we encounter in QM and QFT cannot be solved exactly in closed form, we resort to series expansions. Our goal in this section is to strip away all the technical details associated with path integrals by considering a one-dimensional integral which has an exact analytical expression. We focus on studying two different series expansions of the integral. The advantage here is that we can compare the two series to the exact answer. There is a lot to learn from the series expansion of a one-dimensional integral and the results and conclusions are very likely to be pertinent to the series expansions of path integrals in QM or QFT. The qualitative features of a series expansion such as whether it is an asymptotic series, an absolutely convergent series, how fast it converges (if it does) for different couplings or most importantly, whether it is reliable at strong coupling or not, may not depend too sensitively on whether one is expanding a one-dimensional, multi-dimensional or infinite-dimensional integral.

We begin our study by looking at the following quadratic plus quartic integral 
\begin{align}
 \label{BI}
I=\int_{-\infty}^{\infty} e^{-a\,x^2 -\lambda\, x^4}\,dx
\end{align}
where $a$ and $\lambda$ are positive real numbers. The above integral has an exact analytical expression given by
\begin{align}
\label{Bess}
I= \dfrac{1}{2} e^{\frac{a^2}{8 \lambda }} \sqrt{\frac{a}{\lambda }} \,\text{BesselK}\left[\frac{1}{4},\frac{a^2}{8 \lambda }\right]
\end{align}
where BesselK$\,[n,z]$ is the modified Bessel function of the second kind. This means that we can obtain an exact numerical value for this integral for a given $a$ and $\lambda$ (to arbitrary decimal places). We will perform two different series expansions of this integral and compare the $n$th order result to the exact numerical value given by \reff{Bess}. The first one is a series expansion of the quartic term in powers of $\lambda$. This can be viewed as the analog to the familiar series expansion in perturbative QFT of the interaction term in powers of the coupling constant. In the second one, we leave the quartic term alone and perform a series expansion of the quadratic term (the analog to a series expansion of the quadratic term in QFT which typically includes kinetic and mass terms). This second series expansion is less common in QFT. Most Standard Model calculations are carried out in the weak coupling regime where the first series provides precise results, converges more quickly and where Feynman diagrams are a great tool for both computational and visualization purposes.  Moreover, with the first series, since it is built around perturbations of the free field, one has a clear notion of a particle which is encapsulated in the Feynman propagator. However, as we will clearly see, the first series breaks down completely at strong coupling $\lambda$.  ``Completely breaks down'' means it never comes close to the correct answer at any order (there is no region where it plateaus to a value before diverging as an asymptotic series). In contrast, the second series is very well suited for strong coupling. Not only does it converge to the correct answer at large  $\lambda$, it converges faster the larger $\lambda$ is. Moreover, the second series converges to the correct answer also at weak coupling except that the convergence is much slower. 

\subsection{Series expansion of quartic term in powers of $\lambda$} \label{PL}
A series expansion in powers of $\lambda$ of the quartic term in integral \reff{BI} to order $n$ is given by 
\begin{align}
F_1(n)&=\int_{-\infty}^{\infty} dx\, e^{-a\,x^2} \sum_{j=0}^n \dfrac{(-\lambda\, x^4)^j}{j!}\nonumber\\
&= \sum_{j=0}^n \dfrac{(- \lambda)^j}{j!}\int_{-\infty}^{\infty} dx\, e^{-a\,x^2}x^{4\,j}\nonumber\\
&= \sum_{j=0}^n\, \dfrac{(-1)^j}{j!}\,\Big(\dfrac{\lambda}{a^2}\Big)^j\,a^{-1/2}\,\, \Gamma[1/2 + 2 j]\,.
\label{F1}
\end{align}

We set $a=1$ and consider three different values for $\lambda$: 0.01, 0.1 and 1.0. We evaluate $F_1(n)$ at a given $n$ and compare the answer to the exact value $I$ given by \reff{Bess}. $F_1(n)$ and the exact value $I$ are quoted to eight digit accuracy. Below, we present our results for the three cases of $\lambda$.   

\vspace{14 pt}
\newpage
$\boldsymbol{F_1(n): \lambda=0.01}$

\begin{figure}
\centering
\begin{subfigure}[]{0.48\textwidth}
\centering
	\includegraphics[scale=0.7]{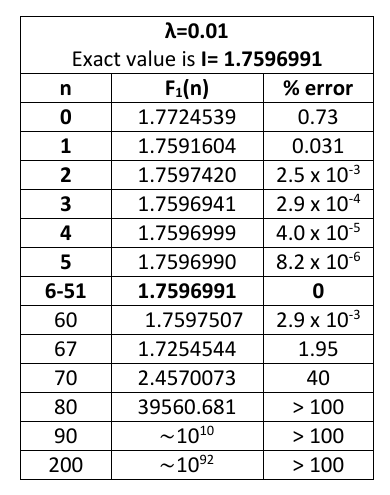}
		\caption{The table contains the numerical values of the series $F_1(n)$ at a given $n$ for $\lambda=0.01$ quoted to eight digit accuracy. The percentage error is equal to the percentage difference between $F_1(n)$ and the exact answer $I$. The series converges quickly to $I$; the error reaches zero to eight digit accuracy at $n=6$ and remains at zero all the way to $n=51$.}       
	\label{Table1}
\end{subfigure}
\hfill
\begin{subfigure}[]{0.48\textwidth}
	\centering
		\includegraphics[scale=0.75]{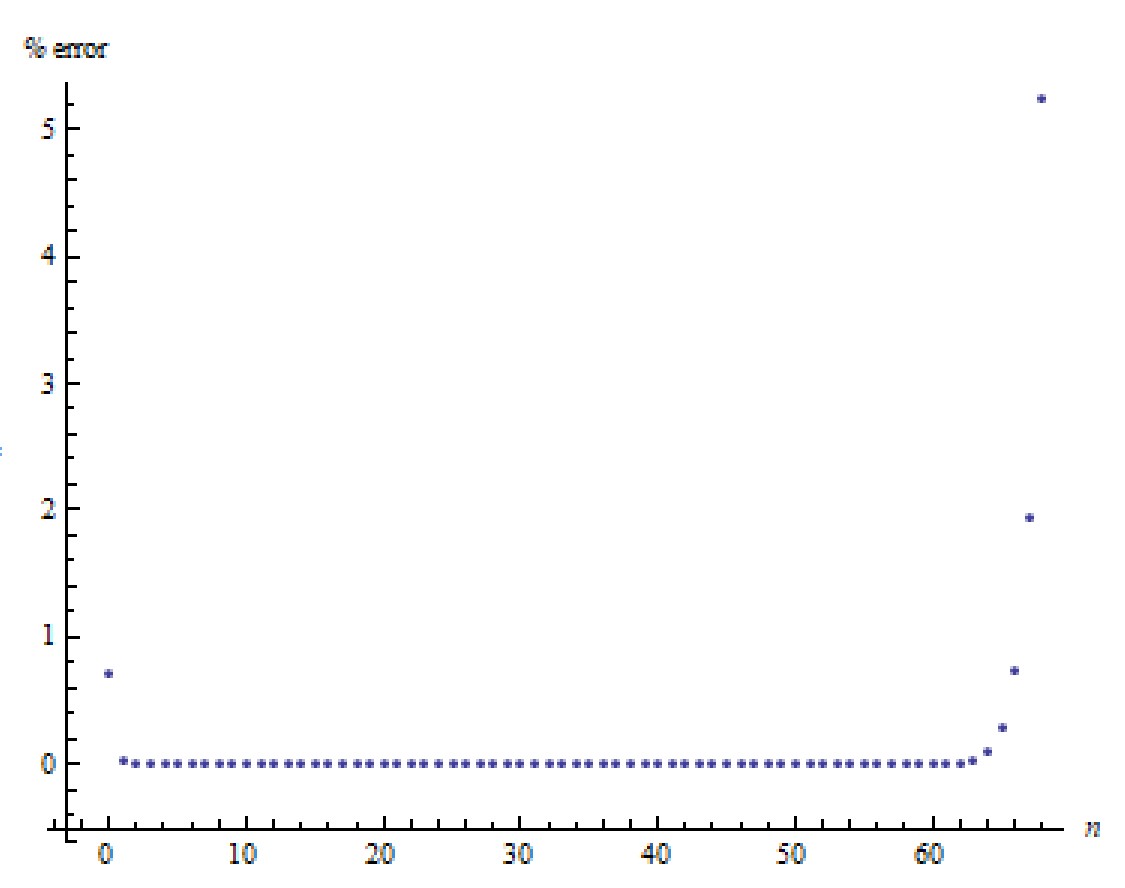}
		\caption{The percentage error vs. $n$ is plotted. It starts off small (0.73 $\%$) and decreases rapidly to zero where it remains over a long range of $n$ before it begins to deviate around $n=65$. It ultimately diverges at larger $n$ values as expected. The distictive feature of this plot is its long plateau region and how early this plateau is reached.}       
	\label{Plot1}
\end{subfigure}
\caption{}
\label{A1}
\end{figure}

The exact value of the original integral \reff{Bess} to eight digit accuracy is I= 1.7596991. The numerical values of $F_1(n)$ as well as the perecntage error (deviation from  $I$) are quoted in Table \ref{Table1}.  At $n=0$, the error is $0.72\%$ and decreases until one reaches $n=6$, where the error is zero ( to within eight digit accuracy). From $n=6$ to $n=51$ the answer remains exactly the same with no error! Then things start to slowly deviate from the correct answer. At $n=60$ the error is still tiny at $0.003 \%$ and at $n=67$ the error is $1.97\%$. However, at $n=70$, the error reaches $40 \%$ and for $n=80, 90$ and $200$ the series is seen to completely deviate from the correct answer and to ultimately diverge (we prove this below).  A plot of $\%$ error vs. n (see figure \ref{Plot1}) shows a long plateau (flat line) region at zero before it starts to deviate from the correct answer. This means that the series $F_1(n)$ is highly reliable at the small value of $\lambda=0.01$. In particular,  the error \textit{decreases} with $n$ before reaching the exact answer and remains  at that value over a long range of $n$. So as long as one does not venture past the plateau (i.e. $n > 70$), $F_1(n)$ matches the original integral to high accuracy.    

The series $F_1(n)$ ultimately diverges since the $n$th term in the summand does not approach zero as $n \to \infty$ i.e. 
\begin{align*}
\lim_{n\to \infty} \dfrac{|(-1)^n|}{n!}\,\Big(\dfrac{\lambda}{a^2}\Big)^n\,a^{-1/2}\, \Gamma[1/2 + 2 n] \to \infty \,.
\end{align*} 
Not only does it not approach zero, but its (absolute) value approaches infinity.  

$\boldsymbol{F_1(n): \lambda=0.1}$

\begin{figure}
\centering
 \begin{subfigure}[]{0.48\textwidth}
	\centering
		\includegraphics[scale=0.8]{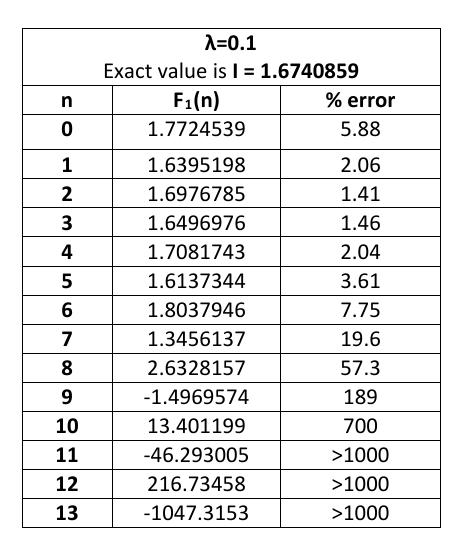}
		\caption{Table of numerical values of $F_1(n)$ for $\lambda=0.1$. For each $n$, the percentage error/deviation from the exact answer $I$ is quoted. The error is a minimum at $n=2$ but never reaches zero. Looking only at the set of numerical values (and not the associated error), there is a cluster of values between 1.6 and 1.7 but it is hard to really say anything more definitive. }       
	\label{Table2}
\end{subfigure}
\hfill
\begin{subfigure}[]{0.48\textwidth}
	\centering
		\includegraphics[scale=0.75]{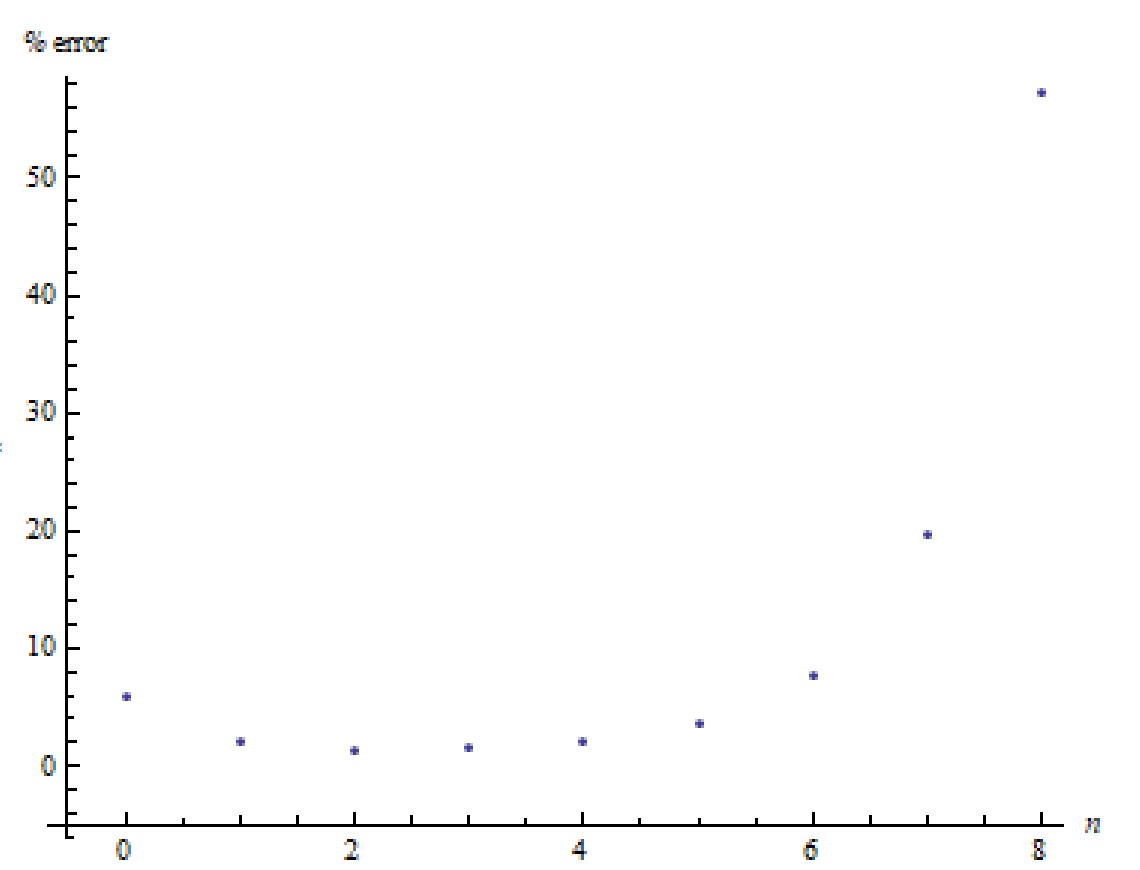}
		\caption{Plot of percentage error vs. $n$ for $F_1(n)$ at $\lambda=0.1$.   The error initially decreases but quickly increases afterwards. There is no plateau region compared to the previous case so that the series cannot be seen to be converging to a particular value.}       
	\label{Plot2}
\end{subfigure}
\caption{}
\label{A2}
\end{figure}

The exact value of the original integral \reff{Bess} to eight digit accuracy is  $I=1.6740859$. The numerical values of $F_1(n)$ as well as the percentage error are quoted in Table \ref{Table2}.  At $n=0$, the error is $5.88\%$ and it decreases to its lowest value of $1.41\%$ at $n=2$. It is less than $5\%$ from $n=1$ to $n=5$ inclusively.  So the series does not completely break down since it gets close to the correct answer near $n=2$ but on the other hand, there is no plateau (flat) region where the error remains small over a long range of $n$ in contrast to the $\lambda=0.01$ case above. Most importantly, the situation does not improve after $n=2$; the error begins to increase and then ultimately diverges. A plot of $\%$ error vs. n (see figure \ref{Plot2}) shows an initial decrease followed immediately by an increase. Therefore, as one goes beyond $n=2$ the results do not improve, they get worse.  Overall, the series is not too reliable at this intermediate value of $\lambda=0.1$.  

\vspace{16pt}
$\boldsymbol{F_1(n): \lambda=1.0}$

\begin{figure}
	\centering
	\begin{subfigure}[]{0.48\textwidth}
	\centering
		\includegraphics[scale=0.8]{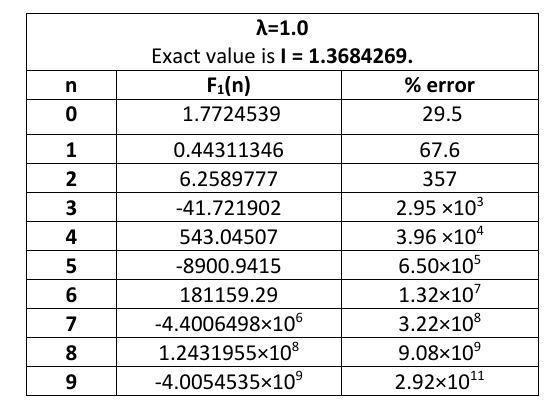}
		\caption{Table of numerical values of $F_1(n)$ for $\lambda=1.0$. Right from the start, the numerical values are way off the expected $I$ value. Things get only worse as $n$ increases.}       
	\label{Table3}
\end{subfigure}
\hfill
\begin{subfigure}[]{0.48\textwidth}
	\centering
		\includegraphics[scale=0.75]{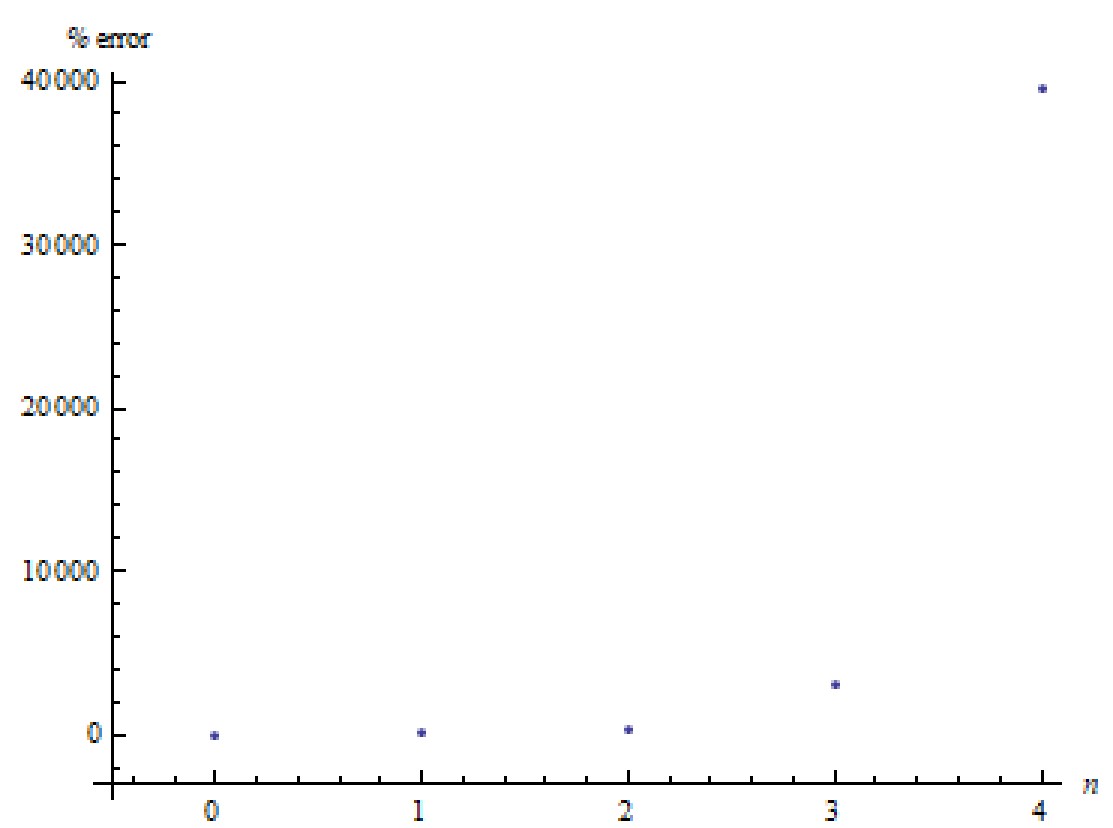}
		\caption{ Plot of percentage error vs. $n$ for $F_1(n)$ at $\lambda=1.0$. The percentage error increases monotonically and drastically so that the series is completely unreliable at $\lambda=1.0$ i.e. at strong coupling. It breaks down completely.}       
	\label{Plot3}
\end{subfigure}
\caption{}
\label{A3}
\end{figure}

The exact value of the original integral \reff{Bess} to eight digit accuracy is  $I=1.3684269$. The numerical values of $F_1(n)$ as well as the percentage error are quoted in Table \ref{Table3}.  At $n=0$, the error is already large at $29.5\%$. The error then just keeps \textit{increasing} so that the series diverges rapidly from the correct answer right from the start. At $n=2$ the error is already over $100\%$.  We plot  $\%$ error vs. n (see figure \ref{Plot3}) only up to $n=4$ because the error is already huge(close to $40,000 \%$). We therefore see that for $\lambda=1$, the series $F_1(n)$ is totally unreliable.

\subsection{Series expansion of quadratic term} \label{IL}
We now perform a series expansion of the quadratic term in integral \reff{BI} to order $n$ while leaving the quartic term as is. This yields
\begin{align}
F_2(n)&=\int_{-\infty}^{\infty} dx\, e^{-\lambda\,x^4} \sum_{j=0}^n \dfrac{(-a\, x^2)^j}{j!}\nonumber\\
&= \sum_{j=0}^n \dfrac{(- a)^j}{j!}\int_{-\infty}^{\infty} dx\, e^{-\lambda\,x^4}\,x^{2\,j}\nonumber\\
&= \sum_{j=0}^n\, \dfrac{(-1)^j}{j!}\,\Big(\dfrac{a^2}{\lambda}\Big)^{j/2}\,\dfrac{1}{2 \,\lambda^{1/4}}\,\, \Gamma[1/4 + j/2]\,.
\label{F2}
\end{align}
The above series $F_2(n)$ is an absolutely convergent series. This can be checked via the ratio test. Let $d_n$ be the $n$th term in the summand. We then obtain that      
\begin{align}
\lim_{n\to \infty}\left|\dfrac{d_{n+1}}{d_n}\right|
=\lim_{n\to \infty}\,\dfrac{1}{n+1}\dfrac{ \Gamma[1/4 + (n+1)/2]}{ \Gamma[1/4 + n/2]}\Big(\dfrac{a^2}{\lambda}\Big)^{1/2} = 0 \,.
\end{align}
Since the limit is zero, which is less than 1, the series passes the ratio test for absolute convergence.  

Again, we set $a=1$ and consider the same three different values for $\lambda$: 0.01, 0.1 and 1.0. The exact value of the integral $I$, given by \reff{Bess}, for each $\lambda$ is of course the same as before (we continue to work at eight digit accuracy).  $F_2(n)$, which is based on a series expansion of the quadratic term, is very different from  $F_1(n)$.  First, $F_2(n)$ always converges to the exact value $I$ asymptotically whereas $F_1(n)$ always diverges asymptotically. Simply put, $F_2(n)$ is an absolutely convergent series whereas $F_1(n)$ is known as an asymptotic series.  Secondly, $F_2(n)$  converges faster at large $\lambda$ than at small $\lambda$. Recall that for $\lambda=1.0$, $F_1(n)$ is completely unreliable: it departs significantly from the correct answer right from the start and then the error \textit{increases} with $n$ so that the series is never close to the correct answer for any $n$.  In contrast, we will see that $F_2(n)$  at $\lambda=1.0$ converges quickly to the exact value $I$ and then remains at that value as $n$ increases. So at large $\lambda$, $F_2(n)$ is very useful whereas $F_1(n)$ breaks down completely. As $\lambda$ decreases, $F_2(n)$ requires more terms to converge to the exact value $I$ i.e. it converges more slowly. We will see that for $\lambda=0.01$, $F_2(n)$ is initially way off target but eventually, at a very large $n$ value, it reaches the exact answer $I$ (as it must, being an absolutely convergent series). So at very small $\lambda$, $F_2(n)$ does not fail but it is not as practical to use as $F_1(n)$, even though the latter is an asymptotic series. $F_1(n)$ at small $\lambda=0.01$ is akin to the series we encounter in perturbative QFT. Very often $n$-loop calculations at weak-coupling involving Feynman diagrams will already match experiments to many decimal places at low values of $n$ (e.g. $1$, $2$ or $3$). The fact that the series is an asymptotic series is inconsequential at weak coupling because we rarely (if ever) require calculations at such high $n$ where the series is no longer reliable. But the most important point here is that at strong coupling (e.g. 
$\lambda=1.0$), the series $F_1(n)$ in powers of $\lambda$ fails completely whereas $F_2(n)$ converges quickly to the correct answer.

\vspace{12pt}
$\boldsymbol{F_2(n): \lambda=0.01}$

\begin{figure}
\centering
\begin{subfigure}[]{\textwidth}
	\centering
		\includegraphics[scale=0.8]{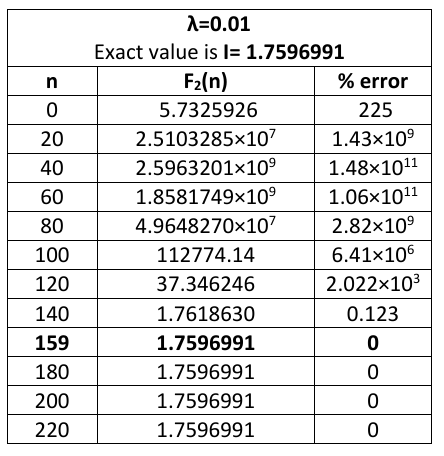}
		\caption{Table of numerical values of $F_2(n)$ for $\lambda=0.01$.  The series converges only at very large $n$; it reaches the correct answer (to eight digit accuracy) at $n=159$ and plateaus at that value for larger $n$.}       
	\label{Table4}
\end{subfigure}
\begin{subfigure}[]{0.3\textwidth}
	\centering
		\includegraphics[scale=0.48]{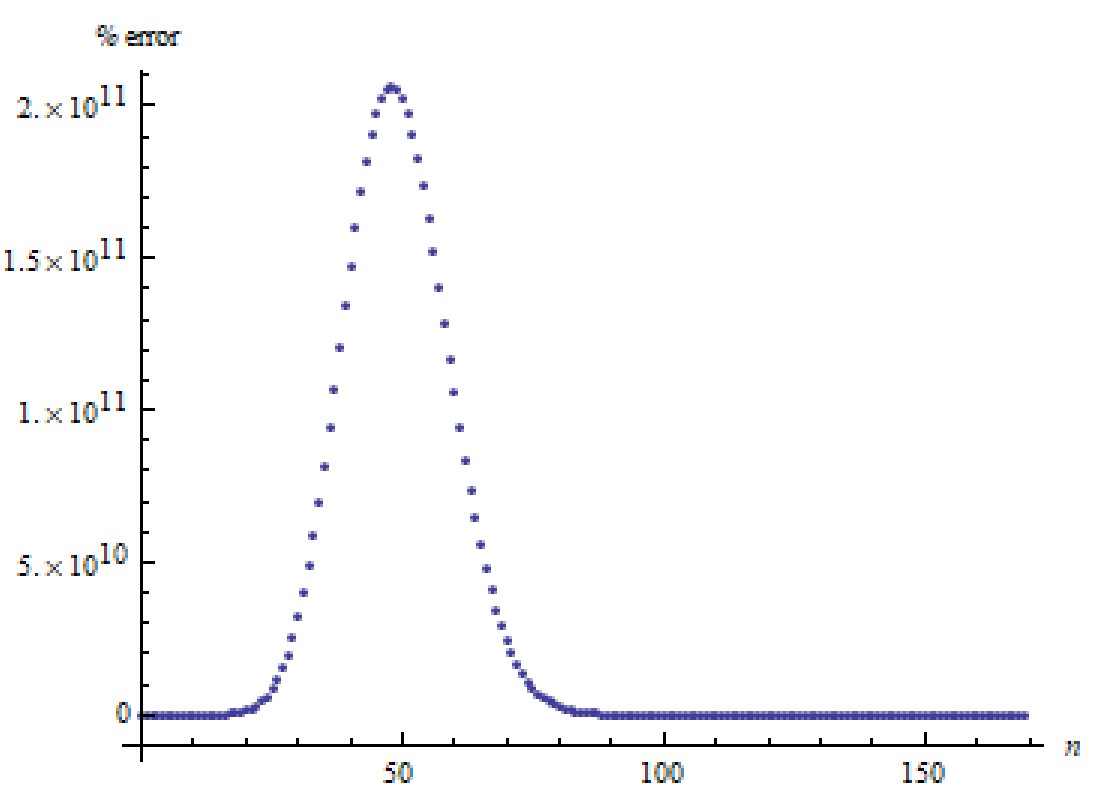}
		\caption{Full plot for $\lambda=0.01$: $\%$ error vs. $n$ over the full range of $n$. The shape of the graph is a hill: the error increases to a peak and then decreases to zero at very large $n$ where it plateaus. The series eventually converges to the correct answer but very slowly and in a non-monotonic fashion.}       
	\label{Plot4A}
\end{subfigure}
\hfill
\begin{subfigure}[]{0.3\textwidth}
	\centering
		\includegraphics[scale=0.46]{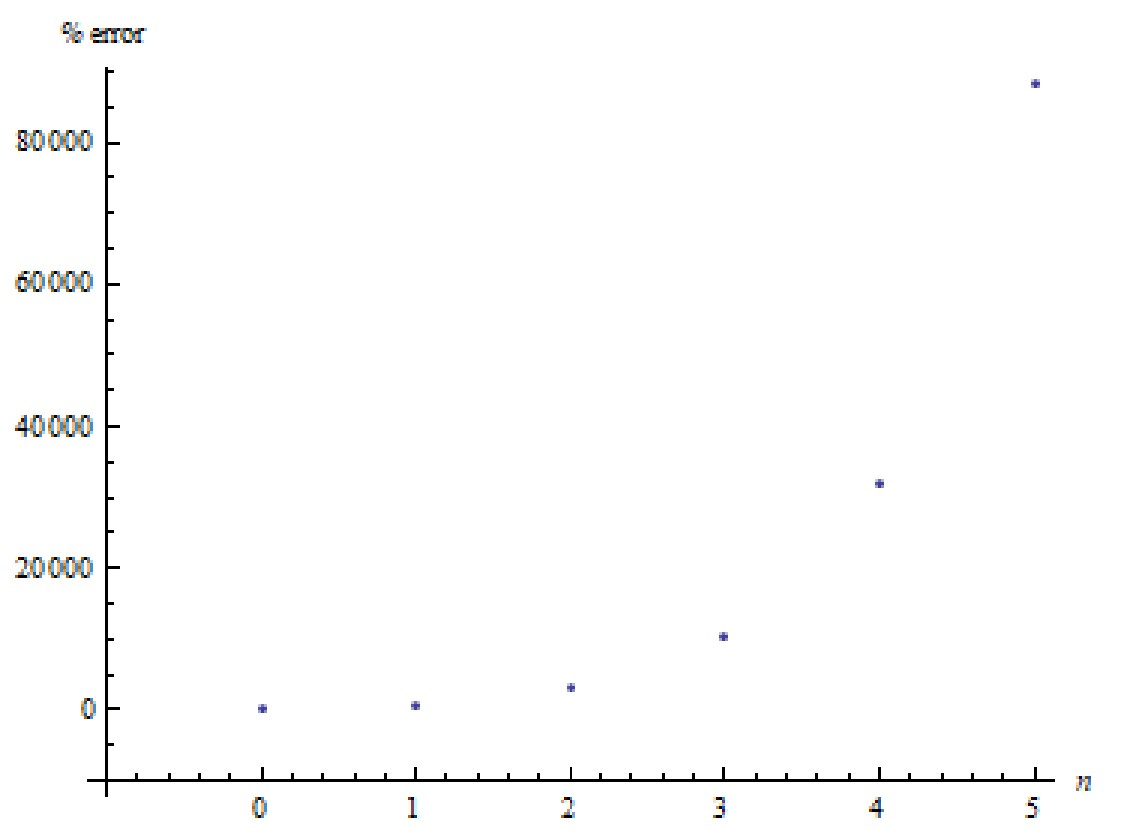}
		\caption{Zooming in on small $n$. Plot of error vs. $n$ up to $n=5$. Initially, the series departs quickly and drastically from the correct answer.}       
	\label{Plot4B}
\end{subfigure}
\hfill
\begin{subfigure}[]{0.3\textwidth}
	\centering
		\includegraphics[scale=0.48]{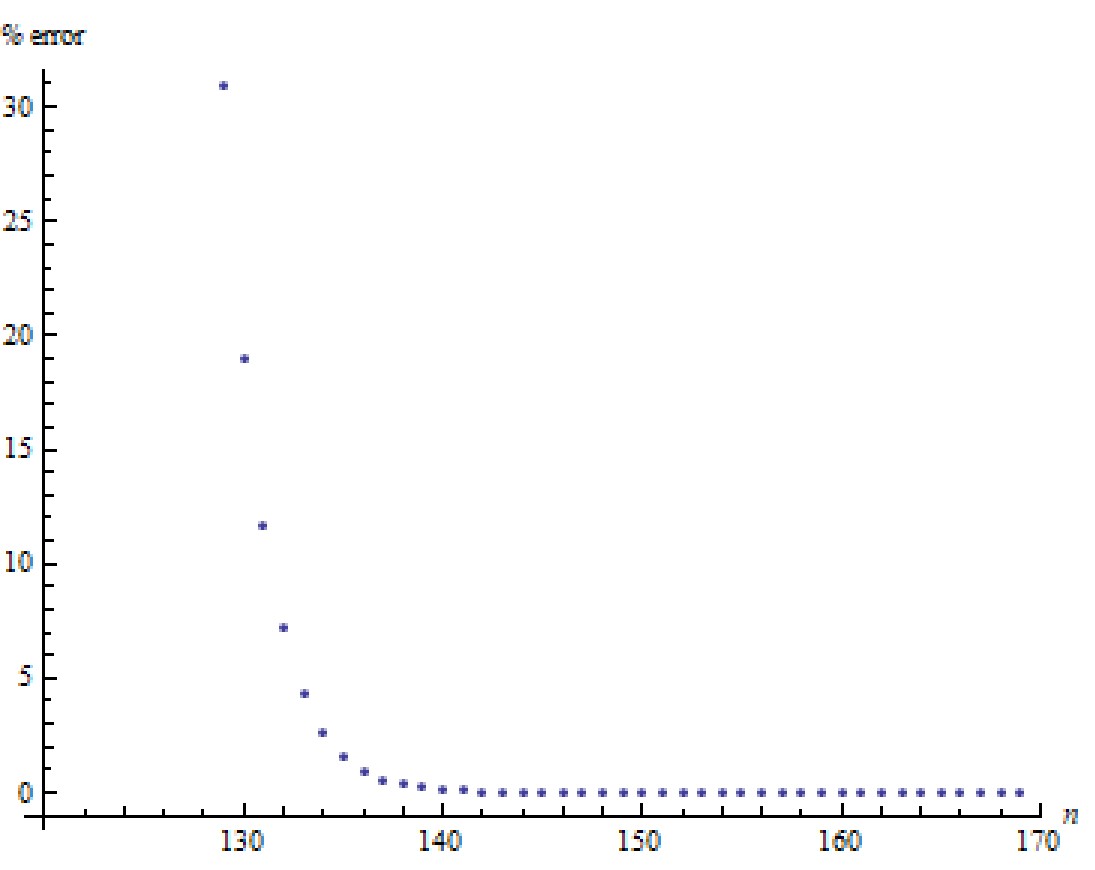}
		\caption{Zooming in at large $n$. Plot of error vs. $n$ for $n>130$. The series eventually converges and plateaus to the correct answer at large $n$.}       
	\label{Plot4C}
\end{subfigure}
\caption{}
\label{B1}
\end{figure}

The exact value to eight digit accuracy is I= 1.7596991. The numerical values of $F_2(n)$ as well as the perecntage error (deviation from  $I$) are quoted in figure \ref{Table4}. The correct answer (to within eight digits) is reached at the large value of $n=159$. The $\%$ error vs. $n $ over the full range of $n$ is plotted in figure \ref{Plot4A}. The graph has the shape of a hill where the error first increases to a peak and then decreases afterwards. In figures \ref{Plot4B} and \ref{Plot4C} we zoom in on small and large $n$ respectively. At small $n$, we see that the error increases quickly and drastically. Eventually, at large $n$,  the error decreases and the correct answer is finally reached. The series eventually converges to the correct value but very slowly and in a non-monotonic fashion. 

\newpage
$\boldsymbol{F_2(n): \lambda=0.1}$

\begin{figure}
	\centering
	\begin{subfigure}[]{0.45\textwidth}
	\centering
		\includegraphics[scale=0.7]{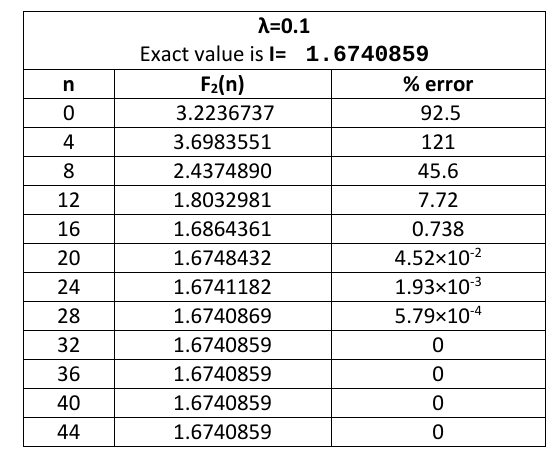}
		\caption{Table of numerical values of $F_2(n)$ for $\lambda=0.1$. The correct answer $I$ is reached at $n=32$ (compared to $n=159$ for $\lambda=0.01$).}       
	\label{Table5}
\end{subfigure}
\hfill
\begin{subfigure}[]{0.45\textwidth}
	\centering
		\includegraphics[scale=0.7]{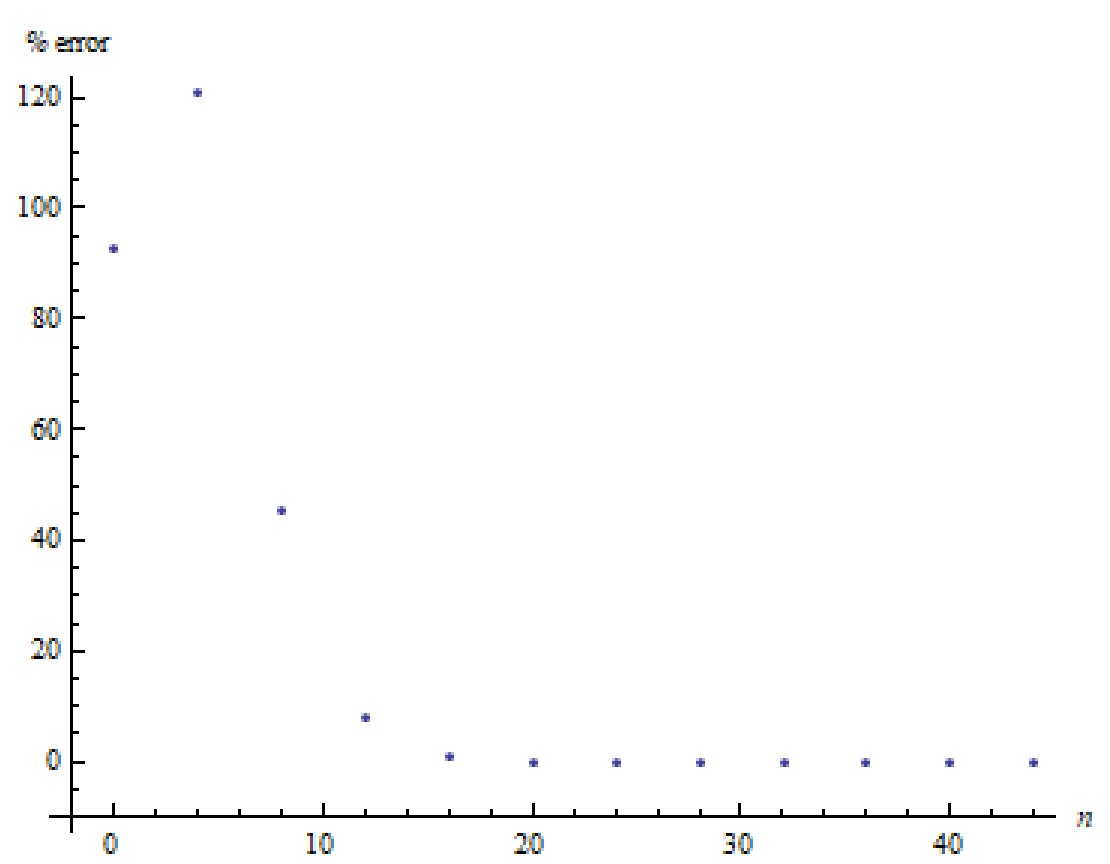}
		\caption{Plot of percentage error vs. $n$ for $F_2(n)$ at $\lambda=0.1$. An important feature of this plot compared to the previous case of $\lambda=0.01$, is the monotonic decrease of the error with $n$ (except for the increase right at the start). The error then plateaus at zero where the correct answer is reached the first time at $n=32$.}
	\label{Plot5}
\end{subfigure}
\caption{}
\label{B2}
\end{figure}

The exact value to eight digit accuracy is  I=1.6740859. The numerical values of $F_2(n)$ as well as the percentage error are quoted in figure \ref{Table5}. The series converges quicker at this larger $\lambda$ compared to the previous case of $\lambda=0.01$; it reaches the correct answer $I$ at $n=32$ (in contrast to $n=159$ for $\lambda=0.01$).  Though the error increases in the small interval from $n=0$ to $n=4$, the trend is a smooth monotonic decrease afterwards. A plot of $\%$ error vs. n (see figure \ref{Plot5}) shows the initial increase followed by an overall monotonic decrease. 

\vspace{12pt}
$\boldsymbol{F_2(n): \lambda=1.0}$

\begin{figure}
	\centering
	\begin{subfigure}[]{0.45\textwidth}
	\centering
		\includegraphics[scale=0.7]{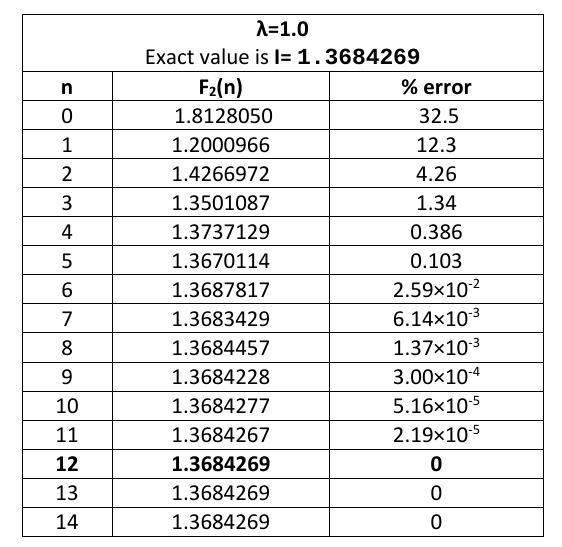}
		\caption{Table of numerical values of $F_2(n)$ for $\lambda=1.0$. The percentage error decreases quickly and reaches zero at $n=12$. }       
	\label{Table6}
\end{subfigure}
\hfill
\begin{subfigure}[]{0.45\textwidth}
	\centering
		\includegraphics[scale=0.7]{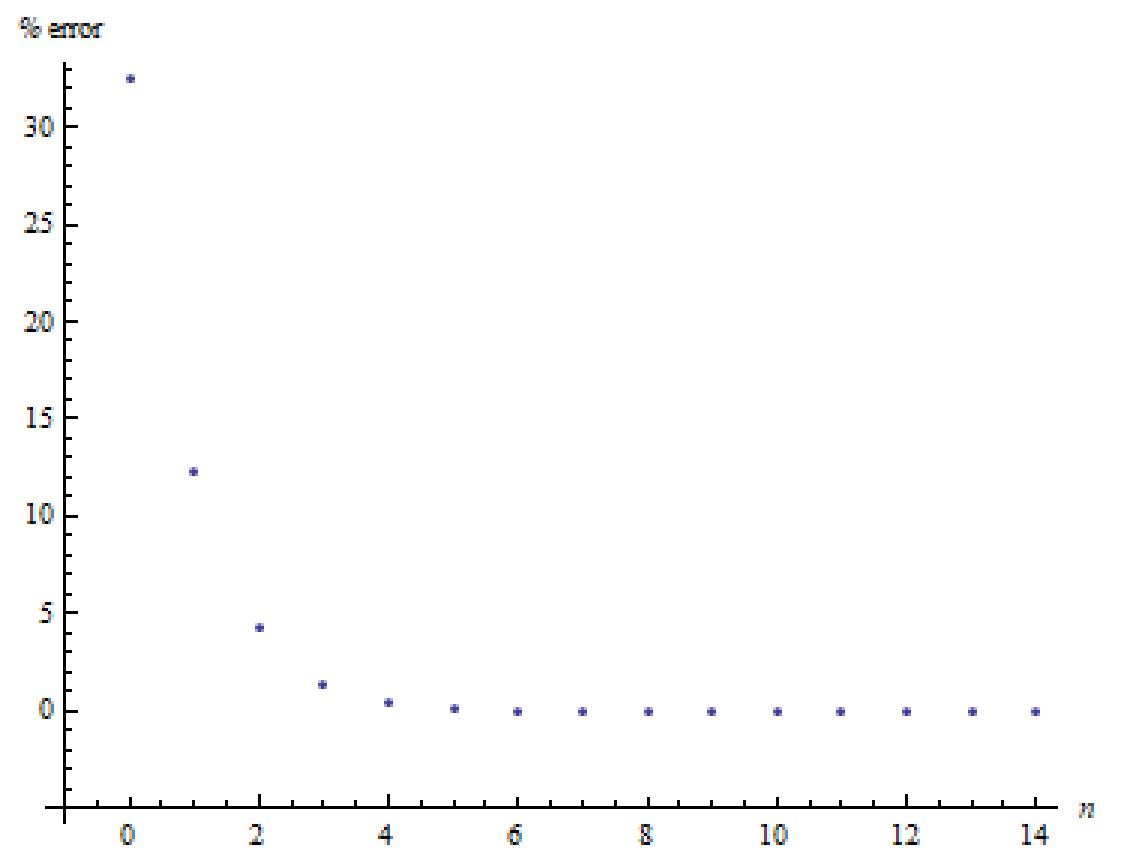}
		\caption{Plot of percentage error vs. $n$ for $F_2(n)$ at $\lambda=1.0$. The error decreases purely monotonically until it reaches zero at $n=12$ where it plateaus.}       
	\label{Plot6}
\end{subfigure}
\caption{}
\label{B3}
\end{figure}

The exact value to eight digit accuracy is  I=1.3684269. The numerical values of $F_2(n)$ as well as the percentage error are quoted in figure \ref{Table6}. The correct answer is already reached at $n=12$  (compared with $n=32$ and $n=159$ for $\lambda=0.1$ and $0.01$ respectively). We see that the series converges fastest at this largest $\lambda$ value (strong coupling). The plot of $\%$ error vs. $n$ (figure \ref{Plot6}) shows a purely monotonic decrease in error.  

\subsection{Asymptotic series: resolving a paradox and the incomplete gamma function $\gamma(z,\alpha)$} \label{CL}
The original integral \reff{BI} yields the finite analytical result \reff{Bess}.  So it is a worthwhile task to try to understand why $F_1(n)$ given by \reff{F1}, which is supposed to represent this original integral, is an asymptotic series.  In other words, it diverges in the limit $n \to \infty$ instead of yielding the expected finite result. In this section we identify the origin of the problem and resolve this long-standing paradox. 

Consider the original integral \reff{BI}. Note that the quartic part $\lambda \,x^4$ dominates over the quadratic part $a\, x^2$ asymptotically as $x \to\pm\infty$. Therefore the integrand is governed by the quartic part asymptotically. The series $F_1(n)$ is precisely a series expansion of this quartic part (in contrast to $F_2(n)$ which is an expansion of the quadratic part).      

We now rewrite the original integral showing explicitly its defined limit: 
\begin{align}
 \label{BI3}
I&=\int_{-\infty}^{\infty} e^{-a\,x^2 -\lambda\, x^4}\,dx\nonumber \\
& =\lim_{\beta \to \infty}\,\int_{-\beta}^{\beta} e^{-a\,x^2 -\lambda\, x^4}\,dx
\end{align}
where $\beta$ is a finite real number so that the integral is performed over finite $x$ before the limit is taken. We now substitute the exact series representation $e^{-\lambda\, x^4}=\lim\limits_{n \to \infty}\sum\limits_{j=0}^n \frac{(-\lambda\, x^4)^j}{j!}$ into \reff{BI3}:
\begin{align}
 \label{BI4}
I_1&=\lim_{\beta \to \infty}\,\int_{-\beta}^{\beta}\, e^{-a\,x^2}\,\lim_{n \to \infty}\sum_{j=0}^n \dfrac{(-\lambda\, x^4)^j}{j!}\,dx\,.
\end{align}
The above expression $I_1$ is exactly the same as the original integral $I$. Note that the infinite $n$ limit is performed \textit{before} the infinite $\beta$ limit. The sum and integral are over finite values and there is no issue if we switch their order but we must maintain the order of the infinite limits. We can therefore rewrite $I_1$ into the equivalent expression;
\begin{align}
 \label{BI5}
I_3&=\lim_{\beta \to \infty}\,\lim_{n \to \infty}\sum_{j=0}^n\,\int_{-\beta}^{\beta}\, e^{-a\,x^2}\, \dfrac{(-\lambda\, x^4)^j}{j!}\,dx\,.
\end{align}
The above expression $I_3$ is again an exact representation of the original integral but it is not $F_1(n)$. The series $F_1(n)$ is defined with the infinite $\beta$ limit moved past the sum to the front of the integral without taking the infinite $n$ limit i.e. 
\begin{align}
F_1(n)&=\sum_{j=0}^n\,\lim_{\beta \to \infty}\,\int_{-\beta}^{\beta}\, e^{-a\,x^2}\, \dfrac{(-\lambda\, x^4)^j}{j!}\nonumber\\
&= \sum_{j=0}^n \dfrac{(- \lambda)^j}{j!}\int_{-\infty}^{\infty} dx\, e^{-a\,x^2}x^{4\,j}\nonumber\\
&= \sum_{j=0}^n\, \dfrac{(-1)^j}{j!}\,\Big(\dfrac{\lambda}{a^2}\Big)^j\,a^{-1/2}\,\, \Gamma[1/2 + 2 j]\,.
\label{F1A}
\end{align}
If we now take the infinite $n$ limit of $F_1(n)$, we have effectively switched the order of the infinite limits in \reff{BI5}. The $n$th term of the series $F_1(n)$ is $\frac{(-1)^n}{n!}\,\Big(\frac{\lambda}{a^2}\Big)^n\,a^{-1/2}\,\, \Gamma[1/2 + 2 n]$ and as we already saw, this diverges as $n\!\to\! \infty$. This implies that $F_1(n)$ is an asymptotic series and its infinite $n$ limit is not equal to our original integral. Why has this happened? The answer is that in switching the order of the infinite limits-- effectively integrating to infinity before summing to infinity-- one cannot capture the asymptotic behaviour of the original integral which is dominated by the quartic part.   The series expansion of the quartic part up to arbitrarily large but finite order $n$ is valid only for finite $x$. Therefore, the limits of integration after performing the series expansion of the quartic part in the integrand should be \textit{finite} instead of infinite (run from $-\beta$ to $\beta$ where $\beta$ is a finite real number). The series will then converge for any given $\beta$. The large but finite $x$ behaviour will now fall off as  $e^{-\lambda\, x^4}$; this will reproduce the original integral with limits of integration running from $-\beta$ to $\beta$. We can then take the limit as $\beta \!\to\!  \infty$ and this should yield exactly the finite analytical result \reff{Bess}. We prove below, via a calculation, that this is indeed the case. 

Before we show the proof, we can use the above analysis to explain why in QFT, perturbative expansions in powers of the coupling yield asymptotic series and how this can be remedied. The interaction part in QFT is composed of a product of more than two fields e.g. in $\lambda \,\phi^4$ theory, the interaction consists of the product of four fields and in QED the interaction $-e \bar{\psi}\gamma^{\mu}A_{\mu}\psi $ consists of a product of three fields. This implies that the interaction part dominates over the quadratic part asymptotically (as the fields tend to infinity) in the integrand of the original path integral. The series expansion of the interaction part alone is only valid for arbitrarily large but finite field (i.e. if we take the limit as the field tends to infinity and then sum the series, it diverges). Therefore, the perturbative series can capture the asymptotics of the original path integral, which is dominated by the interaction part, only if the limits of integration are finite (fields must run from $-\beta$ to $\beta$ where $\beta$ is a finite real number). The series in powers of the (renormalized) coupling at a given $\beta$ will now be absolutely convergent for any value of the coupling. The correct physical result corresponds formally to the $\beta \to \infty$ limit. In practice, this can be reached by increasing $\beta$ until there is no longer any change in the result up to a desired accuracy. 

We now return to our proof. Consider the series $S(\beta)$ obtained from \reff{F1} by replacing $n$ with $\infty$ and replacing the $-\infty$ and $\infty$  limits of integration by the finite real values $-\beta$ and $\beta$ respectively: 
\begin{align}
S(\beta)&=\int_{-\beta}^{\beta} dx\, e^{-a\,x^2} \sum_{j=0}^{\infty} \dfrac{(-\lambda\, x^4)^j}{j!}\nonumber\\
&= \sum_{j=0}^{\infty} \dfrac{(- \lambda)^j}{j!}\int_{-\beta}^{\beta} dx\, e^{-a\,x^2}\,x^{4\,j}\nonumber\\
&= \sum_{j=0}^{\infty}\, \dfrac{(-\lambda)^j}{j!}\,a^{-2 j-\frac{1}{2}} \,\gamma(2 j+\tfrac{1}{2},\,a\, \beta^2)
\label{SBeta}
\end{align}
where the incomplete gamma function $\gamma(z,\alpha)$ is defined as
\beq
\gamma(z,\alpha)=\int_0^{\alpha} e^{-t} \,t^{z-1}\, dt\,.
\eeq{IGamma}  
As we previously explained, we need to perform the infinite sum over $j$ in \reff{SBeta} \textit{before} taking the $\beta \to \infty$ limit. It turns out that the sum in \reff{SBeta} does not yield any known analytical expression in terms of $\beta$ (and the constants $a$ and $\lambda$). However, there is a way around this by employing a series representation of the incomplete gamma function \cite{Grad8354}:
\beq
\gamma(z,\alpha)=\sum_{p=0}^{\infty} \dfrac{(-1)^p\,\alpha^{z+p}}{p!\,(z+p)}\,.
\eeq{RepG}
It follows that 
\beq
\gamma(2 j+\tfrac{1}{2},\,a\, \beta^2)=\sum_{p=0}^{\infty} \dfrac{(-1)^p}{p!}\dfrac{(a\,\beta^2)^{2\,j+\frac{1}{2}+p}}{(2\,j+\frac{1}{2}+p)}\,.
\eeq{RepG2}
Substituting \reff{RepG2} into \reff{SBeta} we obtain
\begin{align}
S(\beta)&=\sum_{p=0}^{\infty} \dfrac{(-a)^p}{p!} \sum_{j=0}^{\infty}\, \dfrac{(-\lambda)^j}{j!} \,\dfrac{(\beta)^{4\,j+1+2\,p}}{(2\,j+\frac{1}{2}+p)}\,.
\label{SBeta2}
\end{align}
The important point is that sum over $j$ can now be performed and yields
\beq 
\sum_{j=0}^{\infty}\, \dfrac{(-\lambda)^j}{j!} \,\dfrac{(\beta)^{4\,j+1+2\,p}}{(2\,j+\frac{1}{2}+p)}=\frac{1}{2}\, \lambda ^{-\frac{p}{2}-\frac{1}{4}}\, \gamma \left(\frac{p}{2}+\frac{1}{4},\beta ^4 \,\lambda \right)\,.
\eeq{JSum}
Substituting \reff{JSum} into \reff{SBeta2} yields
\begin{align}
S(\beta)=\sum_{p=0}^{\infty} \dfrac{(-a)^p}{2\,p!} \,\, \lambda ^{-\frac{p}{2}-\frac{1}{4}}\,\, \gamma \left(\frac{p}{2}+\frac{1}{4},\beta ^4 \lambda \right)\,.
\label{SBeta3}
\end{align}
Since we have already performed the sum over $j$, we can now take the limit as $\beta \to \infty$. This yields
\begin{align}
\lim_{\beta \to \infty}S(\beta)&=\sum_{p=0}^{\infty} \dfrac{(-a)^p}{2\,p!} \,\, \lambda ^{-\frac{p}{2}-\frac{1}{4}}\,\,\lim_{\beta \to \infty}\, \gamma \left(\frac{p}{2}+\frac{1}{4},\,\beta ^4 \lambda \right)\nonumber\\
&=\sum_{p=0}^{\infty} \dfrac{(-a)^p}{2\,p!} \,\, \lambda ^{-\frac{p}{2}-\frac{1}{4}}\,\,\Gamma\Big(\frac{p}{2}+\frac{1}{4}\Big)\nonumber\\
&= \dfrac{1}{2} \,e^{\frac{a^2}{8\, \lambda }} \,\sqrt{\frac{a}{\lambda }} \,\,\text{BesselK}\left[\frac{1}{4},\frac{a^2}{8 \,\lambda }\right]
\label{SBeta4}
\end{align}
where we used
\beq
\lim_{\beta \to \infty} \gamma \left(\frac{p}{2}+\frac{1}{4},\,\beta ^4 \lambda \right)= \Gamma\Big(\frac{p}{2}+\frac{1}{4}\Big)
\eeq{BetaLimit}
with $\Gamma(z)$ the gamma function. The analytical result in the last line of \reff{SBeta4} containing a Bessel function of $a$ and $\lambda$ matches exactly the result \reff{Bess} obtained by integrating directly the original integral.   

\subsection{A series $S(n,\beta)$ valid at both strong and weak coupling $\lambda$}\label{SNB}
If the sum in $S(\beta)$ appearing in \reff{SBeta} goes up to $n$ instead of infinity we obtain the series
\begin{align}
S(n,\beta)&= \sum_{j=0}^{n}\, \dfrac{(-\lambda)^j}{j!}\,a^{-2 j-\frac{1}{2}} \,\gamma(2 j+\tfrac{1}{2},\,a\, \beta^2)
 \nonumber\\
&=\, \dfrac{\gamma(\tfrac{1}{2},\,a\, \beta^2)}{a^{\frac{1}{2}}} - \lambda\,\dfrac{\gamma(\tfrac{5}{2},\,a\, \beta^2)}{a^{\frac{5}{2}}}  +\dfrac{\lambda^2}{2}\,\dfrac{\gamma(\tfrac{9}{2},\,a\, \beta^2)}{a^{\frac{9}{2}}} -\dfrac{\lambda^3}{3!}\,\dfrac{\gamma(\tfrac{13}{2},\,a\, \beta^2)}{a^{\frac{13}{2}}} + ...
\label{SnBeta}
\end{align}

As we did for the series $F_1(n)$, we make a table of values for $S(n,\beta)$ as a function of $n$ for $\lambda=0.01$, 
$\lambda=0.1$ and $\lambda=1$ (again, we set $a=1$). We show results for $\beta=1,2,3$ and $4$. It turns out that we do not in practice require to have $\beta$ go all the way to infinity (or very large) to obtain convergence;  for $\lambda\ge 0.01$ going up to $\beta=4$ was sufficient  (i.e. larger $\beta$ simply gave the same result). 
\begin{table}
	\includegraphics[scale=0.685]{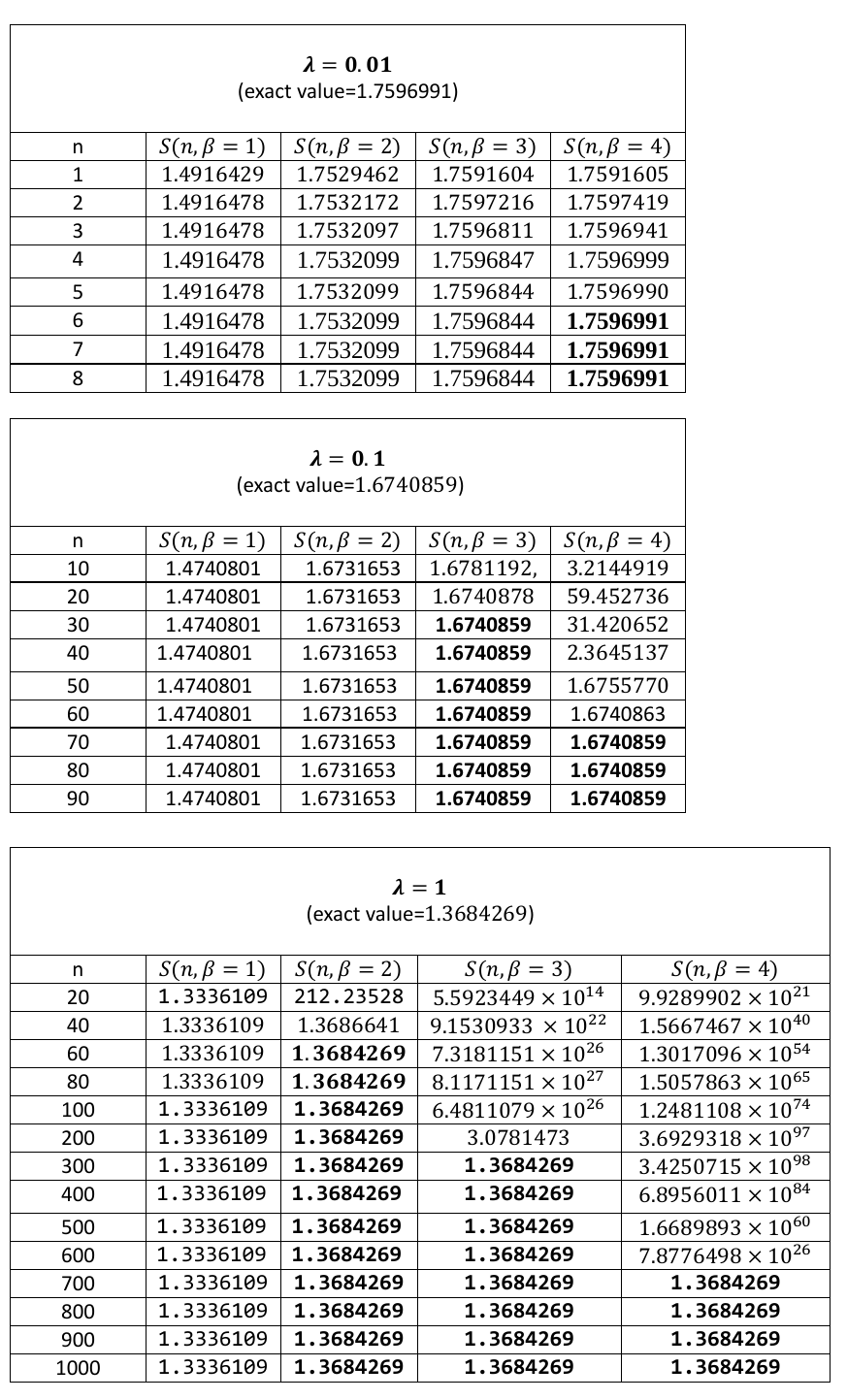}
		\caption{Table of values for the series $S(n,\beta)$ for $\lambda=0.01$, $\lambda=0.1$ and $\lambda=1$ (the constant $a$ is set to unity and we quote results for $\beta=1,2,3$ and $4$.). In all three cases, the series converges to the value of the original integral \reff{BI} (to within eight digits). As $\lambda$ increases, one needs to sum more terms in the series to reach convergence and it occurs at lower values of $\beta$.}
	\label{Incomplete}
\end{table}
The series $S(n,\beta)$ given by \reff{SnBeta} has a remarkable property: though the series is in powers of $\lambda$ and hence looks likes a conventional perturbative expansion, it yields the correct finite result regardless of how large $\lambda$ is. As one can see from Table \ref{Incomplete} the series converged to the correct value (to eight digit accuracy) for all three values of $\lambda$: at small coupling $\lambda=0.01$, at intermediate coupling  $\lambda=0.1$ and at large coupling $\lambda=1$ . In other words, the series is valid for both ``weak'' and ``strong'' coupling.  This is in contrast to the more conventional perturbative series $F_1(n)$ that we studied previously in section \ref{PL} which was also a series in powers of $\lambda$. $F_1(n)$ was an asymptotic series, but nonetheless, for $\lambda=0.01$ it had a plateau region where it converged to the correct value before it diverged asymptotically (see 
figure \ref{Plot1}).  However, for the larger values of $\lambda=0.1$ and $\lambda=1$ there is no plateau region where the series converged to the correct value (see figures \ref{Plot2} and \reff{Plot3} respectively). In particular, for $\lambda=1$, the series departed significantly from the correct value before it diverges.

The series $S(n,\beta)$ is an absolutely convergent series and from Table \ref{Incomplete}, we see that as $\lambda$ increases, one needs to sum more terms (i.e. the order $n$ has to be larger) to converge to the correct value. For $\lambda=0.01$ the $n$'s are in the single digits,  for $\lambda=0.1$ the $n$'s are in the ten digits, whereas for $\lambda=1$ the $n$'s are in the hundred digits. The important point is that at large coupling $\lambda=1$, the series does eventually converge (albeit slowly).  So going from an asymptotic series to an absolutely convergent series has an important physical ramification: one can do strong coupling physics with a series expressed in powers of the coupling. 

\subsection{Weak-strong coupling duality}

We saw that though the series $S(n,\beta)$ given by \reff{SnBeta} is expressed in powers of the coupling $\lambda$, it is an absolutely convergent series valid at any value (weak or strong) of the coupling $\lambda$. From Table \ref{Incomplete}, we saw that as $\lambda$ increased from $0.01$ to $1$, one needed to sum considerably more terms (i.e. larger $n$) in order for the series to converge to the correct value . This means it converges quickly at weak coupling but slowly at strong coupling $\lambda$. Recall that $F_2(n)$ given by \reff{F2} was was also an absolutely convergent series but it is expressed in inverse powers of $\lambda$ so that the reverse happens: it converges quickly at strong coupling and slowly at weak coupling $\lambda$.  The series $S(\beta)$ in powers of $\lambda$ given by \reff{SBeta} is simply $S(n,\beta)$ with $n \to \infty$. By using the series representation  \reff{RepG} of the incomplete gamma function we then showed  in \reff{SBeta3} that $S(\beta)$ can be expressed in inverse powers of $\lambda$. Taking the limit as $\beta \to \infty$ in \reff{SBeta4}, we obtained the folowing series:
\begin{align}
\sum_{p=0}^{\infty} \dfrac{(-a)^p}{2\,p!} \,\, \lambda ^{-\frac{p}{2}-\frac{1}{4}}\,\,\Gamma\Big(\frac{p}{2}+\frac{1}{4}\Big)\,.
\label{F2Copy}
\end{align}
But the above series in inverse powers of $\lambda$ corresponds exacly to $F_2(n)$ given by \reff{F2} (assuming we sum $p$ 
in \reff{F2Copy}to $n$ instead of infinity). So we have shown that a series in powers of $\lambda$, which converges more quickly at weak coupling, is dual to a series in inverse powers of $\lambda$, which converges more quickly at strong coupling. We therefore have a weak-strong coupling duality. 

\subsection{How Dyson's argument on convergence was circumvented}
\begin{figure}[t]
\centering
\begin{subfigure}[]{0.48\textwidth}
\centering
	\includegraphics[scale=0.45]{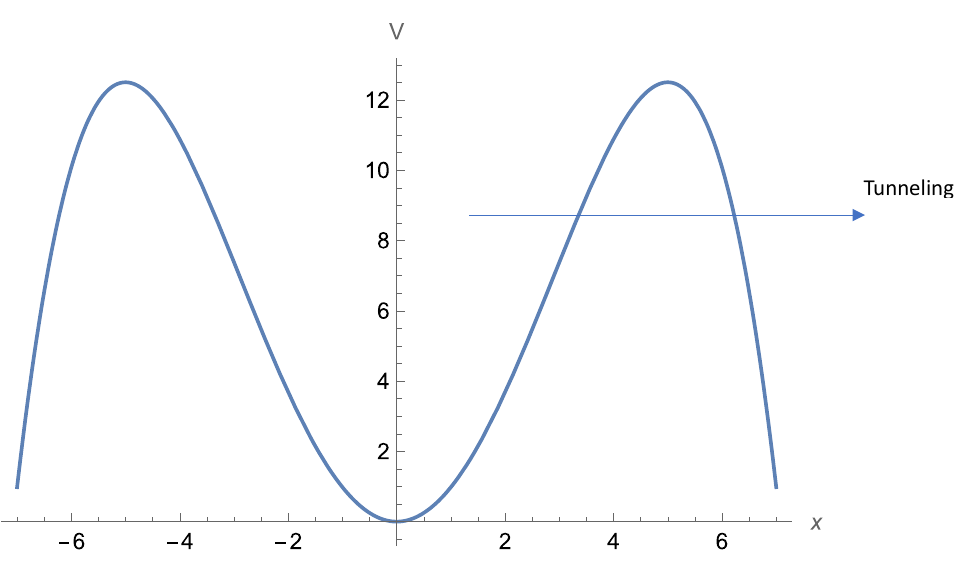}
		\caption{Potential  $V(x) =\lambda\, x^4 +a \,x^2$ with negative $\lambda$ ($a$ is positive). This leads to tunneling and the so-called resonant state does not correspond to a regular real energy. Therefore, the series with negative $\lambda$ about $\lambda=0$ cannot be convergent regardless of how small is its absolute value. Dyson's argument then implies that the series about $\lambda=0$ with positive $\lambda$ cannot be convergent either (which agrees with the fact that it is an an asymptotic series).}    \label{Tun}
\end{subfigure}
\hfill
\begin{subfigure}[]{0.48\textwidth}
	\centering
		\includegraphics[scale=0.45]{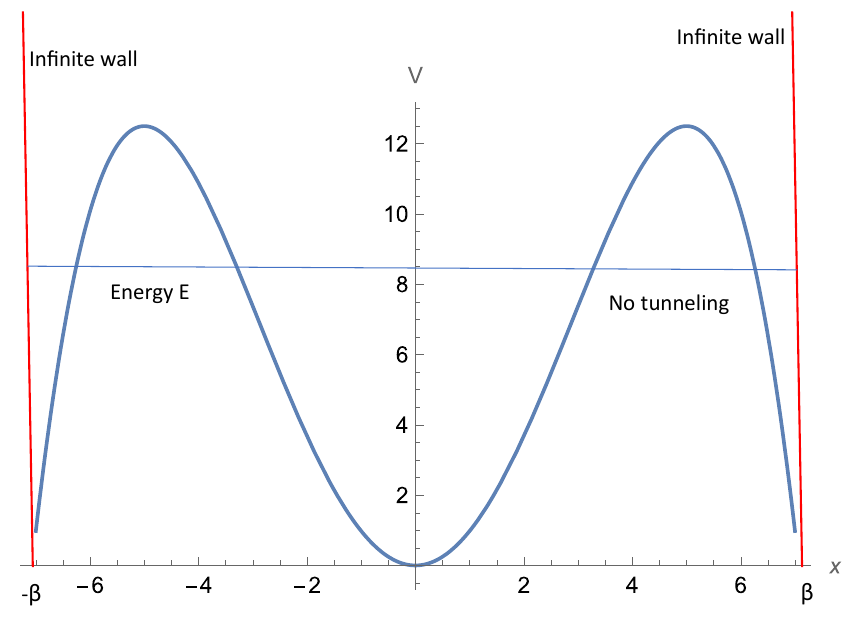}
		\caption{Potential  $V(x) =\lambda\, x^4 +a \,x^2$ with negative $\lambda$ ($a$ is positive) but with infinite walls at $x=\pm \beta$ where $\beta$ is a real number. There is no tunneling and the potential yields a regular real energy regardless of how large $\beta$ is. Therefore, the series with negative $\lambda$ about $\lambda=0$ can converge now to a real energy so that there is no longer any argument against the convergence of the series with a positive $\lambda$. The presence of the walls circumvents Dyson's argument since there is no longer any tunneling.}       
	\label{Tun2}
\end{subfigure}
\begin{subfigure}[]{0.48\textwidth}
		\includegraphics[scale=0.45]{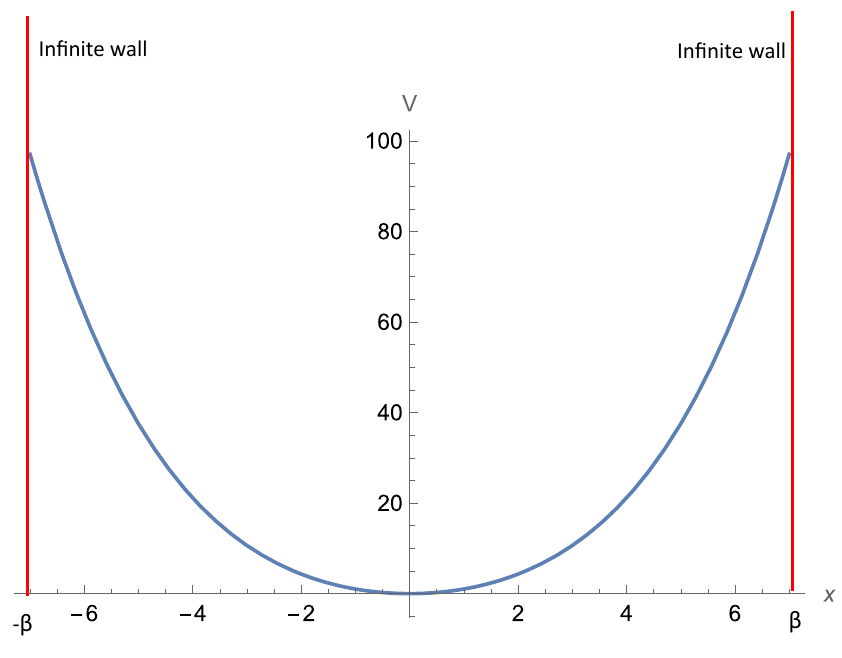}
		\caption{Potential  $V(x) =\lambda\, x^4 +a \,x^2$ with positive $\lambda$ ($a$ is positive) but with infinite walls at $x=\pm \beta$ where $\beta$ is a real number. This yields a convergent series and we recover the energies of the original potential (no walls) as $\beta$ is made arbitrarily large.}       
	\label{PotWalls}
\end{subfigure}
\caption{}
\label{Potun}
\end{figure}

There is a well-know argument made by Dyson \cite{Dyson} as to why a perturbative series expansion about $\lambda=0$ in powers of $\lambda$ should yield an asymptotic series regardless of how small $\lambda$  is (see \cite{Strocchi,Marino1} for a discussion). Dyson's argument explains why the series $F_1(n)$ is an asymptotic series. The argument is that if it were  absolutely convergent then it would follow that the series about $\lambda=0$  would also be convergent for negative $\lambda$ assuming its absolute value is sufficiently small. However,  this clearly cannot occur because the original integral given by \reff{BI} clearly diverges for negative $\lambda$ \textit{regardless of how small is its absolute value}. Hence, the series $F_1(n)$ with positive $\lambda$ must be an asymptotic series which is what we found. In quantum mechanics, Dyson's argument would be that a negative $\lambda$ corresponds to a potential that leads to a qualitatively different physical system. The potential $V(x) =\lambda\, x^4 +a \,x^2$  now yields tunneling (see fig. \reff{Tun}) and this does not lead to a regular energy state but to a resonant state (this could be viewed as having an energy with both real and imaginary parts \cite{Marino1}). So the series with a negative $\lambda$, regardless of how small is its absolute value, cannot be expected to converge or else it would yield a regular real energy. Hence, it must diverge and this implies that the small positive $\lambda$ case must diverge also. How does the series $S(n,\beta)$ circumvent Dyson's argument? The answer is that $x$ ranges between the finite values of $-\beta$ and $\beta$ so that our original integral \reff{BI} confined to those limits does not diverge when $\lambda$ is negative. So we can no longer use the negative $\lambda$ case to argue that the series expansion with positive $\lambda$ should diverge; in fact we saw that $S(n,\beta)$ is an absolutely convergent series. The quantum mechanical case is interesting. Since $x$ ranges between the finite values of $-\beta$ and $\beta$, the particle must be confined to that region. This requires placing infinite walls at $x=\pm \beta$ in the potential $V(x)$. In the negative $\lambda$ case, the infinite walls prevent tunneling from occuring (see fig. \ref{Tun2}) and this is true \textit{regardless of how large $\beta$ is}. So a negative $\lambda$ no longer corresponds to a tunneling scenario and the series can now converge to a real energy. One can no longer use the negative $\lambda$ case to argue that the series with positive $\lambda$ should diverge; in fact we expect it to converge absolutely. The potential $V(x)$ with positive $\lambda$ must have walls also (see fig. \ref{PotWalls}) but the walls do not cause a problem since $\beta$ can be made arbitrarily large; we then recover the same energies as the original potential (which has no walls). 

\section{Quantum mechanical path integrals containing quadratic and quartic terms} \label{QMPI}

In the previous section, we explored two different series expansions of a basic (one-dimensional) integral whose integrand was the  exponential of a quadratic plus quartic term. In one case, we expanded the quartic term and in the second case we expanded the quadratic term. The basic integral highlighted the salient features of the two series expansions with minimal technical detail. We now consider a more physical context by evaluating path integrals in ordinary (non-relativistic) quantum mechanics (QM) in one spatial dimension $x$. The kinetic term in the Lagrangian is automatically quadratic and we will consider a potential $V(x)$ that contains both a quadratic term (the harmonic oscillator) as well as a quartic term (the quantum system is commonly referred to as the quartic anharmonic oscillator).  Both terms in the potential are of significant physical interest. The QM harmonic oscillator is at the heart of QFT; it is often said that ``Quantum field theory is just quantum mechanics with an infinite number of harmonic oscillators'' \cite{Schwartz} (see also \cite{Zee}). The quartic term can be viewed as the QM analogue of a $\lambda\,\phi^4$ interaction in QFT which is of interest because it appears in the Higgs sector of the standard model and is usually the first interacting theory one encounters in studying QFT and in applying Feynman diagrams. 

One difference between QM and QFT path integrals are the boundary conditions. In QFT, the fields are usually assumed to approach zero asymptotically. In QM, the path starts at a finite $x_a$ (at time $t_a$) and ends at a finite $x_b$ (at time $t_b$). A common method used for doing path integrals in QM is to solve first for the classical path $x_c(t)$ and consider fluctuations $y(t)$ about it. The spatial end points $x_a$ and $x_b$ then appear only in the classical action $S_c$ which is factored out of the path integral. In particular, this is how the amplitude for the harmonic oscillator is usually derived. However, we will not be using this method and will solve the harmonic oscillator and other path integrals directly without extracting the classical action. The spatial end points $x_a$ and $x_b$ then appear in a linear term $f\,x$ where $f$ can be $x_a$ or $x_b$. This method then lends itself naturally to the Gaussian case of the forced harmonic oscillator (harmonic oscillator plus additional term $J(t) \,x$ in the action) because we can absorb $J(t)$ into $f$. We are not interested in the forced harmonic oscillator as a physical system per say since at the end of calculations we set $J(t)$ to zero. It simply serves as a generating functional for the first series and $J(t)\,x$ can be viewed simply as the source term that we are familiar with from QFT. 

For the second series, the path integral for the quartic part plus source term yields a generating functional $Z[J]$ expressed as products of generalized hypergeometric functions. To generate the terms in the second series we therefore take functional derivatives of hypergeometric functions (in contrast to Gaussians for the first series). The time interval is divided into $N$ equal parts and we obtain a discretized $N-1$-dimensional path integral. We derive analytical expressions for the $n$th order terms in the series as functions of $N$. Expressions for the first three orders are worked out explicitly. The second series in the QM case is absolutely convergent for any given $N$ and any coupling $\lambda$.     

In contrast to the basic integral of the last section, we do not have a closed form analytical solution of the path integral to compare the series to. The exact path integral will therefore be evaluated numerically to a given precision and the series will then be compared to this numerical value. The original path integral is not well suited for direct numerical computations because it has a highly oscillating integrand.  The Euclidean path integral, obtained after a Wick rotation, converges faster and there are no computational issues that arise due to highly oscillating integrands. To compare the series to an exact numerical answer (to a given precision), it is therefore better to use the Euclidean path integral. Nonetheless, we begin with the original path integral and evaluate the amplitude for the harmonic oscillator directly. We then consider the Euclidean version with quartic term and consider two different series expansions of it. The series expansions are then compared to a direct numerical computation of the Euclidean path integral.

\subsection{Expression for amplitude for a potential containing harmonic oscillator plus quartic term} \label{AnHarmonic}
  
Consider a non-relativistic particle of mass $m$ moving in one spatial dimension $x$ under a potential $V(x,t)$. The particle moves from an initial position $x_a$ at time $t_a$ to its final position $x_b$ at time $t_b$. The quantum mechanical path integral or amplitude $K$ is defined as 
\beq
K= \int \mathcal{D} x(t) \, e^{i\,S/\hbar}
\eeq{Ordinary}
where $ \mathcal{D} x(t)$ means we integrate over all possible paths that start at point $(x_a,t_a)$ and end at point  $(x_b,t_b)$. The action is given by
\beq
S=\int_{t_a}^{t_b} \Big( \dfrac{1}{2}\,m\,\dot{x}^2-V(x,t)\Big)\,dt 
\eeq{S1}
where $\dot{x}=dx/dt$. We will consider in this section a time-independent potential composed of a harmonic oscillator plus a quartic term: 
\beq
V(x)=\dfrac{1}{2}\,m\,\omega^2\,x^2 +\lambda\,x^4
\eeq{Pot}
where $\lambda$ and $\omega$ are positive constants. To evaluate the path integral, we dicretize the time. We divide the time interval $T=t_b-t_a$ into $N$ equal intervals $\epsilon$. We therefore have the relation 
\beq
\epsilon=\dfrac{t_b-t_a}{N}=\dfrac{T}{N}\,.
\eeq{eps}   
We replace $dt$ by $\epsilon$ and this becomes exact in the contimuum limit where $N$ approaches infinity. Let $i$ be an integer that runs from $0$ to $N$ inclusively so that $t_i=t_a + i\,\epsilon$. Let $x_i=x(t_i)$.  The endpoints are $x_0=x_a$ and $x_N=x_b$. 
The discretized version of the action \reff{S1} with potential \reff{Pot} is then given by 
\begin{align}
S&=\int_{t_a}^{t_b} \Big( \dfrac{1}{2}\,m\,\dot{x}^2-\dfrac{1}{2}\,m\,\omega^2\,x^2 -\lambda\,x^4\Big)\,dt \nonumber\\
&=\sum_{i=1}^{N}\Big( \dfrac{1}{2} \,m\,\dfrac{(x_i-x_{i-1})^2}{\epsilon^2} -\dfrac{1}{2} \,m\,\omega^2\,x_i^2 - \lambda \,x_i^4\Big) \epsilon\nonumber\\
&=\dfrac{m}{2 \,\epsilon}\sum_{i=1}^{N} \Big(\,(x_i - x_{i-1})^2 -\omega^2\,x_i^2\,\epsilon^2-\dfrac{2 \lambda}{m} \,x_i^4\,\epsilon^2\,\Big)
\label{DAction}
\end{align}
where we replaced $\dot{x}^2$ by its discretized version $\tfrac{(x_i-x_{i-1})^2}{\epsilon^2}$. In the path integral, we do not integrate over the fixed end points $x_0$ and $x_N$. We therefore expand the sums to extract out terms containing the end points: 
\begin{align}
&\sum_{i=1}^{N} (x_i - x_{i-1})^2=x_0^2+x_N^2 -2 x_0 \,x_1-2\,x_N\,x_{N-1} +\sum_{i=1}^{N-1} 2 \,x_i ^2 -\sum_{i=2}^{N-1} 2x_i\,x_{i-1}\nonumber\\
&\sum_{i=1}^{N} \omega^2\,x_i^2\,\epsilon^2=\omega^2\,x_N^2\,\epsilon^2+ \sum_{i=1}^{N-1}\omega^2\,x_i^2\,\epsilon^2\nonumber\\
&\sum_{i=1}^{N} \dfrac{2 \lambda}{m} \,x_i^4\,\epsilon^2= \dfrac{2 \lambda}{m} \,x_N^4\,\epsilon^2+\sum_{i=1}^{N-1} \dfrac{2 \lambda}{m} \,x_i^4\,\epsilon^2\,.
\label{SumEnd}
\end{align}
Using the above expanded sums, the discretized action \reff{DAction} can be written as
\begin{align}
&S=\dfrac{m}{2 \,\epsilon}\,\Big[\,\big(x_0^2+x_N^2 -\omega^2 x_N^2\,\epsilon^2- \dfrac{2 \lambda}{m} \,x_N^4\,\epsilon^2\big)
+ \sum_{i=1}^{N-1} \big((2-\omega^2\,\epsilon^2)  \,x_i ^2- \dfrac{2 \lambda}{m} \,x_i^4\,\epsilon^2\big)\nonumber\\
&\qquad\qquad-\sum_{i=2}^{N-1} (2\,x_i\,x_{i-1}) -2 x_0 \,x_1-2\,x_N\,x_{N-1}\,\Big]\nonumber\\
&=\dfrac{m}{2 \,\epsilon}\,\Big[\,\big(x_0^2+x_N^2 -\omega^2 x_N^2\,\epsilon^2- \dfrac{2 \lambda}{m} \,x_N^4\,\epsilon^2\big)
+ \sum_{i=1}^{N-1} \big(x_i \,A_{i\,j}\,x_j - \dfrac{2 \lambda}{m} \,x_i^4\,\epsilon^2 +f_{i}\,x_{i}\big)\Big]\nonumber\\
&=\dfrac{m}{2 \,\epsilon}\,\Big[\,\big(x_0^2+x_N^2 -\omega^2 x_N^2\,\epsilon^2- \dfrac{2 \lambda}{m} \,x_N^4\,\epsilon^2\big)
+ \big(\vec{x}\,A\,\vec{x} +\vec{f}\cdot \vec{x}- \dfrac{2 \lambda}{m}\,\epsilon^2 \sum_{i=1}^{N-1} x_i^4 \big)\Big]\,.
\label{S2}
\end{align}
In the last line, the quadratic plus linear part of the action is expressed in matrix form without summation sign where $\vec{x}=[x_1,x_2,...,x_{N-1}]$ and $\vec{f}=[f_1,f_2,...,f_{N-1}]=[-2x_0,0,....,0,-2x_N]$ i.e. $f_1=-2\,x_0$, $f_{N-1}=-2\,x_N$ and all other $f_i$'s  zero. The end points $x_0=x_a$ and $x_N=x_b$ do not appear in $\vec{x}$. They appear in the first round brackets and in the vector $\vec{f}$.  The sum over $x_i^4$ has to be written explicitly as it cannot readily be expressed in terms of $\vec{x}$.  The matrix $A$ is an $(N-1) \times (N-1)$ symmetric matrix with elements $A_{i\,i}=2 -\omega^2\,\epsilon^2 \text{ for } N-1\ge i \ge1$ and $A_{i -1\,i}=A_{i\,i-1}=-1 \text{ for } N-1\ge i \ge 2$. All other entries are zero. As an example, the matrix for $N=6$ is given by
\beq
A=
\begin{bmatrix} 
2 -\omega^2\,\epsilon^2&-1&0&0&0\\
-1&2 -\omega^2\,\epsilon^2&-1&0&0\\
0&-1&2 -\omega^2\,\epsilon^2&-1&0\\
0&0&-1&2 -\omega^2\,\epsilon^2&-1\\
0&0&0&-1&2 -\omega^2\,\epsilon^2
\end{bmatrix}\,.
\eeq{Mat}
The amplitude \reff{Ordinary} after the path is discretized is given by \cite{Feynman}
\beq
K(x_b,t_b;x_a,t_a)= \Big(\dfrac{m}{2\, \pi\, i \hbar\,\epsilon}\Big)^{N/2}\int_{-\infty}^{\infty} e^{i\,S/\hbar}\,dx_1\,dx_2...dx_{N-1}
\eeq{K}
where the coefficient in front of the integral is a normalization factor valid for any system whose Lagrangian is given by $ \dfrac{1}{2}\,m\,\dot{x}^2-V(x,t)$ (which corresponds to our situation).  We will not derive this normalization factor as it is well known (see \cite{Feynman} for derivation). For the above action \reff{S2} the amplitude is given by 
\begin{align}
K(x_b,t_b;x_a,t_a)= \lim_{N\to \infty}\,\bigg(\frac{m}{2\, \pi\, i \hbar\,\epsilon}\bigg)^{\text{\tiny{N/2}}}&\resizebox{6cm}{!}{$e^{\frac{i\,m}{2 \,\epsilon\,\hbar}\,\big(x_0^2+x_N^2 -\omega^2 x_N^2\,\epsilon^2- \frac{2 \lambda}{m} \,x_N^4\,\epsilon^2\big)}$}\nonumber\\
&\int_{-\infty}^{\infty}\resizebox{5cm}{!}{$e^{\frac{i\,m}{2 \,\epsilon\,\hbar}\,\big(\vec{x}\,A\,\vec{x}\, + \,\vec{f}\cdot \vec{x}\,\,- \frac{2\,\lambda}{m}\epsilon^2\sum\limits_{i=1}^{\text{\tiny{N-1}}}x_i^4 \big)}$}\,dx_1\,dx_2...dx_{N-1}\,.
\label{Amplitude}
\end{align}
In the above limit $\epsilon=T/N$ tends to zero. There is no exact analytical solution to the above amplitude or path integral when $\lambda \ne 0$. However, it can be calculated exactly for the case of the harmonic oscillator (i.e. $\lambda=0$). As already mentioned, our technqiue does not require solving for the classical path and moreover, is naturally adapted for solving the forced harmonic oscillator.  We will therefore obtain exact analytical formulas for both the harmonic and forced harmonic oscillator with no reference to the classical action. 

%
%
\subsubsection{Amplitude of harmonic oscillator without use of classical action} \label{AHarmonic}

The harmonic oscillator corresponds to setting $\lambda=0$ in the potential \reff{Pot}. Its amplitude, which we label $K_H$, can be wrritten as 
\beq
K_H= \int \mathcal{D} x(t) \, e^{i\,S_H/\hbar}
\eeq{KH1}
where $S_H$ is the action for the harmonic oscillator:
\beq
S_H=\int_{t_a}^{t_b} ( \dfrac{1}{2}\,m\,\dot{x}^2-\dfrac{1}{2}\,m\,\omega^2\,x^2)\,dt\,.
\eeq{SH1}
The discretized version of the amplitude is obtained by setting $\lambda=0$ in \reff{Amplitude}:
\begin{align}
K_H&= \bigg(\frac{m}{2\, \pi\, i \hbar\,\epsilon}\bigg)^{\text{\tiny{N/2}}}\resizebox{4cm}{!}{$e^{\frac{i\,m}{2 \,\epsilon\,\hbar}\,\big(x_0^2+x_N^2 -\omega^2 x_N^2\,\epsilon^2\big)}$}
\int_{-\infty}^{\infty}\resizebox{3cm}{!}{$e^{\frac{i\,m}{2 \,\epsilon\,\hbar}\,\big(\,\vec{x}\,A\,\vec{x}\, + \,\vec{f}\cdot \vec{x}\big)}$}\,d\vec{x}\nonumber\\
&= \bigg(\frac{m}{2\, \pi\, i \hbar\,\epsilon}\bigg)^{\text{\tiny{N/2}}}\resizebox{4cm}{!}{$e^{\frac{i\,m}{2 \,\epsilon\,\hbar}\,\big(x_0^2+x_N^2 -\omega^2 x_N^2\,\epsilon^2\big)}$}
\int_{-\infty}^{\infty}\resizebox{2.5cm}{!}{$e^{-\frac{1}{2}\vec{x}\,A'\,\vec{x}\, + \,\vec{f}^{\prime}\cdot \vec{x}}\,$}d\vec{x}
\label{AmpH}
\end{align}
where we have defined $A'=-\frac{i\,m}{\hbar\,\epsilon} A$ and $\vec{f}^{\prime}=\frac{i\,m}{2\,\hbar\,\epsilon} \vec{f}$. We will take the infinite $N$ limit at the end.  The integral in \reff{AmpH} is a multi-dimensional Gaussian integral which can readily be evaluated \cite{Schwartz}:
\beq
\int_{-\infty}^{\infty}e^{-\frac{1}{2}\vec{x}\,A'\,\vec{x}\, + \,\vec{f}^{\prime}\cdot \vec{x}}\,d\vec{x}=\sqrt{\dfrac{(2 \pi)^{N-1}}{\text{det}A^{\prime}}}\,e^{\frac{1}{2} \vec{f}^{\prime} \,A^{{\prime}^{-1}}\vec{f}^{\prime}}\,.
\eeq{MultiGauss}
We have the following relations:
\begin{align}
\text{det}A'&=\text{det}\Big(-\frac{i\,m}{\hbar\,\epsilon} A\Big)=\Big(\frac{m}{i\,\hbar\,\epsilon}\Big)^{N-1} \text{det}A\,;\nonumber\\
A^{{\prime}^{-1}}&=\Big(\frac{\hbar\,\epsilon}{-i\,m}\Big)\, A^{-1}\,;\nonumber\\
\vec{f}^{\prime} A^{{\prime}^{-1}}\vec{f}^{\prime}&=\frac{m}{4\,i\,\hbar\,\epsilon}\,\vec{f} A^{-1}\vec{f}\,.
\label{Relations}
\end{align}
Substituting  \reff{MultiGauss} into the amplitude \reff{AmpH} and using the relations \reff{Relations} yields 
\begin{align}
K_H= \bigg(\frac{m}{2\,  \pi\,i\, \hbar\,\epsilon\,\text{det} A}\bigg)^{1/2}\,\resizebox{6cm}{!}{$e^{\frac{m}{8 \,i\,\hbar\,\epsilon} \vec{f} A^{-1}\vec{f}}\,e^{\frac{i\,m}{2 \,\epsilon\,\hbar}\,\big(x_0^2 + x_N^2 - \omega^2 x_N^2\,\epsilon^2\big)}$}\,.
\label{AmpH2}
\end{align}
To evaluate $K_H$ above, we need to obtain expressions for det A and $\vec{f} A^{-1}\vec{f}$.  Matrix A has the form given by \reff{Mat}.  We can obtain its determinant as a function of $N$ via a recursion relation. Let $D(N)$ be the det A for a given $N$. Recall that A is an \resizebox{3cm}{!}{$(N-1)\times (N-1)$} matrix. Let $P= 2- \omega^2\,\epsilon^2$. For $N=3$, we have the $2\times 2$ matrix at the bottom right hand corner of \reff{Mat} whose determinant is $D(3)=P^2-1$. The determinant of the $3 \times 3$ matrix at the bottom right hand corner of \reff{Mat} is $D(4)=P \,(P^2-1)-P= P\, D(3) - D(2)$ since $D(2)=P$. It is easy to verify that $D(5)= P\, D(4) - D(3)$. This pattern continues and we obtain the following recursion relation:
\beq
D(N)= P \,D(N-1) - D(N-2) \text{ with } D(2) = P \text{ and } D(1) = 1\,.
\eeq{Recur}  
This yields an alternating  series in even or odd powers of $P$ (even powers if $N$ is odd and odd powers if $N$ is even). The alternating powers can be expressed as $P^{N-2 j+1}\,(-1)^{j+1}$ where $j$ runs from 1 to $\lfloor(N+1)/2\rfloor$. Excluding the alternating sign, the coefficient of the first term ($j=1$) is $1$, of the second term is $N - 2$, of the third term is $(N - 3)(N - 4)/2!$, of the fourth term is $(N - 4)(N - 5)(N - 6)/3!$, etc. By induction, one can show that the coefficient at a given $j$ is the binomial coefficient $\binom{N-j}{j-1}$. We obtain the finite series below which can be expressed as a hypergeometric function that has a simple expression in terms of radicals:
\begin{align}
D(N)&=\sum_{j=1}^{\lfloor (N+1)/2 \rfloor} (-1)^{j+1}\,\binom{N-j}{j-1}\,P^{N-2 j+1} \nonumber\\
&=\frac{2^{-N}}{\sqrt{-4+P^2}}\,\big[(P \p \sqrt{-4+P^2}\,)^N-(P \m \sqrt{-4+P^2}\,)^N\big]\,.
\label{DN}
\end{align}
The above expression is equal to detA as a function of $N$. With $P= 2- \omega^2\,\epsilon^2=2-\omega^2\,T^2/N^2$ we obtain in the large $N$ limit that $\sqrt{-4 +P^2}= \frac{2\, i\, \omega T}{N} + O\left(\frac{1}{N}\right)^2$. In \reff{AmpH2} we need to evaluate $\epsilon\,$detA= $\frac{T}{N}\, D(N)$. In the large $N$ limit we obtain
\begin{align}
\frac{T}{N} \,D(N)&=\frac{T}{N}\frac{2^{-N}}{\sqrt{-4+P^2}}\,\big[(P \p \sqrt{-4+P^2}\,)^N-(P \m \sqrt{-4+P^2}\,)^N\big]\nonumber\\
&=\frac{\big(1+\frac{i\,\omega\,T}{N}\big)^N- \big(1-\frac{i\,\omega\,T}{N}\big)^N}{2\,i\,\omega}
\end{align}
where terms of order $\epsilon^2=O\left(\frac{1}{N}\right)^2$ are omitted since we will be taking the infinite $N$ limit. We finally obtain 
\beq 
\lim_{N\to \infty} \frac{\big(1+\frac{i\,\omega\,T}{N}\big)^N- \big(1-\frac{i\,\omega\,T}{N}\big)^N}{2\,i\,\omega}= \frac{e^{i\,\omega \,T}-e^{-i\,\omega \,T}}{2\, i\, \omega}= \frac{\sin(\omega\,T)}{\omega}\,.
\eeq{Sin}
So $\epsilon\,$detA in \reff{AmpH2} gets replaced by $\frac{\sin(\omega\,T)}{\omega}$. We now determine $\vec{f} A^{-1}\vec{f}$. Recall that $f_1=-2 x_0$, $f_{N-1}=-2 x_N$ and all other $f_i$'s are zero. We therefore obtain  $\vec{f} A^{-1}\vec{f}=f_i A_{ik}^{-1} f_k$. This yields $4 \,x_0^2 \,A^{-1}_{11}+4 \,x_N^2 \,A^{-1}_{N-1 N-1}+8 \,x_0\,x_N \,A^{-1}_{1 N-1}$. Using the cofactor matrix, we find  $A^{-1}_{11}=\frac{D(N-1)}{D(N)}$, $A^{-1}_{N-1 N-1}=A^{-1}_{11}=\frac{D(N-1)}{D(N)}$ and $A^{-1}_{1 N-1}= \frac{1}{D(N)}$. Using the above results, the product of the two exponentials in \reff{AmpH2} yields
\begin{align}
&\resizebox{5.5cm}{!}{$e^{\frac{m}{8 \,i\,\hbar\,\epsilon} \vec{f} A^{-1}\vec{f}}\,e^{\frac{i\,m}{2 \,\epsilon\,\hbar}\,\big(x_0^2+x_N^2 -\omega^2 x_N^2\,\epsilon^2\big)}$}\nonumber\\
&= \resizebox{7.5cm}{!}{$e^{\frac{i\,m}{2 \,\hbar} \big[(x_0^2+x_N^2)\frac{D(N) - D(N-1)}{D(N) \,\epsilon} - \frac{2\, x_0 \,x_N} {D(N) \,\epsilon} - \omega^2 x_N^2\,\epsilon\,\big]}$}\,.
\label{Exp2}
\end{align}
We want to evaluate the last line above in the limit as $N \to \infty$. We replace $\epsilon$ by $T/N$.  We already worked out that $\lim_{N\to \infty} D(N)\, \epsilon=\frac{\sin(\omega\,T)}{\omega}$. We also have that $\lim_{N\to \infty}\omega^2 x_N^2\,\epsilon=0$. The only term left to evaluate is $D(N) - D(N-1)$ in the large $N$ limit which we calculate below:
\begin{align}
 D(N) - D(N-1)&=\frac{2^{-N}}{\sqrt{-4+P^2}}\,\big[\,(P \p \sqrt{-4+P^2}\,)^N-2 (P \p \sqrt{-4+P^2}\,)^{N-1}\nonumber\\&\qquad\qquad\qquad
 -(P \m \sqrt{-4+P^2}\,)^N  +2 (P \m \sqrt{-4+P^2}\,)^{N-1}\big]\nonumber\\
 &=\frac{N}{2 \,i\,\omega\,T}\Big[\big(1+\frac{i\,\omega\,T}{N}\big)^N-\big(1+\frac{i\,\omega\,T}{N}\big)^N \big(1-\frac{i\,\omega\,T}{N}\big)-\big(1-\frac{i\,\omega\,T}{N}\big)^N \nonumber\\&\qquad\qquad\qquad+ \big(1-\frac{i\,\omega\,T}{N}\big)^N\big(1+\frac{i\,\omega\,T}{N}\big)\Big]\nonumber\\
 &=\frac{1}{2}\,\Big[\big(1+\frac{i\,\omega\,T}{N}\big)^N+\big(1-\frac{i\,\omega\,T}{N}\big)^N\Big]\,.
 \end{align}
 In the infinite $N$ limit we obtain
 \begin{align}
 D(N) - D(N-1)&=\lim_{N\to \infty}\frac{1}{2}\,\Big[\big(1+\frac{i\,\omega\,T}{N}\big)^N+\big(1-\frac{i\,\omega\,T}{N}\big)^N\Big]\nonumber\\
 &=\frac{e^{i\,\omega \,T}+e^{-i\,\omega \,T}}{2}\nonumber\\
 &=\cos(\omega\,T)\,.
 \label{cos1}
 \end{align}
 Substituting the above results into \reff{Exp2} yields 
 \begin{align}
&\resizebox{6cm}{!}{$e^{\frac{m}{8 \,i\,\hbar\,\epsilon} \vec{f} A^{-1}\vec{f}}\,e^{\frac{i\,m}{2 \,\epsilon\,\hbar}\,\big(x_0^2+x_N^2 -\omega^2 x_N^2\,\epsilon^2\big)}$}\nonumber\\
&=\resizebox{6cm}{!}{$e^{\frac{i\,m\,\omega}{2 \hbar\sin(\omega\,T)} \big[(x_a^2+x_b^2)\cos(\omega\,T) - 2\, x_a\,x_b\big]}$}\,.
\label{Exp3}
\end{align}
Substituting \reff{Exp3} into \reff{AmpH2} and replacing $\epsilon\,$detA by $\frac{\sin(\omega\,T)}{\omega}$ yields our final result for the amplitude of the harmonic oscillator:
\begin{align}
K_H= \Big(\frac{m\,\omega}{2\,\pi\,  i\, \hbar\,\sin(\omega\,T)}\Big)^{1/2}\,\resizebox{7cm}{!}{$e^{\frac{i\,m\,\omega}{2 \hbar\sin(\omega\,T)} \big[(x_a^2+x_b^2)\,\cos(\omega\,T) - 2\, x_a\,x_b\big]}$}\,.
\label{AmpH3}
\end{align}
The above result is in agreement with the amplitude obtained for the harmonic oscillator using the more common method of solving for the classical path and extracting out the classical action from the path integral \cite{Feynman}. 

\subsubsection{Spin off: amplitude for forced harmonic oscillator} \label{AFHO}

In the forced harmonic oscillator (FHO) we add an extra term $J(t)\,x$ to the action (equivalent  to adding a term $-J(t) x$ in the potential). The amplitude of the FHO (subscript $FH$) is given by 
\beq
K_{FH}= \int \mathcal{D} x(t) \, e^{i\,S_{FH}/\hbar}
\eeq{FH1}
where $S_{FH}$ is the action for the FHO:
\beq
S_{FH}=\int_{t_a}^{t_b} ( \dfrac{1}{2}\,m\,\dot{x}^2-\dfrac{1}{2}\,m\,\omega^2\,x^2 +J(t) x)\,dt\,.
\eeq{FH2}
One of the advantages of our previous technique used to obtain $K_{H}$ is that the FHO can be naturally incorporated. The reason is that  $J(t)x$, which is  $\vec{J}\cdot \vec{x}$ when discretized, can be absorbed readily into the  term $\vec{f}\cdot \vec{x}$ appearing in the integral on the first line of \reff{AmpH}. Recall that $\vec{f} =[-2x_0,0,....,0,-2x_N]$ is what incorporates the end points $x_0=x_a$ and $x_N=x_b$. From \reff{DAction}, we see that in the discretized action, the terms appearing in the potential when factored outside $m/(2 \epsilon)$ are multiplied by a factor of $-\frac{2 \,\epsilon^2}{m}$. The FHO potential term is $-J(t)x$ so it is $\frac{2 \,\epsilon^2}{m} \vec{J}\cdot \vec{x}$  that needs to be added to  $\vec{f}\cdot \vec{x}$. This yields  $(\tfrac{2 \,\epsilon^2}{m} \,\vec{J}+ \vec{f}\,)\cdot \vec{x}$ which we write as $\vec{F}\cdot \vec{x}$. So when discretized, the amplitude for the FHO is still given by \reff{AmpH2} but with $\vec{f}$ replaced by $\vec{F}$ i.e. 
\begin{align}
K_{FH}= \bigg(\frac{m \,\omega}{2\,  \pi\,i\, \hbar\,\sin (\omega\,T)}\bigg)^{1/2}\,\resizebox{5cm}{!}{$e^{\frac{m}{8 \,i\,\hbar\,\epsilon} \vec{F} A^{-1}\vec{F}}\,e^{\frac{i\,m}{2 \,\epsilon\,\hbar}\,\big(x_0^2 + x_N^2 \big)}$}
\label{AmpFH}
\end{align}
where 
\beq
\vec{F}=\dfrac{2 \,\epsilon^2}{m} \,\vec{J}+ \vec{f}\,.
\eeq{FF}
In \reff{AmpFH}, we have already substituted the result \reff{Sin} that $\epsilon \,$det A= $\epsilon\,D(N)= \tfrac{\sin(\omega\,T)}{\omega}$  (in the limit $N\to \infty$ where $\epsilon=T/N \to 0$).  We also removed the term $\omega^2 \,x_N^2\,\epsilon^2$ appearing in \reff{AmpH2} since it makes no contribution in that limit. In component form, we have $F_1=\tfrac{2 \,\epsilon^2}{m} \,J_1- 2x_0$ , $F_{N-1}=\tfrac{2 \,\epsilon^2}{m} \,J_{N-1}- 2x_N$ with all other components given by $F_i=\tfrac{2 \,\epsilon^2}{m} \,J_i$ (where  $i\ne 1$ and $i\ne N-1$). Using $\vec{F}$ above we obtain 
\begin{align}
\vec{F} A^{-1}\vec{F}&= F_{\ell}\, A^{-1}_{\ell\, m}
\, F_{m} 
=(4 x_0^2 +4x_N^2)\,A^{-1}_{11} + 8 \,x_0\,x_N \,A^{-1}_{1 \,N-1}\nonumber\\&- \frac{8\,\epsilon^2}{m}\,x_0\,A^{-1}_{1\, m} \,J_m- \frac{8\,\epsilon^2}{m}\,x_N\,A^{-1}_{N-1\, m}\,J_m + \frac{4\,\epsilon^4}{m^2}\,J_\ell\, A^{-1}_{ \ell\,m}\, J_{m}
\label{JA}
\end{align}
where implicit summation is asummed for repeated indices and $\ell$ and $m$ run through all components from $1$ to $N-1$ inclusively. The matrix A has the form \reff{Mat} and the components  $A^{-1}_{ \ell\,m}$ of its inverse are given by $ \tfrac{D(N-\ell)\,D(m)}{D(N)}$ if $\ell\ge m$ (if $m\ge \ell$ simply switch $\ell$ and $m$ in $D$). Recall that $D(N)$ is the determinant of the $N-1$ dimensional matrix A and is given by the expression \reff{DN}. We obtain that $A^{-1}_{11} = \tfrac{D(N-1)}{D(N)}$, $A^{-1}_{1 \,N-1}= \tfrac{1}{D(N)}$, $A^{-1}_{1\, m}=\tfrac{D(N-m)}{D(N)}$ and $A^{-1}_{N-1\, m}=\tfrac{D(m)}{D(N)}$. Note that since $A^{-1}_{ \ell\,m}= \tfrac{D(N-\ell)\,D(m)}{D(N)}$ for $\ell\ge m$ we have that $J_\ell\, A^{-1}_{ \ell\,m}\, J_{m}=2\,\sum_{\ell=1}^{N-1}\sum_{m=1}^{\ell}\tfrac{D(N-\ell)\,D(m)}{D(N)}\,J_{\ell}\,J_{m}$ . By summing $m$ only up to $\ell$ we ensure that $\ell\ge m$ and we multiply by two because the original implicit summation is over all $\ell$ and $m$. Inserting the above components of the inverse matrix into \reff{JA} and then substituting the result into the exponential part of \reff{AmpFH} yields
\begin{align}
&\exp\Big(\frac{m}{8 \,i\,\hbar\,\epsilon} \vec{F} A^{-1}\vec{F}\Big)\,\exp\Big(\frac{i\,m}{2 \,\epsilon\,\hbar}\,\big(x_0^2+x_N^2\Big)\nonumber\\
&= \exp\Big(\frac{i\,m}{2 \,\hbar} \,\big[(x_0^2+x_N^2)\,\frac{D(N) \m D(N \m 1)}{D(N) \,\epsilon} - \frac{2\, x_0 \,x_N} {D(N) \,\epsilon} + \frac{2 x_0}{m}\,J_m\,\epsilon \frac{D(N\m m)\,\epsilon}{D(N) \,\epsilon}   \nonumber\\&\qquad\qquad+ \frac{2 x_N}{m}\,J_m\,\epsilon \frac{D(m)\,\epsilon}{D(N) \,\epsilon}-\frac{2}{m^2}\sum_{\ell=1}^{N-1}\sum_{m=1}^{\ell} J_{\ell}\,J_m\,\epsilon^2 \frac{D(N\m \ell)\,D(m)\,\epsilon^2}{D(N)\,\epsilon}\,\big]\Big)\nonumber\\
&= \exp\Big(\frac{i\,m\,\omega}{2 \,\hbar\,\sin (\omega\,T)}\, \big[(x_0^2+x_N^2) \,\cos (\omega\,T) - 2\, x_0 \,x_N + \frac{2 x_0}{m}\,J_m\,\epsilon \,(D(N \m m)\,\epsilon)  \nonumber\\&\qquad\qquad+ \frac{2 x_N}{m}\,J_m \epsilon \,(D(m)\,\epsilon) -\frac{2}{m^2}\sum_{\ell=1}^{N-1}\sum_{m=1}^{\ell}\,J_{\ell}\,J_m\epsilon^2\,(D(N \m \ell)\,D(m)\,\epsilon^2)\big]\Big)
\label{ExpFH}
\end{align}
where we substituted our previous result \reff{Sin} that $\epsilon\,D(N)$ in the limit as $\epsilon \to 0$ is equal to $\tfrac{\sin (\omega\,T)}{\omega}$. The index $N$ corresponds to the time $t_N=t_b$. Since the start time is $t_a$ (i.e. $t_0=t_a$),  the time interval from start (index 0)  to $N$ is $T=t_b-t_a$. An arbitrary index $m$, corresponds to a time $t_m$ which we simply write as $t$ and this corresponds to a time interval $t-t_a$ whereas the index $N-m$ corresponds to a time interval $t_b-t$. Therefore $\epsilon=\tfrac{T}{N}=\tfrac{t-t_a}{m}=\tfrac{t_b-t}{N-m}$.  To evaluate $\epsilon\,D(m)$ and $\epsilon\,D(N-m)$ in \reff{ExpFH}, we simply replace $\tfrac{T}{N}$ inside the large round brackets in \reff{Sin} by $\tfrac{t-t_a}{m}$ and $\tfrac{t_b-t}{N-m}$ respectively and also replace $N$ by $m$ and $N-m$ respectively. It follows that  $\epsilon\,D(m)$  and $\epsilon\, D(N-m)$ get replaced  by $\tfrac{\sin (\omega\,(t-t_a)}{\omega}$ and $\tfrac{\sin (\omega\,(t_b-t))}{\omega}$ respectively in the $\epsilon \to 0$ limit. The quantities $J_m \epsilon \,(D(m)\,\epsilon)$  and $J_m\epsilon\,(D(N-m)\,\epsilon)$ contain an implicit sum over all $m$ from $1$ to $N-1$. In the continuum limit, the sum gets replaced by an integral and $J_m\,\epsilon \to J(t)\,dt$ so that we obtain
\begin{align}
& J_m\,\epsilon\, D(m)\,\epsilon \to \int_{t_a}^{t_b} \dfrac{1}{\omega}\,J(t)\, \sin (\omega\,(t-t_a))\,dt\nonumber\\
& J_m\,\epsilon\,D(N-m)\,\epsilon \to \int_{t_a}^{t_b} \dfrac{1}{\omega}\,J(t)\, \sin (\omega\,(t_b-t))\,dt \,.
\label{Cont}
\end{align}
The last term in \reff{ExpFH} has two time labels: let $s$ correspond to $m$ and $t$ to $\ell$. In the continuum limit, the double sum is replaced by a double integral and we obtain 
\begin{align}
\sum_{\ell=1}^{N-1}\sum_{m=1}^{\ell}\!J_{\ell}\,J_m\,\epsilon^2 \,D(N-\ell)\,D(m)\,\epsilon^2\to \dfrac{1}{\omega^2}\int_{t_a}^{t_b}J(t)\,\sin (\omega\,(t_b-t))  \int_{t_a}^{t} J(s)\,\sin (\omega\,(s-t_a))\,ds\,dt.
\label{Cont2}
\end{align}
Substituting \reff{Cont2} and \reff{Cont} into \reff{ExpFH} and replacing $x_0$ by $x_a$ and $x_N$ by $x_b$, the final expression  for the amplitude \reff{AmpFH} of the forced harmonic oscillator is
\begin{align}
K_{FH}&= \bigg(\frac{m \,\omega}{2\,  \pi\,i\, \hbar\,\sin (\omega\,T)}\bigg)^{1/2}\nonumber\\&\exp\bigg\{\frac{i\,m\,\omega}{2 \,\hbar\,\sin (\omega\,T)} \Big[(x_a^2+x_b^2) \cos (\omega\,T) - 2\, x_a \,x_b + \frac{2 x_a}{m\,\omega}\int_{t_a}^{t_b}\,J(t) \,\sin (\omega\,(t_b-t))\,dt \nonumber\\&+ \frac{2 x_b}{m\,\omega}\int_{t_a}^{t_b}\,J(t) \,\sin (\omega\,(t-t_a))\,dt\nonumber\\&-\frac{2}{m^2 \,\omega^2}\int_{t_a}^{t_b}J(t)\,\sin (\omega\,(t_b-t))  \int_{t_a}^{t} J(s)\,\sin (\omega\,(s-t_a))\,ds\,dt\,\Big]\bigg\}\,.
\label{FH}
\end{align}
The FHO corresponds to the time-dependent potential $V(x,t)=\frac{1}{2}\,m\,\omega^2\,x^2- J(t)\,x$. When $J=0$, we recover the potential for the harmonic oscillator and we see that the above amplitude $K_{FH}$ reduces to the amplitude $K_{H}$ of the harmonic oscillator given by \reff{AmpH3}. 

The above expression for $K_{FH}$ is very useful because it will enable us to evaluate terms in a perturbative series by taking functional derivatives with respect to $J(t)$ and then setting $J$ to zero.    

\section{Euclidean path integral}\label{Euclid}

The ordinary path integral has a highly oscillatory integrand which is not ideal for numerical integration.  The Euclidean path integral is more suitable for numerical integration as it converges more rapidly and can be obtained with greater numerical precision.  For this reason, we will obtain two different series expansions for the Euclidean path integral. We will then be able to compare our analytical results order by order with the exact numerical answer. 

The Euclidean path integral or amplitude is defined as
\beq
K_E= \int \mathcal{D} x(\tau) \, e^{- S_E/\hbar}
\eeq{Euclidean}
where $S_E$ is the Euclidean action given by
\beq
S_E=\int_{\tau_a}^{\tau_b} \Big(\dfrac{1}{2} \,m \dot{x}^2 + V(x,\tau)\,\Big)\, d\tau\,.
\eeq{ActionE}
The time derivatives are with respect to $\tau$.  This differs from the ordinary action because the integrand now has the form ``kinetic plus potential'' instead of ``kinetic minus potential''.  The Euclidean path integral can be obtained by analytically continuing the ordinary amplitude $K$ given by \reff{Ordinary} to imaginary time i.e. by performing a Wick rotation $t \to - i \,\tau$.  The Euclidean and ordinary amplitudes are therefore related to each other via 
\beq
K_E(x_b,\tau_b;x_a,\tau_a)=K(x_b,t_b=-i \tau_b;x_a,t_a=-i\,\tau_a)\,.
\eeq{KEK}
The potential we will consider is $V(x,\tau)=\dfrac{1}{2}\,m\,\omega^2\,x^2 +\lambda\,x^4 -J(\tau)\,x$.  We include $J(\tau)\,x$ for future convenience; we will set $J=0$ at the end of calculations so that the physical system is composed only of the harmonic oscillator plus quartic term.  The Euclidean path integral with the above potential is given by
\beq
K_E= \int \mathcal{D} x(\tau) \, \exp\Big[\,\frac{-1}{\hbar}\int_{\tau_a}^{\tau_b} \Big ( \dfrac{1}{2} \,m \dot{x}^2 +\dfrac{1}{2}\,m\,\omega^2\,x^2 +\lambda\,x^4 - J(\tau)\,x\Big)\, d\tau\,\Big]\,.
\eeq{AmplitudeE}
When $\lambda=0$, this reduces to the Euclidean amplitude $K_{{FH}_E}$ of the forced harmonic oscillator 
\beq
K_{{FH}_E}= \int \mathcal{D} x(\tau) \,\exp\Big[\,\frac{-1}{\hbar}\int_{\tau_a}^{\tau_b} \Big ( \dfrac{1}{2} \,m \dot{x}^2 +\dfrac{1}{2}\,m\,\omega^2\,x^2 - J(\tau)\,x\Big)\, d\tau\,\Big]\,.
\eeq{FullE}
 An exact analytical expression can readily be obtained for the above Gaussian integral by using the relation  \reff{KEK} between the Euclidean and ordinary amplitudes. We simply replace $t$ by $-i\,\tau$ (and $s$ by $-i\sigma$) in the ordinary amplitude \reff{FH} and we define $\mathcal{T}=\tau_b-\tau_a$. This yields 
\begin{align}
K_{{FH}_E}&=\bigg(\frac{m \,\omega}{2\,  \pi\, \hbar\,\sinh(\omega\,\mathcal{T})}\bigg)^{1/2}\nonumber\\&\exp\bigg\{\frac{-\,m\,\omega}{2 \,\hbar\,\sinh (\omega\,\mathcal{T})} \Big[(x_a^2+x_b^2) \cosh (\omega\,\mathcal{T}) - 2\, x_a \,x_b - \frac{2 x_a}{m\,\omega}\int_{\tau_a}^{\tau_b}\,J(\tau) \,\sinh(\omega\,(\tau_b-\tau))\,d\tau \nonumber\\&- \frac{2 x_b}{m\,\omega}\int_{\tau_a}^{\tau_b}\,J(\tau) \,\sinh (\omega\,(\tau-\tau_a))\,d\tau\nonumber\\&-\frac{2}{m^2 \,\omega^2}\int_{\tau_a}^{\tau_b}J(\tau)\,\sinh (\omega\,(\tau_b-\tau))  \int_{\tau_a}^{\tau}\,J(\sigma)\,\sinh (\omega\,(\sigma-\tau_a))\,d\sigma\,d\tau\,\Big]\bigg\}\,.
\label{FHE}
\end{align}
where we used $\sin(-i \,\omega \mathcal{T})=-i\,\sinh(\omega\,\mathcal{T})$ and $\cos(-i \omega \mathcal{T})=\cosh(\omega\,\mathcal{T})$. We see that the Euclidean version has hyperbolic sines and cosines instead of sines and cosines.  

When $\lambda \ne 0$, there is no exact analytical expression for the Euclidean amplitude \reff{AmplitudeE}. We will therefore develop two different series expansions of \reff{AmplitudeE} and obtain analytical expressions for the terms in the series, order by order. In the first series we expand the quartic term ( the ``interaction'') in powers of the coupling $\lambda$ which is analogous to the usual perturbative series in QFT in powers of the coupling. This will be evaluated via functional derivatives of \reff{FHE} with respect to $J$. In other words, $K_{{FH}_E}$, originating from a Gaussian integral, is the generating functional for the first series; it can be viewed as the central building block for the first series. Since we will be setting $J$ to zero at the end of calculations, we will be in effect evaluating the amplitude for the quartic anharmonic oscillator. In the second series, we leave the quartic and linear $J(\tau) x$ term alone and expand the quadratic part consisting of the kinetic plus harmonic oscillator term. The integral consisting of the quartic plus linear term alone can be evaluated exactly analytically and it yields products of generalized  hypergeometric functions. This will be the generating functional, the building block for the second series as we will be taking functional derivatives of it to obtain the terms in the expansion. This is less chartered territory.   

The Euclidean path integral \reff{AmplitudeE} when $\lambda \ne 0$, is \textit{finite} for any positive value of $\lambda$, and can be evaluated exactly numerically to within a certain accuracy (depending on one's computational resources). A path integral is a numerically intensive comuputation as it involves multiple integrals (formally an infinite number of them) whose limits run formally from $-\infty$ to $\infty$. Since the integrand is a decreasing exponential, the limits of integration can be reduced substantially with basically no loss of accuracy. However, the integrals form nested loops and the time needed to run a program increases significantly as the number of integrals increases. If you have say ten loops and you decrease the step size by a factor of two, you increase the run time by roughly a factor of $2^{10}=1024$. With our computational resources and reasonable time constraints we were able to reach convergence to within four or three decimal places and this usually required evaluating up to eight integrals. 

\section{First series: expansion of quartic term in powers of $\lambda$} \label{SQuart}

We now expand the quartic term in \reff{AmplitudeE} in powers of $\lambda$. The result can be conveniently expressed in terms of functional derivatives with respect to $J$ of $ K_{{FH}_E}$. We set $J=0$ at the end of the calculation which excludes the linear $J(\tau)\,x$ term from the physical system which is composed only of the quartic anharmonic oscillator. The series expansion yields
\begin{align}
K_E&=  \int \mathcal{D} x(\tau) \,\exp\Big[\,\frac{-1}{\hbar}\int_{\tau_a}^{\tau_b} \Big ( \dfrac{1}{2} \,m \dot{x}^2 +\dfrac{1}{2}\,m\,\omega^2\,x^2 - J(\tau)\,x\Big)\, d\tau\,\Big]\nonumber\\&\qquad\qquad\Big(1-\dfrac{\lambda}{\hbar} \int_{\tau_a}^{\tau_b} x^4(\tau) \,d\tau+\dfrac{\lambda^2}{\hbar^2}\,\dfrac{1}{2}\int_{\tau_a}^{\tau_b}d\tau \int_{\tau_a}^{\tau_b}x^4(\beta)\, x^4(\tau)\, d\beta +...\,\Big)\nonumber\\
&= K_{{FH}_E}\Bigr\rvert_{J = 0}- \dfrac{\lambda}{\hbar} \int_{\tau_a}^{\tau_b} d\tau (\hbar^4)\dfrac{\delta^4\,K_{{FH}_E}}{\delta J(\tau)^4}\Bigr\rvert_{J = 0}\nonumber\\
&\qquad\qquad\qquad\qquad\qquad\qquad\qquad+ \dfrac{\lambda^2}{2\,\hbar^2}\int_{\tau_a}^{\tau_b}d\tau\int_{\tau_a}^{\tau_b}d\beta\,(\hbar^8)\dfrac{\delta^8\,K_{{FH}_E}}{\delta J(\tau)^4\,\delta J(\beta)^4}\Bigr\rvert_{J = 0}+...\nonumber\\
&= K_{H_E} \Big(1- a_1\,\dfrac{\lambda}{\hbar} + \frac{a_2}{2!}\,\dfrac{\lambda^2}{\hbar^2} +...+ (-1)^n\frac{a_n}{n!}\dfrac{\lambda^n}{\hbar^n}\Big)
\label{Series1}    
\end{align}
where $K_{H_E}=K_{{FH}_E}\Bigr\rvert_{J = 0}$ is the Euclidean amplitude of the harmonic oscillator obtained by setting $J=0$ in \reff{FHE}:
\begin{align}
K_{H_E}= \bigg(\frac{m \,\omega}{2\,  \pi\, \hbar\,\sinh(\omega\,\mathcal{T})}\bigg)^{1/2}\exp\Big[\frac{-\,m\,\omega}{2 \,\hbar\,\sinh (\omega\,\mathcal{T})} \big((x_a^2+x_b^2) \cosh (\omega\,\mathcal{T}) - 2\, x_a \,x_b \big)\Big]\,.
\label{KHE}
\end{align}
The series expansion \reff{Series1} can be viewed as quartic corrections to the harmonic oscillator with the nth order correction proportional to $\lambda^n$. The goal now is to obtain analytical expressions for the coefficients $a_n$ which are functions of $x_a$, $\tau_a$, $x_b$ and $\tau_b$. The coefficients $a_n$ are defined as 
\begin{align}
a_n=\dfrac{1}{K_{H_E}} \int_{\tau_a}^{\tau_b}d\tau_1d\tau_2...d\tau_n\,(\hbar^{4\,n})\,\dfrac{\delta^{4n}K_{{FH}_E}}{\delta J(\tau_1)^4 \delta J(\tau_2)^4...\delta J(\tau_n)^4}\Bigr\rvert_{J = 0}
\label{an}
\end{align}
where the integral is an $n-$dimensional integral. Since we have an exact analytical expression for $K_{{FH}_E}$ given by \reff{FHE}, one can readily obtain analytical expressions for all the coefficients $a_n$ since this involves only taking functional derivatives of a known functional. Though a software package can calculate the derivatives quickly at any order,  we illustrate below an explicit calculation for the coefficient $a_1$ given by
\begin{align}
a_1=\dfrac{1}{K_{H_E}}\int_{\tau_a}^{\tau_b} d\tau_1 \,(\hbar^4)\,\dfrac{\delta^4\,K_{{FH}_E}}{\delta J(\tau_1)^4}\Bigr\rvert_{J = 0}\,. 
\label{a11}
\end{align}
We can express $K_{{FH}_E}$ as $C \,e^{G[J]}$ where $C$ is independent of $J$ and 
\begin{align}
G[J]&=\frac{-\,m\,\omega}{2 \,\hbar\,\sinh (\omega\,\mathcal{T})} \Big[(x_a^2+x_b^2) \cosh (\omega\,\mathcal{T}) - 2\, x_a \,x_b - \frac{2 x_a}{m\,\omega}\int_{\tau_a}^{\tau_b}\,J(\tau) \,\sinh(\omega\,(\tau_b-\tau))\,d\tau \nonumber\\&- \frac{2 x_b}{m\,\omega}\int_{\tau_a}^{\tau_b}\,J(\tau) \,\sinh (\omega\,(\tau-\tau_a))\,d\tau\nonumber\\&-\frac{2}{m^2 \,\omega^2}\int_{\tau_a}^{\tau_b}J(\tau)\,\sinh (\omega\,(\tau_b-\tau))  \int_{\tau_a}^{\tau}\,J(\sigma)\,\sinh (\omega\,(\sigma-\tau_a))\,d\sigma\,d\tau\,\Big]\,.
\label{GJ}
\end{align}
Let  $\dfrac{\delta^n\,G[J]}{\delta J(\tau_1)^n}$ be labeled as $G_{(n)}$.  Since $G$ has a maximum of two powers of $J$, the only non-zero derivatives are $G_{(1)}$ and $G_{(2)}$. A derivative of $C \,e^G$ yields itself multiplied by $G_{(1)}$. It follows then that two derivatives yields $G_{(1)}^2 +G_{(2)}$, three yields $G_{(1)}^3 + 3 \,G_{(1)}\,G_{(2)}$ and four yields $G_{(1)}^4 + 6\, G_{(1)}^2\, G_{(2)} +3\, G_{(2)}^2$. We therefore obtain 
\beq
\dfrac{\delta^4\,K_{{FH}_E}}{\delta J(\tau_1)^4}\Bigr\rvert_{J = 0}=K_{{FH}_E}\,\big(G_{(1)}^4 + 6\, G_{(1)}^2 \,G_{(2)}  +3\,G_{(2)}^2\Big)\Bigr\rvert_{J = 0}= K_{H_E}\,(\,G1^4 +6\, G1^2\, G2 +3\,G2^2\,)
\eeq{FourFunc}
where 
\begin{align} 
G1&= G_{(1)}\Bigr\rvert_{J = 0}=\frac{x_a\,\sinh (\omega  (\tau_b-\tau _1))+ x_b\, \sinh (\omega  (\tau_1-\tau _a))}{\hbar  \sinh (\omega \mathcal{T})}\nonumber\\\text{and}\nonumber\\
G2&= G_{(2)}\Bigr\rvert_{J = 0}=\frac{ \sinh (\omega  (\tau_1-\tau _a))\sinh (\omega  (\tau_b-\tau _1))}{m\, \omega\,  \hbar  \sinh ( \omega \mathcal{T})}\,.
\label{G1G2}
\end{align}
Substituting \reff{G1G2} and \reff{FourFunc} into \reff{a11} and performing the integration over $\tau_1$, yields the following analytical expression for the $a_1$ coefficient:
\begin{align}
a_1=&\dfrac{1}{16\, \omega\,\sinh^4(\omega \mathcal{T}) }\Big\{
\big(x_a^4+x_b^4\big)\,\big( \tfrac{1}{2}\,\sinh(4\,\omega\, \mathcal{T}) -4\,\sinh(2\,\omega\, \mathcal{T}) + 6\,\omega\,\mathcal{T}\big) \nonumber\\&\quad\quad\quad\quad+\big(x_a^3\,x_b+x_a\,x_b^3\big)\,\big(-24\,\omega\,\mathcal{T}\,\cosh(\omega\, \mathcal{T}) +18\,\sinh(\omega\, \mathcal{T}) + 2 \,\sinh(3\,\omega\, \mathcal{T})\, \big)
\nonumber\\&\quad\quad\quad\quad+x_a^2\,x_b^2\,\big(-18 \sinh(2\,\omega\, \mathcal{T}) +12\,\omega\,\mathcal{T}\cosh(2\,\omega\, \mathcal{T})+ 24\,\omega\,\mathcal{T}\big) \Big\}\nonumber\\
+&\dfrac{3 \,\hbar}{8\,m\, \omega^2\,\sinh^3(\omega \mathcal{T}) }\Big\{
\big(x_a^2+x_b^2\big)\,\big(-6 \,\omega\,\mathcal{T}\,\cosh(\omega\, \mathcal{T})+ \tfrac{9}{2}\,\sinh(\omega\, \mathcal{T})+\tfrac{1}{2}\,\sinh(3\,\omega\, \mathcal{T}) \big) \nonumber\\&\quad\quad\quad\quad\quad\quad\quad\quad+ x_a\,x_b\,\big(4\,\omega\,\mathcal{T}\,\cosh(2\,\omega\, \mathcal{T}) -6\,\sinh(2\,\omega\, \mathcal{T}) + 8\,\omega\, \mathcal{T} \big) \Big\}
\nonumber\\
+&\dfrac{3 \,\hbar^2}{16\,m^2\, \omega^3\,\sinh^2(\omega \mathcal{T}) } \Big\{\big(-3 \,\sinh(2\,\omega\, \mathcal{T}) +2\,\omega\,\mathcal{T}\cosh(2\,\omega\, \mathcal{T})+ 4\,\omega\,\mathcal{T}\big) \Big\}\,.
\label{a1a1}
\end{align}
The above expression for $a_1$ is exact and includes its full dependence on the end points $x_a$ and $x_b$ as well as the time interval $\mathcal {T}=\tau_b-\tau_a$. We can readily obtain the expressions for all the other coefficients $a_n$ via \reff{an} since we have the exact expression \reff{FHE} for $K_{{FH}_E}$. The coefficient $a_1$ above is not explicitly derived in the literature on the quartic anharmonic oscillator where the focus usually lies on obtaining corrections to the harmonic oscillator energies or calculating Feynman correlators \cite{Marino1}. In that case, the coefficients in the series expansion of the energy (or correlators) have no dependence on the end points or time interval and are typically calculated using Feynman diagrams. They contain less information than the coefficients $a_n$.

Now that we have an explicit expression for the first coefficient $a_1$ as a function of the end points, it is worthwhile to discuss the physical properties and symmetries of the amplitude $K_E$ given by the series \reff{Series1}. The amplitude is invariant under the exchange of the end points $x_a$ and $x_b$ since the analytical expressions for both $K_{H_E}$ and $a_1$, given by \reff{KHE} and \reff{a1a1} respectively, are invariant under exchange of $x_a$ and $x_b$. This is physically intuitive: since the potential of the physical system has no time dependence, the amplitude to start at  $x_a$ and end at $x_b$ should be the same as the amplitude to start at $x_b$ and end at $x_a$ (assuming of course the same positive time interval). Though we referred here to the first coefficient $a_1$ defined by \reff{a11}, it is easy to see that the coefficients $a_n$ defined by \reff{an} share the same invariance under exchange of $x_a$ and $x_b$.  The amplitude is also invariant under spatial reflections $x_a\to -x_a$ and $x_b \to -x_b$. This stems from the fact that the potential is even under reflections. Since we set $J=0$ at the end, the potential of the physical system is $V(x)=\tfrac{1}{2}\,m\,\omega^2\,x^2 +\lambda\,x^4$ which is even under reflections i.e. $V(x)=V(-x)$. It can easily be seen that the amplitude is not invariant under spatial translations $x_a \to x_a+c$ and $x_b \to x_b+c$ where $c$ is a real constant. This stems from the fact that the potential is not invariant under spatial translations i.e. $V(x)\ne V(x+c)$. In contrast, the amplitude is invariant under time translations $\tau \to \tau +c$ since it depends only on the time difference $\mathcal{T}=\tau_b-\tau_a$. This is due to the time-indepedence of $V(x)$; the origin of time does not matter (only time differences matter). This leads of course to the quantum mechanical version of conservation of energy. Most importantly and less intuitive, the amplitude is \textit{not} invariant under time reversal since it clearly changes when $\mathcal{T}\to -\mathcal{T}$. Though the Euclidean action $S_E$ is invariant under time reversal, the amplitude is not; the path integral is time-ordered. Propagation from point A to point B occurs only if the time at B is later than the time at A. Note that the factor in front of the exponential in $K_{H_E}$ becomes \textit{complex} when $\mathcal{T}\to -\mathcal{T}$.  
  
What we are interested in is comparing the series order by order to the exact numerical integration of the path integral as a function of the values of the coupling $\lambda$. We therefore need to fix numerical values for the end points $x_a$ and $x_b$. Though we are free to choose any values for the end points,  a judicious choice  is $x_a=x_b=0$ as this simplifies greatly the expressions for the coefficients $a_n$ (and even then they are large). It is expected that the end points play a minor role in determing the overall behaviour of the series as a function of the coupling $\lambda$ e.g. if it plateaus over a range of orders at some weak coupling $\lambda$ but does not plateau at strong coupling $\lambda$, the end points will not change that trend ( they might shift slightly the value of $\lambda$ where a plateau occurs but not the trend itself from weak to strong coupling).

We write down below the first three coefficients $a_1$, $a_2$ and $a_3$ when $x_a=x_b=0$:
\begin{align}
a_1&=\dfrac{3 \,\hbar^2}{16\,m^2\, \omega^3\,\sinh^2(\omega \mathcal{T})} \Big\{\big(-3 \,\sinh(2\,\omega\, \mathcal{T}) +2\,\omega\,\mathcal{T}\cosh(2\,\omega\, \mathcal{T})+ 4\,\omega\,\mathcal{T}\big) \Big\}\nonumber\\
a_2&=\dfrac{3 \hbar ^4}{1024\, m^4\, \omega ^6 \,\sinh^4(\omega \mathcal{T} )} \Big\{(12  \omega \mathcal{T} -8 \sinh (2 \omega \mathcal{T})+\sinh (4  \omega \mathcal{T} ))^2\nonumber\\&\qquad\qquad+(288  \omega \mathcal{T} -3 \sinh (2  \omega \mathcal{T} )+144  \omega \mathcal{T}  \cosh (2 \omega \mathcal{T} ))^2\nonumber\\&\qquad\qquad\qquad\qquad+16 (9 \sinh ( \omega \mathcal{T})+\sinh (3  \omega \mathcal{T} )-12 \omega \mathcal{T}  \cosh ( \omega \mathcal{T}))^2\Big\}\nonumber\\
a_3&=\frac{135\, \hbar^6}{65536 \,m^6 \,\omega ^9\,\sinh^6(\mathcal{T} \omega )}\big(115200\, \mathcal{T}^3 \omega ^3+576 \,\mathcal{T}^3 \omega ^3 \cosh (6 \mathcal{T} \omega )-225792 \mathcal{T}^2 \omega ^2 \sinh (2 \mathcal{T} \omega )\nonumber\\&-80640 \mathcal{T}^2 \omega ^2 \sinh (4 \mathcal{T} \omega )-4032 \mathcal{T}^2 \omega ^2 \sinh (6 \mathcal{T} \omega )+12 \mathcal{T} \omega  \big(10800 \mathcal{T}^2 \omega ^2+791\big) \cosh (2 \mathcal{T} \omega )\nonumber\\&+768 \mathcal{T} \omega  \big(27 \mathcal{T}^2 \omega ^2+101\big) \cosh (4 \mathcal{T} \omega )-95928 \mathcal{T} \omega +39858 \sinh (2 \mathcal{T} \omega )-11856 \sinh (4 \mathcal{T} \omega )\nonumber\\&-5427 \sinh (6 \mathcal{T} \omega )+40 \sinh (8 \mathcal{T} \omega )-5 \sinh (10 \mathcal{T} \omega )+8934 \mathcal{T} \omega  \cosh (6 \mathcal{T} \omega )-72 \mathcal{T} \omega  \cosh (8 \mathcal{T} \omega )\nonumber\\&+6 \mathcal{T} \omega  \cosh (10 \mathcal{T} \omega )\big)\,.
\label{AnZero}
\end{align}
To obtain a numerical value for the coefficients, one needs to set values for the parameters. A convenient choice is to set all parameters to unity: $\omega=\mathcal{T}=\hbar=m=1$. We generated the numerical value of the first five coefficients which we quote below:\footnote{The coefficients $a_n$ are generated via \reff{an} which involves calculating $4\,n$ functional derivatives of $K_{{FH}_E}$ given by \reff{FHE}. This becomes computationally intensive as $n$ increases both in terms of time and memory.  We nonetheless succeeded in generating the first five coefficients which was sufficient to see the basic trend of the series and whether it plateaued or departed from the value obtained via direct numerical integration.}     
\begin{align}
a_1=0.0874047 \quad;\quad a_2=0.314084\quad;\quad a_3=1.67012 \quad;\quad a_4=21.9119\quad;\quad a_5=443.183\,.
\label{Coeff}
\end{align}
Substituting the values of the above parameters into \reff{KHE} and setting $x_a=x_b=0$ yields 
\beq
K_{H_E}=  \bigg(\frac{1}{2\,  \pi\, \,\sinh(1)}\bigg)^{1/2}=0.368005\,.
\eeq{KHEN}
Using the numerical values \reff{Coeff} for the coefficients, we evaluate the series \reff{Series1} up to fifth order for four different cases: $\lambda=0.01$, $\lambda=0.1$, $\lambda=1$ and $\lambda=10$. The results are listed below and in Table \ref{KE_Table}. 
\begin{table}
	\includegraphics[scale=0.9]{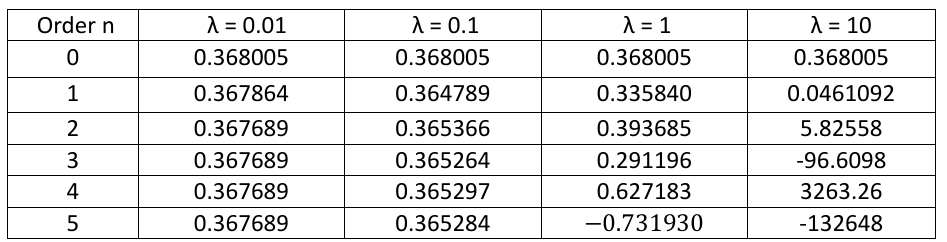}
		\caption{}       
	\label{KE_Table}
\end{table}

\textbf{Case 1: $\lambda$=0.01}
\begin{align}
K_{E_{(0)}}&=K_{H_E}=0.368005\nonumber\\
K_{E_{(1)}}&=K_{H_E}\,\Big(1- a_1\,\dfrac{\lambda}{\hbar}\Big)=0.367864\nonumber\\
K_{E_{(2)}}&= K_{H_E}\, \Big(1- a_1\,\dfrac{\lambda}{\hbar} + \frac{a_2}{2!}\,\dfrac{\lambda^2}{\hbar^2}\Big)= 0.367689\nonumber\\
K_{E_{(3)}}&= K_{H_E}\, \Big(1- a_1\,\dfrac{\lambda}{\hbar} + \frac{a_2}{2!}\,\dfrac{\lambda^2}{\hbar^2}- \frac{a_3}{3!}\,\dfrac{\lambda^3}{\hbar^3}\Big)= 0.367689\nonumber\\
K_{E_{(4)}}&= K_{H_E}\, \Big(1- a_1\,\dfrac{\lambda}{\hbar} + \frac{a_2}{2!}\,\dfrac{\lambda^2}{\hbar^2}- \frac{a_3}{3!}\,\dfrac{\lambda^3}{\hbar^3}+ \frac{a_4}{4!}\,\dfrac{\lambda^4}{\hbar^4}\Big)= 0.367689\nonumber\\
K_{E_{(5)}}&= K_{H_E}\, \Big(1- a_1\,\dfrac{\lambda}{\hbar} + \frac{a_2}{2!}\,\dfrac{\lambda^2}{\hbar^2}- \frac{a_3}{3!}\,\dfrac{\lambda^3}{\hbar^3}+ \frac{a_4}{4!}\,\dfrac{\lambda^4}{\hbar^4}- \frac{a_5}{5!}\,\dfrac{\lambda^5}{\hbar^5}\Big)=0.367689\,.
\end{align}
\\
\textbf{Case 2: $\lambda$=0.1}

$K_{E_{(0)}}=0.368005\,$;$\, K_{E_{(1)}}=0.364789\,$;$\, K_{E_{(2)}}= 0.365366\,$;$\, K_{E_{(3)}}= 0.365264\,$;$\,  K_{E_{(4)}}= 0.365297\,$;$\\K_{E_{(5)}}= 0.365284\,.$

\textbf{Case 3: $\lambda$=1}

$K_{E_{(0)}}=0.368005\,$;$\, K_{E_{(1)}}=0.335840\,$;$\, K_{E_{(2)}}= 0.393685\,$;$\, K_{E_{(3)}}= 0.291196\,$;$\,  K_{E_{(4)}}= 0.627183\,$;$\\K_{E_{(5)}}=  - 0.731930\,.$

\textbf{Case 4: $\lambda$=10}

$K_{E_{(0)}}=0.368005\,$;$\, K_{E_{(1)}}=0.0461092\,$;$\, K_{E_{(2)}}= 5.82558\,$;$\, K_{E_{(3)}}= -96.6098\,$;$\,  K_{E_{(4)}}= 3263.26\,$;$\\K_{E_{(5)}}= -132648\,.$

We see that the series is predictive at the weak coupling of $\lambda=0.01$ because the first few terms plateau at the value of $0.367689$. The exact value for $\lambda=0.01$ from direct numerical integration was obtained to four decimal accuracy and is equal to $0.3677$. The series and exact value for $\lambda=0.01$ therefore match at least to four decimal places. At the intermediate value of $\lambda=0.1$, the series plateaus to within three decimal places to $0.365$. The exact value for $\lambda=0.1$ from direct numerical integration was obtained to four decimal accuracy and is equal to $0.3653$. The series and exact value for $\lambda=0.1$ therefore match to within three decimal places. The series clearly fails to be predictive for the strong coupling values of $\lambda=1$ and $\lambda=10$ as it does not plateau to any value (and even becomes negative) during the first five terms. The exact value for $\lambda=1$ and $\lambda=10$ from direct numerical integration was obtained to within three decimal places to be $0.342$ and $0.237$ respectively. Clearly, the first series does not approach those values and is therefore not reliable at strong coupling. 

The fact that the series plateaus/converges at low orders for $\lambda=0.01$ and $\lambda=0.1$ but not at the strong couplings of $\lambda=1$ and $\lambda=10$ suggests that this is an asymptotic series that can be predictive for small $\lambda$ at lower orders but that it ultimately diverges at higher orders. This is in accord with Dyson's argument. If you switch the sign of the coupling constant $\lambda$ from positive to negative, the quartic term $-\lambda x^4$ becomes positive and dominates over any quadratic term asymptotically (as $x \to \infty$) regardless of how small the absolute value of $\lambda$ is. The original Euclidean path integral will therefore clearly diverge for negative $\lambda$. So a power series expansion about $\lambda=0$ for positive value of $\lambda$, regardless of how small it is, must also ultimately diverge at higher orders. More interestingly, Dyson's argument can be viewed from a physical point of view. Physically, the situation changes qualitatively if $\lambda$ is negative (regardless again of how small is its absolute value). The potential now yields tunneling (see fig. \ref{Tun}) so that one has an unstable vacuum. This is not consistent with a well behaved absolutely convergent series. This implies that a power series expansion about $\lambda=0$ for positive $\lambda$ must also ultimately diverge. 

In the next section we will see that the second series produces better results at the strong coupling values of $\lambda=1$ and $\lambda=10$. In contrast to the first series, the second series is an absolutely convergent series for any positive value of $\lambda$. Dyson's argument does not apply to the second series because it is not based on a series expansion of the interaction in powers of the coupling i.e. it is not an expansion about zero coupling. 

It is interesting to note that if the limits of integration in the Euclidean path integral were actually finite (ran from $-\beta$ to $\beta$ where $\beta$ is a real positive finite number), the first series in powers of the coupling would also yield an absolutely convergent series (this can be viewed as the QM path integral version of the series $S(n,\beta)$ we obtained for the basic integral in section \reff{SNB}). Dyson's argument would be circumvented here because tunneling at negative $\lambda$ would no longer occur since the potential would now have infinite walls located at $x=\pm \beta$ (see fig. \ref{Tun2}). With no unstable vacuum at negative $\lambda$, Dyson's argument no longer holds. This is discussed further in the conclusion. 

\section{Second series: expansion of quadratic term yields inverse powers of $\lambda$} \label{SQuad}
Expanding the quadratic part in the Euclidean path integral \reff{AmplitudeE} but leaving the quartic part alone yields
\begin{align}
K_E&= \int \mathcal{D} x(\tau)\, e^{-S_E/\hbar}\nonumber\\
&= \int \mathcal{D} x(\tau)\exp\Big[\,\frac{-1}{\hbar}\int_{\tau_a}^{\tau_b} \Big ( \dfrac{1}{2} \,m \dot{x}^2 +\dfrac{1}{2}\,m\,\omega^2\,x^2 +\lambda\,x^4 - J(\tau)\,x\Big)\, d\tau\,\Big]\nonumber\\
&= \int \mathcal{D} x(\tau)\, e^{\frac{1}{\hbar} \int_{\tau_a}^{\tau_b} \,\big(- \lambda\, x^4 +J(\tau)\,x\,\big)\,d\tau}\nonumber\\
&\qquad\Bigg(1- \frac{1}{\hbar} \int_{\tau_a}^{\tau_b}\big ( \dfrac{1}{2} \,m \dot{x}^2 +\dfrac{1}{2}\,m\,\omega^2\,x^2\big) d\tau + \frac{1}{\hbar^2}\,\frac{1}{2!}
\bigg( \int_{\tau_a}^{\tau_b}\Big ( \dfrac{1}{2} \,m \dot{x}^2 +\dfrac{1}{2}\,m\,\omega^2\,x^2\Big)\,d\tau \bigg)^2+...\Bigg)\,.
\label{Series2}
\end{align}
We will ultimately want to compare our series expansion to a numerical value for the exact path integral. This comparison requires us to choose numerical values for the end points. In developing expressions for our series, one can keep the end points $x_0$ and $x_N$ general but it becomes somewhat cumbersome to do so. We will therefore set from the start the end points to the values $x_0=x_N=0$ in order to match those chosen in our previous expansion. This choice simplifies things without of course affecting the structure of the series expansion. 

As usual, to evaluate the above series we have to discretize. The quadratic part (kinetic plus harmonic oscillator) becomes  upon discretization 
\begin{align}
&- \frac{1}{\hbar} \int_{\tau_a}^{\tau_b}\Big ( \dfrac{1}{2} \,m \dot{x}^2 +\dfrac{1}{2}\,m\,\omega^2\,x^2\Big) d\tau= - \frac{1}{\hbar}\, \sum_{i=1}^{N} \Big(\dfrac{1}{2} m \dfrac{ (x_i-x_{i-1})^2}{\epsilon^2}  +\dfrac{1}{2}\,m\,\omega^2\,x_i^2\Big)\,\epsilon\nonumber\\
&=- \frac{m}{2\,\epsilon\,\hbar}\Big( (2 + \omega^2\,\epsilon^2)\sum_{i=1}^{N-1}x_i^2 - 2 \sum_{i=2}^{N-1} x_i\,x_{i-1} \Big)\nonumber\\
\label{DisQuad}
\end{align}
and the quartic part is
\begin{align}
\frac{1}{\hbar} \int_{\tau_a}^{\tau_b} \big(- \lambda\, x^4 +J(\tau)\,x\big)\,d\tau
=\sum_{i=1}^{N-1}-\frac{\lambda\, \epsilon}{\hbar}\, x_i^4+ \vec{J}\cdot\vec{x}
\label{QuartDis}
\end{align}
where $\vec{x}=(x_1,x_2,...,x_{N-1})$. We have absorbed factors of $\epsilon$ and $\hbar$ into a redefinition of $J$ for convenience (this redefinition has no effect since $J$ is not physical and is set to zero in the end. It will simply be used to extract $x_i$ by taking a functional derivative with respect to $J_i$). The path integral measure is 
\begin{align}
\int \mathcal{D} x(\tau)= \Big( \dfrac{m}{2\,\pi\,\epsilon\, \hbar}\Big)^{N/2}\int_{-\infty}^{\infty} dx_1\,dx_2...dx_{N-1}
\label{MeDis}
\end{align}
where the integral is an $N-1$-dimensional integral.  Substituting \reff{DisQuad}, \reff{QuartDis} and \reff{MeDis} into the series \reff{Series2} we obtain
\begin{align}
K_E&=\Big( \dfrac{m}{2\,\pi\,\epsilon\, \hbar}\Big)^{N/2}\int_{-\infty}^{\infty}\exp\Big(-\dfrac{\lambda\, \epsilon}{\hbar}\sum_{i=1}^{N-1}\, x_i^4 + \vec{J}\cdot\vec{x} \Big) \nonumber\\&\qquad\qquad\Bigg(1- \frac{m}{2\,\epsilon\,\hbar}\Big( (2 + \omega^2\,\epsilon^2)\sum_{i=1}^{N-1}x_i^2 - 2 \sum_{i=2}^{N-1} x_i\,x_{i-1} \Big) \nonumber\\
&\qquad\qquad\qquad+ \Big[-\frac{m}{2\,\epsilon\,\hbar} \Big( (2 + \omega^2\,\epsilon^2)\sum_{i=1}^{N-1}x_i^2 - 2 \sum_{i=2}^{N-1} x_i\,x_{i-1} \Big)\Big]^2 \dfrac{1}{2!}+...\Bigg)\,dx_1\,dx_2...dx_{N-1}\,.
\label{SeriesDis}
 \end{align}
We define the generating functional 
\begin{align}
Z[\vec{J}]&=\int_{-\infty}^{\infty}\exp\Big(-\dfrac{\lambda\, \epsilon}{\hbar}\sum_{i=1}^{N-1}\, x_i^4 + \vec{J}\cdot\vec{x} \Big) \,dx_1\,dx_2...dx_{N-1}\nonumber\\&=\int_{-\infty}^{\infty}dx_1\,\exp\Big(-\dfrac{\lambda\, \epsilon}{\hbar} \,x_1^4+ J_1\,x_1\Big)
\int_{-\infty}^{\infty}dx_2\,\exp\Big(-\dfrac{\lambda\, \epsilon}{\hbar} \,x_2^4+ J_2\,x_2\Big)\nonumber\\
&\qquad\qquad...\int_{-\infty}^{\infty}dx_{N-1}\,\exp\Big(-\dfrac{\lambda\, \epsilon}{\hbar} \,x_{N-1}^4+ J_{N-1}\,x_{N-1}\Big)\nonumber\\
&=\prod_{i=1}^{N-1}\, I[J_i]
\label{ZJ}
\end{align}
where $I[J_i]$ is a one-dimensional integral which can be expressed in terms of generalized hypergeometric functions:
\begin{align}
I[J_i]&=\int_{-\infty}^{\infty}dx_i\,\exp\Big(-\dfrac{\lambda\, \epsilon}{\hbar} \,x_i^4+ J_i\,x_i\Big)\nonumber\\
&=2\,\Gamma\big(\frac{5}{4}\big) \,\Big(\frac{\hbar}{\epsilon\,\lambda}\Big)^{1/4} \, _0F_2\left(;\frac{1}{2},\frac{3}{4};\frac{J_i^4 \,\hbar }{256\, \epsilon \, \lambda }\right)+ \frac{1}{4}\,\Gamma\big(\frac{3}{4}\big)\, \Big(\frac{\hbar}{\epsilon\,\lambda}\Big)^{3/4} \,  _0F_2\left(;\frac{5}{4},\frac{3}{2};\frac{J_i^4 \,\hbar }{256\, \epsilon\,  \lambda }\right)\,J_i^2\,.
\label{IJ}
\end{align}
Here $_0F_2\left(;\frac{1}{2},\frac{3}{4};\frac{J_i^4 \,\hbar }{256\, \epsilon \, \lambda }\right)$ and $ _0F_2\left(;\frac{5}{4},\frac{3}{2};\frac{J_i^4 \,\hbar }{256\, \epsilon\,  \lambda }\right)$ are generalized hypergoemetric functions\\ 
 $_pF_q(a ; b ; z)$ with $p=0$, $q=2$ and $z=\frac{J_i^4 \,\hbar }{256\, \epsilon\,  \lambda }$. The $a$ and $b$ in the function  $_pF_q(a ; b; z)$ are short-hand for a set of coefficients $\{a_1,...,a_p\}$ and $\{b_1,...,b_q\}$.  The function $_pF_q(a ; b; z)$ can be defined via its series expansion:
\begin{align}
 _pF_q(a ; b ; z)&=\sum_{k=0}^{\infty}\dfrac{(a_1)_k...(a_p)_k}{(b_1)_k...(b_q)_k}\,\dfrac{z^k}{k!}\nonumber\\
 &=1+ \dfrac{a_1...  a_p}{ b_1 ... b_q}\,z+ \dfrac{a_1(a_1+1) ... a_p(a_p+1)}{ b_1(b_1+1) ... b_q(b_q+1)}\,\dfrac{z^2}{2!}+...
\label{Fq}
\end{align}
where $(a)_0 \!=1$ and $(a)_k\! =a(a+1)...(a+k-1)$ are the Pochammer symbols. Its derivative with respect to $z$ yields again a hypergeometric function but with $a \to a+1$ and $b \to b+1$ i.e. 
\begin{align} 
\dfrac{\partial}{\partial z}\, _pF_q(a ; b; z)= \dfrac{a_1...  a_p}{ b_1 ... b_q}\, _pF_q(a+1; b+1; z)\,.
\label{Prop}
\end{align}
Since $ _pF_q(a; b ; 0)=1$, it follows from \reff{IJ} and \reff{ZJ} that
 \begin{align}
I[0]&=2\,\Gamma\big(\frac{5}{4}\big) \,\Big(\frac{\hbar}{\epsilon\,\lambda}\Big)^{1/4}
\label{I0}\\
Z[0]&= I[0]^{N-1}=\Big[2\,\Gamma\big(\frac{5}{4}\big) \,\Big(\frac{\hbar}{\epsilon\,\lambda}\Big)^{1/4}\Big]^{N-1}\,.
\label{Z0}
\end{align}
Note that the above expression \reff{IJ} has inverse powers (or negative powers) of $\lambda$. We will see that the expansion \reff{SeriesDis} is a series in inverse powers of $\lambda$ in contrast to the positive powers of $\lambda$ that one encounters in usual perturbation theory. This is a tell-tale sign that this series is suited for strong coupling $\lambda$ and hence non-perturbative phenomena.

Note that a functional derivative with respect to $J_i$ of $Z[\vec{J}]$  brings down a factor of $x_i$. The series \reff{SeriesDis} is in powers of the quadratic term \reff{DisQuad}. It is therefore convenient to replace $x_i$ in \reff{DisQuad} by a functional derivative. This yields the operator
\begin{align}
\hat{Q}= - \frac{m}{2\,\epsilon\,\hbar}\Bigg((2 + \omega^2\,\epsilon^2) \sum_{i=1}^{N-1}\big(\frac{\delta}{\delta J_i}\big)^2 -2 \sum_{i=2}^{N-1} \frac{\delta}{\delta J_i}\,\frac{\delta}{\delta J_{i-1}} \Bigg)\,.
\label{QHat}
\end{align}
We can therefore express the series \reff{SeriesDis} as
\begin{align}
K_E&= C \, \sum_{n=0}^{\infty} \,\frac{1}{n!} \,\hat{Q}^n \, Z[\vec{J}]\,\Bigr\rvert_{\vec{J}=0}\nonumber\\
&=C \Big[ Z[0] +  \hat{Q} \, Z[\vec{J}]\,\Bigr\rvert_{\vec{J}=0}+ \frac{1}{2!} \,\hat{Q}^2 \, Z[\vec{J}]\,\Bigr\rvert_{\vec{J}=0} +...\Big]
\label{HypSeries}
\end{align}
where the prefactor $C$ is
\beq
C=\Big( \dfrac{m}{2\,\pi\,\epsilon\, \hbar}\Big)^{N/2}
\eeq{CC}
 and we set $\vec{J}=0$ at the end as it is not part of the physical system. $Z[\vec{J}]$, given by \reff{ZJ}, is a product of the $I[J_i]\,$s. To evaluate the above series \reff{HypSeries}, we therefore need to determine the derivatives with respect to $J_i$ of $I[J_i]$. By looking at the integral in the first line of \reff{IJ}, derivatives of $I[J_i]$ with respect to $J_i$ bring down a factor of $x_i$ in the integral. It follows that odd derivatives bring down odd powers of $x_i$ so that after $J_i$ is set to zero, the integrand is an odd function and the integral is  zero. So only even derivatives survive and this is given by the simple expression  
\begin{align}
\Big(\frac{\delta}{\delta J_i}\Big)^{2n} I[J_i] \,\Bigr\rvert_{J_i=0} &=\int_{-\infty}^{\infty}dx_i\,\exp\Big(-\dfrac{\lambda\, \epsilon}{\hbar} \,x_i^4\Big)\,x_i^{2n}\nonumber\\&=\frac{1}{2} \,\Gamma\big(\frac{2n+1}{4}\big)\,\Big(\frac{\hbar}{\epsilon\,\lambda}\Big)^{\frac{2n+1}{4}}
\label{Central}
\end{align}
where $n$ is any non-negative integer. When $n=0$ we recover \reff{I0} since $\Gamma\big(\frac{1}{4}\big)=4\,\Gamma\big(\frac{5}{4}\big)$. The above result \reff{Central} is central to evaluating the series \reff{HypSeries}.  

\subsection{First order result} \label{FirstO}

Using \reff{Central}, the first order ($n=1$) contribution to the above series \reff{HypSeries} is given by
\begin{align}
\hat{Q} \, Z[\vec{J}]\,\Bigr\rvert_{\vec{J}=0}&=- \frac{m}{2\,\epsilon\,\hbar}\Bigg( (2 + \omega^2\,\epsilon^2) \sum_{i=1}^{N-1}\big(\frac{\delta}{\delta J_i}\big)^2 -2 \sum_{i=2}^{N-1} \frac{\delta}{\delta J_i}\,\frac{\delta}{\delta J_{i-1}} \Bigg)\,\prod_{i=1}^{N-1}\, I[J_i]\,\Bigr\rvert_{\vec{J}=0}\nonumber\\
&=- \frac{m}{2\,\epsilon\,\hbar}\, (2 + \omega^2\,\epsilon^2)\,\sum_{i=1}^{N-1}\big(\frac{\delta}{\delta J_i}\big)^2 \,\prod_{i=1}^{N-1}\, I[J_i]\,\Bigr\rvert_{\vec{J}=0}\nonumber\\
&=- \frac{m}{2\,\epsilon\,\hbar}\, (2 + \omega^2\,\epsilon^2)\sum_{i=1}^{N-1}
 I[J_1]\,I[J_2]...\big(\frac{\delta}{\delta J_i}\big)^2 I[J_i]...I[J_{N-1}] \,\Bigr\rvert_{\vec{J}=0}\nonumber\\
 &= - \frac{m}{2\,\epsilon\,\hbar}\, (2 + \omega^2\,\epsilon^2)\,\sum_{i=1}^{N-1}I[0]^{N-2}\, \big(\frac{\delta}{\delta J_i}\big)^2 I[J_i]\,\Bigr\rvert_{\vec{J}=0}\nonumber\\
 &= - \frac{m}{2\,\epsilon\,\hbar}\, (2 + \omega^2\,\epsilon^2) (N-1)\,I[0]^{N-2}\,\frac{1}{2} \,\Gamma\big(\frac{3}{4}\big)\,\Big(\frac{\hbar}{\epsilon\,\lambda}\Big)^{\frac{3}{4}}\nonumber\\
 &= -Z[0]\, (N-1)\,\dfrac{1}{\lambda^{1/2}}\,\frac{\Gamma\big(\frac{3}{4}\big)}{\Gamma\big(\frac{5}{4}\big)}\,(2 + \omega^2\,\epsilon^2)\,\frac{m}{8\, \hbar^{1/2}\,\epsilon^{3/2}}\,.
 \label{Order1}
\end{align}
The second term in the large round brackets of the first line above makes a zero contribution because $\tfrac{\delta}{\delta J_i}$ and $\tfrac{\delta}{\delta J_{i-1}}$ are each single derivatives and odd derivatives acting on $I[J_i]$ make a zero contribution after $J_i$ is set to zero. Since $Z[0]=I[0]^{N-1}$, we replaced $I[0]^{N-2}$ by $Z[0]\,I[0]^{-1}=Z[0]\, \tfrac{\big(\frac{\epsilon\,\lambda}{\hbar}\big)^{1/4}}{2\,\Gamma\big(\frac{5}{4}\big)}$ where \reff{I0} was used for $I[0]$. This is convenient since $Z[0]$ can be pulled out as a common factor for all orders. 

The series \reff{HypSeries} up to first order (subscript (1)) is then given by the analytical formula 
\begin{align}
K_{E_{(1)}}& =C \, Z[0]\,\Bigg(1-  \dfrac{1}{\lambda^{1/2}}\,(N-1)\,\frac{\Gamma\big(\frac{3}{4}\big)}{\Gamma\big(\frac{5}{4}\big)}\,(2 + \omega^2\,\epsilon^2)\,\frac{m}{8\, \hbar^{1/2}\,\epsilon^{3/2}}\Bigg)
\label{FirstOrder}
\end{align}
where $C$ is given by \reff{CC}.  The expression \reff{FirstOrder} is a function of $N$,  the coupling constant $\lambda$ and the parameters $\omega$ and $m$ as well as the constant $\hbar$. It depends also on the time interval $\mathcal{T}$ via $\epsilon=\mathcal{T}/N$. Having an expression as a function of $N$ is very useful since numerically, $N$ is the number of integrations required in the original path integral and this can become computationally intensive in the continuum limit where $N$ is large and formally infinite. 

As a simple check on our calculation \reff{Order1}, we compared the analytical formula  \reff{FirstOrder} to a first order numerical integration of the series \reff{SeriesDis} for the case of $N=4$ which involves $N-1=3$ integrals. We used the following numerical values for the parameters: $m=\hbar=\omega=\mathcal{T}=1$. Hence  $\epsilon=\mathcal{T}/N=1/4$. The numerical value of $\lambda$ was not specified. The analytical formula and first order numerical integration \textit{matched} and gave the following result: 
\beq
K_{E_{(1)}}= \frac{64 \,\sqrt{2}\, \Gamma \left(\frac{5}{4}\right)^3}{\pi ^2 \,\lambda ^{3/4}}-\frac{99 \,\Gamma \left(\frac{1}{4}\right)}{2 \,\pi \, \lambda ^{5/4}} \qquad \text{for } N=4 \text{ and } m=\hbar=\omega=\mathcal{T}=1.
\eeq{FirstOrderNumeric}

The inverse powers of $\lambda$ above illustrates again that this series is outside the usual perturbative regime and is well suited to the strong coupling non-perturbative regime. 

\subsection{Second order result}\label{SecondO}

To obtain the second order contribution to \reff{HypSeries} we need to evaluate the operator $\hat{Q}^2$: 
\begin{align}
\hat{Q}^2&=\Bigg[- \frac{m}{2\,\epsilon\,\hbar}\Big((2 + \omega^2\,\epsilon^2) \sum_{i=1}^{N-1}\big(\frac{\delta}{\delta J_i}\big)^2 -2 \sum_{i=2}^{N-1} \frac{\delta}{\delta J_i}\,\frac{\delta}{\delta J_{i-1}} \Big)\Bigg]^2\nonumber\\
&= \frac{m^2}{4\,\epsilon^2\,\hbar^2}\Bigg[(2 + \omega^2\,\epsilon^2)^2 \sum_{i_1=1}^{N-1}\sum_{i_2=1}^{N-1}\big(\frac{\delta}{\delta J_{i_1}}\big)^2 \big(\frac{\delta}{\delta J_{i_2}}\big)^2 -4 (2 + \omega^2\,\epsilon^2)\sum_{i_1=1}^{N-1}\big(\frac{\delta}{\delta J_{i_1}}\big)^2  \sum_{i_2=2}^{N-1} \frac{\delta}{\delta J_{i_2}}\,\frac{\delta}{\delta J_{i_2-1}} \nonumber\\
&\qquad\qquad\qquad+4  \sum_{i_1=2}^{N-1} \, \sum_{i_2=2}^{N-1} \frac{\delta}{\delta J_{i_2}}\,\frac{\delta}{\delta J_{i_2-1}}  \frac{\delta}{\delta J_{i_1}}\,\frac{\delta}{\delta J_{i_1-1}} \Bigg]\,.
\label{Q2}
\end{align}
The second term in the square brackets contains an odd number of derivatives and makes a zero contribution when acting on $Z[\vec{J}]$. There is an odd number because if $i_1\ne i_2$ then there is a single derivative with label $i_2$ and  a single derivative with separate label $i_2-1$ (each of them is odd) and if $i_1=i_2$ (or $i_1=i_2-1$) then there are three derivatives with label $i_1$ and either one derivative with label $i_1-1$ or one derivative with label $i_1+1$ (the derivatives on each label are odd). So the part of $\hat{Q}^2$ that yields a non-zero result is 
\begin{align}
\hat{Q}^2&= \frac{m^2}{4\,\epsilon^2\,\hbar^2}\Bigg[(2 + \omega^2\,\epsilon^2)^2 \sum_{i_1=1}^{N-1}\sum_{i_2=1}^{N-1}\big(\frac{\delta}{\delta J_{i_1}}\big)^2 \big(\frac{\delta}{\delta J_{i_2}}\big)^2 +4  \sum_{i_1=2}^{N-1} \, \sum_{i_2=2}^{N-1} \frac{\delta}{\delta J_{i_2}}\,\frac{\delta}{\delta J_{i_2-1}}  \frac{\delta}{\delta J_{i_1}}\,\frac{\delta}{\delta J_{i_1-1}} \Bigg]\,.
\label{Q22}
\end{align}
The double sum in the first term can be broken into a sum where $i_1=i_2$ plus another sum where $i_1\ne i_2$ i.e.
\begin{align}
\sum_{i_1=1}^{N-1}\sum_{i_2=1}^{N-1}\big(\frac{\delta}{\delta J_{i_1}}\big)^2 \big(\frac{\delta}{\delta J_{i_2}}\big)^2=\sum_{i=1}^{N-1} \big(\frac{\delta}{\delta J_{i}}\big)^4 
+\sum_{\substack{ i_1,i_2=1\\ i_1\ne i_2}}^{N-1}\,\big(\frac{\delta}{\delta J_{i_1}}\big)^2 \big(\frac{\delta}{\delta J_{i_2}}\big)^2 \,.
\label{Order2A}
\end{align}
Above, we have derivatives to the power of $4$ and $2$. When they act on $I[J_i]$, using \reff{Central},  they yield 
\begin{align}
\Big(\frac{\delta}{\delta J_i}\Big)^{4} I[J_i] \,\Bigr\rvert_{J_i=0} &=\frac{1}{2} \,\Gamma\big(\frac{5}{4}\big)\,\Big(\frac{\hbar}{\epsilon\,\lambda}\Big)^{\frac{5}{4}}\label{J4}\\
\Big(\frac{\delta}{\delta J_i}\Big)^{2} I[J_i] \,\Bigr\rvert_{J_i=0} &=\frac{1}{2} \,\Gamma\big(\frac{3}{4}\big)\,\Big(\frac{\hbar}{\epsilon\,\lambda}\Big)^{\frac{3}{4}}\label{J2}\,.
\end{align}
Since $Z[\vec{J}]$ is a product of $N\m 1$ $I[J_i]\,$s, it is convenient to think of it as a string of $N\m 1$ boxes each containing one $I[J_i]$ where $i$ runs from $1$ to $N-1$ inclusively. In \reff{Order2A} we have summations of derivatives and in the series they act on $Z[\vec{J}]$. For example, we have to evaluate $ \sum_{i=1}^{N-1} \big(\tfrac{\delta}{\delta J_{i}}\big)^4 \,Z[\vec{J}]\Bigr\rvert_{\vec{J}=0}$. A convenient way to think of this is the following:  $\big(\frac{\delta}{\delta J_{i}}\big)^4$ is a single object (a four-derivative object) and we want to know ``In how many ways can we place a single object among $N\m 1$ boxes?''. The answer is $N\m 1$ ways. Placing the object in a box yields the result \reff{J4} while the remaining $N\m 2$ boxes contribute $I[0]^{N-2}=Z[0]\,I[0]^{-1}=Z[0]\, \tfrac{\big(\frac{\epsilon\,\lambda}{\hbar}\big)^{1/4}}{2\,\Gamma\big(\frac{5}{4}\big)}$. Putting these results together yields
\beq  
\sum_{i=1}^{N-1}  \big(\frac{\delta}{\delta J_{i}}\big)^4\,Z[\vec{J}]\Bigr\rvert_{\vec{J}=0}= (N-1) \,Z[0]\, \frac{1}{2} \,\Gamma\big(\frac{5}{4}\big)\,\Big(\frac{\hbar}{\epsilon\,\lambda}\Big)^{\frac{5}{4}} \frac{\big(\frac{\epsilon\,\lambda}{\hbar}\big)^{1/4}}{2\,\Gamma\big(\frac{5}{4}\big)}=(N-1) \,Z[0]\,\frac{\hbar}{4\,\epsilon\,\lambda}\,.
\eeq{Order2A1}
In the second part of \reff{Order2A} we have a sum over $\big(\frac{\delta}{\delta J_{i_1}}\big)^2 \big(\frac{\delta}{\delta J_{i_2}}\big)^2$ where $i_1\ne i_2$. We now have two distinct objects, each one being a two-derivative object. When acting on  $Z[\vec{J}]$  we now want to know ``In how many ways can we place two distinguishable objects among  $N\m1$ boxes?'' The answer is $\binom{N-1}{2}\,2!= (N\m1)\,(N\m2)$ ways. Placing one object in a box contributes a factor of \reff{J2} and placing the second one in a different box multiplies it by the same factor, yielding a factor squared. After the two objects are placed, the remaining $N\m 3$ boxes contribute $I[0]^{N-3}=Z[0]\,I[0]^{-2}$. We therefore obtain
\beq  
\sum_{\substack{ i_1,i_2=1\\ i_1\ne i_2}}^{N-1}\,\big(\frac{\delta}{\delta J_{i_1}}\big)^2 \big(\frac{\delta}{\delta J_{i_2}}\big)^2 \,Z[\vec{J}]\Bigr\rvert_{\vec{J}=0}= (N-1)\,(N-2) \,Z[0]\, \Bigg(\frac{\Gamma\big(\frac{3}{4}\big)}{\Gamma\big(\frac{5}{4}\big)}\Bigg)^2\,\frac{\hbar}{16\,\epsilon\,\lambda}\,.
\eeq{Order2A2}
We now need to evaluate the second term in \reff{Q22}. When acting on $Z[\vec{J}]$, the result is zero unless $i_1=i_2$. We have to evaluate $\sum_{i=2}^{N-1} \big(\frac{\delta}{\delta J_{i}}\big)^2 \big(\frac{\delta}{\delta J_{i-1}}\big)^2$ acting on $Z[\vec{J}]$.  We have two consecutive objects: $i-1$ followed by $i$. They fill consecutive boxes and hence come in pairs. How many ways are there to place a pair of consecutive objects along the $N\m1$ boxes?  The answer is $N\m 2$. The two filled boxes   
contribute a factor of \reff{J2} squared. The remaining $N\m3$ boxes contribute $I[0]^{N-3}=Z[0]\,I[0]^{-2}$. This results in 
\beq  
\sum_{i=1}^{N-1}\,\big(\frac{\delta}{\delta J_{i}}\big)^2 \big(\frac{\delta}{\delta J_{i-1}}\big)^2 \,Z[\vec{J}]\Bigr\rvert_{\vec{J}=0}= (N-2) \,Z[0]\, \Bigg(\frac{\Gamma\big(\frac{3}{4}\big)}{\Gamma\big(\frac{5}{4}\big)}\Bigg)^2\,\frac{\hbar}{4\,\epsilon\,\lambda}\,.
\eeq{Order2B}
Using the results \reff{Order2A1}, \reff{Order2A2}, \reff{Order2B} and $\hat{Q}^2$ given by \reff{Q22},  the second order contribution to the series \reff{HypSeries} is given by 
\begin{align}
\frac{1}{2!} \,\hat{Q}^2 \, Z[\vec{J}]\,\Bigr\rvert_{\vec{J}=0}&= Z[0]\,\dfrac{1}{\lambda}\dfrac{m^2}{32\,\epsilon^3\,\hbar}
\,\Bigg[(2 + \omega^2\,\epsilon^2)^2\,\bigg((N-1)+\dfrac{1}{4}\, (N-1)\,(N-2) \,\Big(\frac{\Gamma\big(\frac{3}{4}\big)}{\Gamma\big(\frac{5}{4}\big)}\Big)^2 \bigg)\nonumber\\&\qquad\qquad\qquad\qquad\qquad+ (N-2) \, \Big(\frac{\Gamma\big(\frac{3}{4}\big)}{\Gamma\big(\frac{5}{4}\big)}\Big)^2\,\Bigg]\,.
\label{Order2}
\end{align}
The series \reff{HypSeries} up to second order (subscript (2)) is then given by the analytical formula 
\begin{align}
K_{E_{(2)}}& =C \, Z[0]\,\Bigg[1-  \dfrac{1}{\lambda^{1/2}}\,(N-1)\,\frac{\Gamma\big(\frac{3}{4}\big)}{\Gamma\big(\frac{5}{4}\big)}\,(2 + \omega^2\,\epsilon^2)\,\frac{m}{8\, \hbar^{1/2}\,\epsilon^{3/2}}\nonumber\\&\qquad\qquad\quad\quad+\dfrac{1}{\lambda}\,\dfrac{m^2}{32\,\epsilon^3\,\hbar}\bigg((2 + \omega^2\,\epsilon^2)^2\,\Big[(N-1)+\dfrac{1}{4}\, (N-1)\,(N-2) \,\Big(\frac{\Gamma\big(\frac{3}{4}\big)}{\Gamma\big(\frac{5}{4}\big)}\Big)^2 \,\Big]\nonumber\\&\qquad\qquad\qquad\qquad\qquad\qquad\qquad+ (N-2) \, \Big(\frac{\Gamma\big(\frac{3}{4}\big)}{\Gamma\big(\frac{5}{4}\big)}\Big)^2\,\bigg)\Bigg]\,.
\label{SecondOrder}
\end{align}
Note that in the above series, the signs alternate, and the first order term has an inverse power of $\lambda^{1/2}$ while the second order term has an inverse power of $\lambda$. 

The series \reff{HypSeries} is an alternating series in inverse powers of $\lambda^{1/2}$ and has the following form:
\begin{align}
K_E& =C \, Z[0]\,\Bigg[1-  \dfrac{b_1}{\lambda^{1/2}}+\dfrac{b_2}{\lambda}-\dfrac{b_3}{\lambda^{3/2}}+...(-1)^n\,\dfrac{b_n}{\lambda^{n/2}}\Bigg]
\label{SeriesForm}
\end{align}
where the $b_i\,$ are positive and depend on $N$, $\mathcal{T}$ (via $\epsilon$) as well as the parameters $\omega$, $m$ and the constant $\hbar$. 

\subsection{Third order result: first non-zero cross term} \label{ThirdO}

Before discussing the general procedure for generating the $n$th order result, it is important to work out also the third order result explicitly. To see why, let us write the operator $\hat{Q}$ given by \reff{QHat} as a sum of two operators $\hat{A}$ and $\hat{B}$:
\begin{align}
\hat{Q}= - \frac{m}{2\,\epsilon\,\hbar}\Big(\hat{A}+\hat{B}\Big)
\label{AB}
\end{align}
where 
\begin{align}
\hat{A}=(2 + \omega^2\,\epsilon^2) \sum_{i=1}^{N-1}\big(\frac{\delta}{\delta J_i}\big)^2 
\label{AA}
\end{align}
and 
\begin{align}
\hat{B}=-2 \sum_{i=2}^{N-1} \frac{\delta}{\delta J_i}\,\frac{\delta}{\delta J_{i-1}}\,.
\label{BB}
\end{align}

The first order contribution \reff{Order1}, stemming from one power of $\hat{Q}$, received a non-zero contribution from $\hat{A}$ but not from $\hat{B}$. The second order contribution \reff{Order2} stemming from $\hat{Q}^2$, received a non-zero contribution from $\hat{A}^2$ and $\hat{B}^2$ but not from the cross term $\hat{A}\,\hat{B}$. The reason is that any term with an odd power of $\hat{B}$ makes a zero contribution. At third order,  $\hat{Q}^3$ will receive non-zero contributions from  $\hat{A}^3$ and the cross-term $\hat{A}\,\hat{B}^2$. So it is at third order that we encounter for the first time a cross-term that makes a non-zero contribution.  The general $n$th order result stemming from $\hat{Q}^n$ will contain cross terms $\hat{A}^m\,\hat{B}^k$ that will make a non-zero contribution when $k$ is a positive even integer (here $n=m+k$ with $m$ a positive integer). Calculating explicitly the cross-term $\hat{A}\,\hat{B}^2$ that appears at third order will therefore prepare us to follow the more general $n$th order case. Moreover, the $\hat{A}^3$ term itself is considerably more complicated to work out than $\hat{A}^2$ so it will also be beneficial to see this calculation explicitly. 

The third order contribution to the series \reff{HypSeries} is given by 
\begin{align}
 \frac{1}{3!} \,\hat{Q}^3 \, Z[\vec{J}]\,\Bigr\rvert_{\vec{J}=0}\,.
 \label{Third}
\end{align} 
We therefore need to evaluate the operator $\hat{Q}^3$. $\hat{Q}$ is given by \reff{AB} where the operators $\hat{A}$ and $\hat{B}$ are given by \reff{AA} and \reff{BB} respectively:
\begin{align}
\hat{Q}^3&=\Bigg[- \frac{m}{2\,\epsilon\,\hbar}\Big(\hat{A}+\hat{B}\Big)\Bigg]^3\nonumber\\
&= -\frac{m^3}{8\,\epsilon^3\,\hbar^3}\Big(\hat{A}^3 + 3\,\hat{A}^2\,\hat{B} + 3 \hat{A}\,\hat{B}^2 + \hat{B}^3\Big)\nonumber\\
&= -\frac{m^3}{8\,\epsilon^3\,\hbar^3}\Big(\hat{A}^3 + 3 \hat{A}\,\hat{B}^2\Big)\nonumber\\
&= -\frac{m^3}{8\,\epsilon^3\,\hbar^3}\Big[\,(2 + \omega^2\,\epsilon^2)^3 \sum_{i_1,i_2,i_3=1}^{N-1}\big(\frac{\delta}{\delta J_{i_1}}\big)^2 \big(\frac{\delta}{\delta J_{i_2}}\big)^2 \big(\frac{\delta}{\delta J_{i_3}}\big)^2\nonumber\\&\qquad\qquad \qquad+12\,(2 + \omega^2\,\epsilon^2)\,\sum_{i_1=1}^{N-1}\sum_{i_2,i_3=2}^{N-1}\big(\frac{\delta}{\delta J_{i_1}}\big)^2 \frac{\delta}{\delta J_{i_2}}\,\frac{\delta}{\delta J_{i_2-1}}\frac{\delta}{\delta J_{i_3}}\,\frac{\delta}{\delta J_{i_3-1}}\,\Big]
\label{Q3}
\end{align}
where $3\hat{A}^2\hat{B}$ and $\hat{B}^3$ appearing in the second line will make a zero contribution when acting on $Z[\vec{J}]$ since they contain odd powers of $\hat{B}$. In view of this future application, we simply omit those two terms in the third line and in the final result \reff{Q3}. 

The triple sum in the first term of \reff{Q3} stems from $\hat{A}^3$ and  can be broken up into three cases: 1) all three i's are equal: $i_1=i_2=i_3$  2) two i's are equal: $i_1=i_2$ or $i_1=i_3$, or $i_2=i_3$ 3) all three i's are different: $i_1\ne i_2\ne i_3$.

\textit{Case 1}: $i_1=i_2=i_3=i$
\begin{align}
&\big(\frac{\delta}{\delta J_{i_1}}\big)^2 \big(\frac{\delta}{\delta J_{i_2}}\big)^2 \big(\frac{\delta}{\delta J_{i_3}}\big)^2=
\big(\frac{\delta}{\delta J_{i}}\big)^6\nonumber\\
&\sum_{i=1}^{N-1}\big(\frac{\delta}{\delta J_{i}}\big)^6 \,Z[\vec{J}]\Bigr\rvert_{\vec{J}=0}=(N-1)\,Z[0] \frac{1}{4} \,\frac{\Gamma\big(\frac{7}{4}\big)}{\Gamma\big(\frac{5}{4}\big)}\,\Big(\frac{\hbar}{\epsilon\,\lambda}\Big)^{3/2}\,.
\label{A31}
\end{align}
The factor of $N\m 1$ is the number of ways one object (a six-derivative object) can be placed in the $N\m 1$ boxes of $Z[\vec{J}]$ containing each an $I[J_i]$. We then used \reff{Central} with $n=3$ and the remaining $N\m 2$ boxes contributed  $I[0]^{N-2}=Z[0]\,I[0]^{-1}$ with $I[0]$ replaced by \reff{I0}. 

\textit{Case 2}: two i's are equal\\
One can have $i_1=i_2$ or $i_1=i_3$ or $i_2=i_3$. Each makes the same contribution so there is a factor of 3.  
\begin{align}
&\big(\frac{\delta}{\delta J_{i_1}}\big)^2 \big(\frac{\delta}{\delta J_{i_2}}\big)^2 \big(\frac{\delta}{\delta J_{i_3}}\big)^2 \to
3\,\big(\frac{\delta}{\delta J_{i_1}}\big)^4\big(\frac{\delta}{\delta J_{i_2}}\big)^2\nonumber\\
&\sum_{i_1,i_2=1}^{N-1}3\,\big(\frac{\delta}{\delta J_{i}}\big)^4\big(\frac{\delta}{\delta J_{i_2}}\big)^2\,Z[\vec{J}]\Bigr\rvert_{\vec{J}=0}=\frac{3}{16}\,(N\m 1)\,(N\m 2)\,Z[0]  \,\frac{\Gamma\big(\frac{3}{4}\big)}{\Gamma\big(\frac{5}{4}\big)}\,\Big(\frac{\hbar}{\epsilon\,\lambda}\Big)^{3/2}\,.
\label{A32}
\end{align}
The factor of $(N\m 1)\,(N\m2)$ is the number of ways two distinguishable objects (a four-derivative and two-derivative object) can be placed in the $N\m 1$ boxes of $Z[\vec{J}]$ containing each an $I[J_i]$. We then used \reff{Central} with $n=2$ and $n=1$. The $N\m3$ remaining boxes contribute $I[0]^{N-3}=Z[0]\,I[0]^{-2}$ with $I[0]$ given by \reff{I0}. 

\textit{Case 3}:  $i_1\ne i_2\ne i_3$
\begin{align}
 \sum_{\substack{ i_1,i_2,i_3=1\\ i_1\ne i_2\ne i_3}}^{N-1}&\big(\frac{\delta}{\delta J_{i_1}}\big)^2 \big(\frac{\delta}{\delta J_{i_2}}\big)^2 \big(\frac{\delta}{\delta J_{i_3}}\big)^2\,Z[\vec{J}]\Bigr\rvert_{\vec{J}=0}\nonumber\\&=\frac{1}{64} \,(N\m 1)\,(N\m 2)\,(N\m 3)\,Z[0]\,\Big(\frac{\Gamma\big(\frac{3}{4}\big)}{\Gamma\big(\frac{5}{4}\big)}\Big)^3\Big(\frac{\hbar}{\epsilon\,\lambda}\Big)^{3/2}\,.
\label{A33}
\end{align}
The factor of $(N\m 1)\,(N\m2)\,(N \m 3)$ is the number of ways three distinguishable objects ( three two-derivative objects) can be placed in the $N\m 1$ boxes of $Z[\vec{J}]$ containing each an $I[J_i]$. We then used \reff{Central} with $n=1$. The $N\m4$ remaining boxes contribute $I[0]^{N-4}=Z[0]\,I[0]^{-3}$. 

We now consider the triple sum in the second term of \reff{Q3} which stems from the cross-term $3\,\hat{A}\,\hat{B}^2$. There are two cases: I) $i_2=i_3$ with $i_1\ne i_2$ and $i_1 \ne i_2-1$ and II) $i_2=i_3$ with either $i_1=i_2$ or $i_1=i_2-1$. The  important thing to note is that we need to set $i_2=i_3$ so that the $\hat{B}^2$ part has even derivatives.  This is not an issue for $\hat{A}$ since it automatically has even derivatives. 

\textit{Case I}: $i_2=i_3$ with $i_1\ne i_2$ and $i_1 \ne i_2-1$
\begin{align}
&\big(\frac{\delta}{\delta J_{i_1}}\big)^2 \frac{\delta}{\delta J_{i_2}}\,\frac{\delta}{\delta J_{i_2-1}}\frac{\delta}{\delta J_{i_3}}\,\frac{\delta}{\delta J_{i_3-1}}=\big(\frac{\delta}{\delta J_{i_1}}\big)^2\,\big( \frac{\delta}{\delta J_{i_2}}\big)^2\,\big(\frac{\delta}{\delta J_{i_2-1}}\big)^2\nonumber\\
&\sum_{i_1=1}^{N-1}\sum_{i_2=2}^{N-1}\big(\frac{\delta}{\delta J_{i_1}}\big)^2\,\big( \frac{\delta}{\delta J_{i_2}}\big)^2\,\big(\frac{\delta}{\delta J_{i_2-1}}\big)^2\,Z[\vec{J}]\Bigr\rvert_{\vec{J}=0}=\frac{1}{64} \,(N\m 2)\,(N\m 3)\,Z[0]\,
\Big(\frac{\Gamma\big(\frac{3}{4}\big)}{\Gamma\big(\frac{5}{4}\big)}\Big)^3\Big(\frac{\hbar}{\epsilon\,\lambda}\Big)^{3/2}.
\label{B31}
\end{align}
Note that $\big( \frac{\delta}{\delta J_{i_2}}\big)^2\,\big(\frac{\delta}{\delta J_{i_2-1}}\big)^2$ does not contain two independent objects since it always takes up two consecutive boxes. It has only one label ($i_2$) and is treated as a single composite object. The factor of $(N\m2)\,(N \m 3)$ is the number of ways two distinguishable objects (one regular object plus one composite object) can be placed in $N\m 2$ boxes (not $N \m 1$ boxes because the composite object takes up two boxes). We then used \reff{Central} with $n=1$. The $N\m4$ remaining boxes contribute $I[0]^{N-4}=Z[0]\,I[0]^{-3}$. 

\textit{Case II}: $i_2=i_3$ with either $i_1= i_2$ or $i_1= i_2-1$
\begin{align}
&\big(\frac{\delta}{\delta J_{i_1}}\big)^2 \frac{\delta}{\delta J_{i_2}}\,\frac{\delta}{\delta J_{i_2-1}}\frac{\delta}{\delta J_{i_3}}\,\frac{\delta}{\delta J_{i_3-1}}=\big(\frac{\delta}{\delta J_{i_1}}\big)^4\,\big(\frac{\delta}{\delta J_{i_1\pm 1}}\big)^2\nonumber\\
&\sum_{i_1=1}^{N-1}\big(\frac{\delta}{\delta J_{i_1}}\big)^4\,\big(\frac{\delta}{\delta J_{i_1\pm 1}}\big)^2\,Z[\vec{J}]\Bigr\rvert_{\vec{J}=0}=\frac{1}{8} \,(N\m 2)\,Z[0]\,
\Big(\frac{\Gamma\big(\frac{3}{4}\big)}{\Gamma\big(\frac{5}{4}\big)}\Big)\,\Big(\frac{\hbar}{\epsilon\,\lambda}\Big)^{3/2}.
\label{B32}
\end{align}
The factor of $(N\m 2)$ is the number of ways a composite object can be placed in $N\m 2$ boxes. We then used \reff{Central} with $n=2$ and $n=1$. The $N\m3$ remaining boxes contribute $I[0]^{N-3}=Z[0]\,I[0]^{-2}$. In $\big(\frac{\delta}{\delta J_{i_1\pm 1}}\big)^2$, the $+$ and $-$ cases make the same contribution and both are included in the final result of \reff{B32}.

We now have all the necessary results to obtain the third order contribution \reff{ThirdOrder}. In \reff{Q3}, the term  $\hat{A}^3$  is the sum of \reff{A31}, \reff{A32} and \reff{A33} multiplied by the factor of $(2 + \omega^2\,\epsilon^2)^3$ and the term 
$3\,\hat{A}\,\hat{B}^2$ is the sum of \reff{B31} and \reff{B32} multiplied by $12\,(2 + \omega^2\,\epsilon^2)$. Using those results, and substituting $\hat{Q}^3$ given by \reff{Q3} into \reff{ThirdOrder}, the third order contribution is 
\begin{align}
& \frac{1}{3!} \,\hat{Q}^3 \, Z[\vec{J}]\,\Bigr\rvert_{\vec{J}=0}=-\frac{m^3}{192\,\epsilon^{9/2}\,\hbar^{3/2}}\,Z[0]\,\dfrac{1}{\lambda^{3/2}}\nonumber\\&\quad\Bigg[(2 + \omega^2\,\epsilon^2)^3\bigg((N-1)\frac{\Gamma\big(\frac{7}{4}\big)}{\Gamma\big(\frac{5}{4}\big)}\p \frac{3}{4}\,(N\m 1)(N\m 2)\,\frac{\Gamma\big(\frac{3}{4}\big)}{\Gamma\big(\frac{5}{4}\big)}\p \frac{1}{16} (N\m 1)(N\m 2)(N\m 3)\Big(\frac{\Gamma\big(\frac{3}{4}\big)}{\Gamma\big(\frac{5}{4}\big)}\Big)^3\bigg)\nonumber\\&\qquad\qquad+(2 + \omega^2\,\epsilon^2)\bigg(\frac{3}{4} \,(N\m 2)\,(N\m 3)\,
\Big(\frac{\Gamma\big(\frac{3}{4}\big)}{\Gamma\big(\frac{5}{4}\big)}\Big)^3+6 \,(N\m 2)\,
\frac{\Gamma\big(\frac{3}{4}\big)}{\Gamma\big(\frac{5}{4}\big)}\,\bigg)\Bigg]\,.
\label{Order3}
 \end{align}
 
Deriving the above third order contribution \reff{Order3} was considerably more involved than deriving the second order contribution \reff{Order2}. The above expression is also lengthier and more complicated than the one in \reff{Order2}. It should be clear that  obtaining higher order contributions by hand becomes quickly unwieldy. In the next section, we discuss a general procedure that can be implemented into a symbolic software program that generate the analytical expressions for higher-order $n$ contributions.   
 
The series \reff{HypSeries} up to third order (labeled $K_{E_{(3)}}$) is then obtained by adding the above contribution to the second order result \reff{SecondOrder}. This yields the following analytical expression: 
\begin{align}
K_{E_{(3)}}& =C \, Z[0]\,\Bigg[1-  \dfrac{1}{\lambda^{1/2}}\,(N-1)\,\frac{\Gamma\big(\frac{3}{4}\big)}{\Gamma\big(\frac{5}{4}\big)}\,(2 + \omega^2\,\epsilon^2)\,\frac{m}{8\, \hbar^{1/2}\,\epsilon^{3/2}}\nonumber\\&\qquad\qquad\quad\quad+\dfrac{1}{\lambda}\,\dfrac{m^2}{32\,\epsilon^3\,\hbar}\,\bigg[(2 + \omega^2\,\epsilon^2)^2\,\Big(\,(N-1)+\dfrac{1}{4}\, (N-1)\,(N-2) \,\frac{\big[\Gamma\big(\frac{3}{4}\big)\big]^2}{\big[\Gamma\big(\frac{5}{4}\big)\big]^2} \,\Big)\nonumber\\&\qquad\qquad\qquad\qquad\qquad\qquad\qquad+ (N-2) \,\frac{\big[\Gamma\big(\frac{3}{4}\big)\big]^2}{\big[\Gamma\big(\frac{5}{4}\big)\big]^2} \,\bigg]\nonumber\\&\qquad\qquad\quad\quad-\dfrac{1}{\lambda^{3/2}}\,\frac{m^3}{192\,\epsilon^{9/2}\,\hbar^{3/2}}\,\bigg[(2 + \omega^2\,\epsilon^2)^3\Big((N-1)\frac{\Gamma\big(\frac{7}{4}\big)}{\Gamma\big(\frac{5}{4}\big)}\p \frac{3}{4}\,(N\m 1)(N\m 2)\,\frac{\Gamma\big(\frac{3}{4}\big)}{\Gamma\big(\frac{5}{4}\big)}\nonumber\\&\qquad\qquad\qquad\qquad\qquad\qquad\qquad\qquad\qquad\p \frac{1}{16} (N\m 1)(N\m 2)(N\m 3)\frac{\big[\Gamma\big(\frac{3}{4}\big)\big]^3}{\big[\Gamma\big(\frac{5}{4}\big)\big]^3} \,\,\Big)\nonumber\\&\qquad\qquad\qquad\qquad+(2 + \omega^2\,\epsilon^2)\Big(\,\frac{3}{4} \,(N\m 2)\,(N\m 3)\,
\frac{\big[\Gamma\big(\frac{3}{4}\big)\big]^3}{\big[\Gamma\big(\frac{5}{4}\big)\big]^3}+6 \,(N\m 2)\,
\frac{\Gamma\big(\frac{3}{4}\big)}{\Gamma\big(\frac{5}{4}\big)}\,\Big)\,\bigg]\,\,\Bigg]\,.
\label{ThirdOrder}
\end{align}

The series \reff{HypSeries} should be viewed as a function of $N$, where $N$ is the number of steps $\epsilon$ the time interval $\mathcal{T}$ is divided into. It is an absolutely convergent series for a given $N$. In other words, as we sum the terms in the series it converges to a specific value for that given $N$. As $N$ increases, more terms in the series are required to reach convergence. The original path integral is formulated in the continuum, and is formally recovered in the limit as $N$ tends to infinity. Practically, how large $N$ needs to be depends on the desired level of accuracy.  

\subsection{Analytical results for second series at $n\,$th order and integer partitions} \label{norder}

We already mentioned that the series \reff{HypSeries} takes on the following form 
\begin{align}
K_E& =C \, Z[0]\,\Bigg[1-  \dfrac{b_1}{\lambda^{1/2}}+\dfrac{b_2}{\lambda}-\dfrac{b_3}{\lambda^{3/2}}+...(-1)^n\,\dfrac{b_n}{\lambda^{n/2}}\Bigg]\,.
\label{SeriesForm2}
\end{align}
This structure can clearly be seen in our third order result  \reff{ThirdOrder}: the first, second and third order contributions come with a factor of $Z[0]$ multiplied by $\frac{1}{\lambda^{1/2}}$,  $\frac{1}{\lambda}$ and  $\frac{1}{\lambda^{3/2}}$ respectively.  The $b_i\,$s in \reff{SeriesForm2} are positive and a complicated function of $N$ and $\mathcal{T}$ as well as the parameters $\omega$, $m$ and the constant $\hbar$ (though we leave $\epsilon$ in our formulas, it is not really an independent parameter since it is equal to $\mathcal{T}/N$). 

We obtained the analytical formulas \reff{Order1}, \reff{Order2} and \reff{Order3} for the first, second and third order contributions to the series \reff{HypSeries}. In other words, we already obtained $b_1$, $b_2$ and $b_3$. These analytical formulas were worked out by hand and became increasingly more complicated as the order increased. The goal here is to outline a general procedure for obtaining the $n$th order contribution that can be implemented into a symbolic software package (like Mathematica) that would generate the analytical formula for $b_n$.  

The $n\,$th order contribution to the series \reff{HypSeries} is given by 
\begin{align}
C\, \frac{\hat{Q}^n}{n!} \, Z[\vec{J}]\,\Bigr\rvert_{\vec{J}=0}
 \label{Ordern}
 \end{align} 
where $C$ is given by \reff{CC} and
\begin{align}
\hat{Q}^n&=\Bigg[- \frac{m}{2\,\epsilon\,\hbar}\Big(\hat{A}+\hat{B}\Big)\Bigg]^n\nonumber\\
&=(-1)^{n} \frac{m^n}{2^n\,\epsilon^n\,\hbar^n}\sum_{j=0}^{\lfloor n/2 \rfloor}\binom{n}{2\,j}\,\hat{A}^{\,n-2 j}\,\hat{B}^{\,2\,j}\,.
\label{Qn}
\end{align}
Here $\lfloor n/2 \rfloor$ denotes the greatest integer less than or equal to $n/2$ and $\binom{n}{2\,j}$ is a binomial coefficient. The operators $\hat{A}$ and $\hat{B}$ are given by \reff{AA} and \reff{BB} respectively: 
\begin{align}
\hat{A}=(2 + \omega^2\,\epsilon^2) \sum_{i=1}^{N-1}\big(\frac{\delta}{\delta J_i}\big)^2 
\label{AA2}
\end{align}
and 
\begin{align}
\hat{B}=-2 \sum_{i=2}^{N-1} \frac{\delta}{\delta J_i}\,\frac{\delta}{\delta J_{i-1}}\,.
\label{BB2}
\end{align}
The sum in \reff{Qn} contains only even powers of $\hat{B}$ because any term in the binomial series of $\hat{Q}^n$ that has odd powers of $\hat{B}$ yields zero when acting on $Z[\vec{J}]$ (after $\vec{J}$ is set to zero). If $n$ is even, then $\lfloor n/2 \rfloor=n/2$ and the last term in the sum is simply $\hat{B}^n$. If $n$ is odd, then $\lfloor n/2 \rfloor=(n-1)/2$ and the last term in the sum is $n\,\hat{A}\,\hat{B}^{n-1}$. The first term in the sum is $\hat{A}^n$, the second is $\binom{n}{2}\,\hat{A}^{n-2}\,\hat{B}^2$, the third  is $\binom{n}{4}\,\hat{A}^{n-4}\,\hat{B}^4$ etc. until we reach the last term which is either $\hat{B}^n$ or $n\,\hat{A}\,\hat{B}^{n-1}$ depending on whether $n$ is even or odd respectively. 

Any term in \reff{Qn} has the form $\hat{A}^{m}\,\hat{B}^k$  where $m$ is a non-negative integer and $k$ is an even number including zero. Such a term occurs at order $n=m+k$. We created a program that generates the analytical expression stemming from the contribution of any term $\hat{A}^{m}\,\hat{B}^k$. In other words,  inputing a value for $m$ and $k$, the program generates the analytical expression corresponding to $\hat{A}^{m}\,\hat{B}^k\, Z[\vec{J}]\,\Bigr\rvert_{\vec{J}=0}$. The goal here is to outline the procedure for obtaining this result. Instead of tackling the term $\hat{A}^{m}\,\hat{B}^k$ directly all at once, it will prove helpful to first outline the procedure for generating an $\hat{A}^m$ contribution and a $\hat{B}^k$ contribution separately. Once we have treated those two cases, we will combine those results to generate the general term $\hat{A}^{m}\,\hat{B}^k$. 

\subsubsection{Procedure for generating $\hat{A}^m$ contribition and integer partitions} \label{Amm}
Our goal here is to outline the procedure for obtaining the analytical expression given by $\hat{A}^m \, Z[\vec{J}]\,\bigr\rvert_{\vec{J}=0}$ where $\hat{A}$ is given by \reff{AA2} and $m$ is any non-negative integer. We begin by writing out explicitly the operator $\hat{A}^m$ :
\begin{align} 
\hat{A}^m=(2 + \omega^2\,\epsilon^2)^m\!\!\sum_{i_1,i_2,...,i_m=1}^{N-1}\big(\frac{\delta}{\delta J_{i_1}}\big)^2 \big(\frac{\delta}{\delta J_{i_2}}\big)^2 ...\big(\frac{\delta}{\delta J_{i_m}}\big)^2\,.
\label{Am}
\end{align}
We assume that this will act on $ Z[\vec{J}]$ with $\vec{J}$ set to zero at the end. We have already calculated the cases for $m=1$, $m=2$ and $m=3$ which are contained within the results of \reff{Order1}, \reff{Order2} and \reff{Order3}. For $m=1$ we obtained a term with $(N-1)$, for $m=2$, we obtained terms with $(N-1)$ and $(N-1)(N-2)$ respectively and for $m=3$, we obtained terms with $(N-1)$, $(N-1)(N-2)$ and $(N-1)(N-2)(N-3)$ respectively. We will see that this pattern persists and that for a general value of $m$ one obtains terms with $ (N-1)(N-2)...(N-i)$ with $i$ ranging from $1$ to $m$. 

The $m$-dimensional sum in \reff{Am} over $i_j$ where $j$ runs from $1$ to $m$, can be divided into different cases. The first case is when all $m$ of the $i_j\,$s are equal yielding a single $2m$-derivative operator; label this case $(m)$. The second case is when $m-1$ of the $i_j\,$ s are equal, yielding one $(2m -2)$-derivative operator and one two-derivative operator; label this case $(m-1,1)$. The third case is when $m-2$ of the $i_j\,$s are equal, yielding one $(2m-4)$-derivative operator with two possibilities for the remaining two $i_j\,$s: either they are equal yielding a  4-derivative operator (case labeled $(m-2,2)\,)$  or unequal yielding two 2-derivative operators (case labeled $(m-2,1,1)\,)$. It should be clear that the fourth would have the following possible labels: $(m-3,3)$, $(m-3,2,1)$ and $(m-3,1,1,1)$. 

The different cases correspond to the integer partitions of $m$ i.e. the different ways one can split $m$ into a sum of positive integers.  For example, the integer partitions of $m=5$ are $\{5\}$,  $\{4,1\}$, $\{3,2\}$, $\{3,1,1\}$, $\{2,2,1\}$, $\{2,1,1,1\}$ and  $\{1,1,1,1,1\}$. By definition, different orderings are not included (e.g. $\{1,3,1\}$ and  $\{1,1,3\}$ are not included). So there are $7$ partitions in total for $m=5$.  Let  $q$ denote the number of entries (or members) in each partition. For example, the partition  $\{3,1,1\}$ has $q=3$ and $\{2,1,1,1\}$ has $q=4$. In \reff{Am} there is a total of $2 \,m$ derivatives. For $m=5$ the total is $10$ derivatives. Take for example the partition $\{3,2\}$. One needs to multiply each entry by $2$ to obtain the number of derivatives. This yields $\{6,4\}$ and this corresponds to one six-derivative and one 4-derivative operator (for  a total of $10$ derivatives). A partition of  $m$ is denoted as 
\beq
Q=\{c_1,c_2,c_3,...,c_q\}
\eeq{QQ}
and corresponds to the derivative operator
\begin{align}
 \big(\frac{\delta}{\delta J_{i_1}}\big)^{2\,c_1} \big(\frac{\delta}{\delta J_{i_2}}\big)^{2\,c_2} ...\big(\frac{\delta}{\delta J_{i_q}}\big)^{2\,c_q}\,.
 \label{QObject}
 \end{align}
The total number of derivatives is $2 \,(c_1+c_2+,...,+c_q)$ which must be equal to $2\,m$. The number of ways the $q$ distinct objects in \reff{QObject} can be distributed among the $N-1$ boxes of $Z[\vec{J}]$ is given by
\begin{align}
\prod_{i=1}^{q}(N\m i)= (N\m 1)\,(N\m 2)...(N\m q)\,.
\label{Nq}
\end{align}
When an object  $\big(\frac{\delta}{\delta J_{i_{\ell}}}\big)^{2\,c_{\ell}}$ is placed in a box this means it acts on $I[J_{i_{\ell}}]$ and  by \reff{Central} this yields a factor
\begin{align}
\big(\frac{\delta}{\delta J_{i_{\ell}}}\big)^{2c_{\ell}} I[J_{i_{\ell}}] \,\Bigr\rvert_{J_{i_{\ell}}=0} =\frac{1}{2} \,\Gamma\big(\frac{2\,c_{\ell}+1}{4}\big)\,\Big(\frac{\hbar}{\epsilon\,\lambda}\Big)^{\frac{2\,c_{\ell}+1}{4}}\,.
\label{Central2}
\end{align}
Define
\begin{align}
G(k)=\frac{1}{2} \,\Gamma\big(\frac{2\,k+1}{4}\big)\,\Big(\frac{\hbar}{\epsilon\,\lambda}\Big)^{\frac{2\,k+1}{4}}\,.
\label{Gk}
\end{align}
Therefore a partition $Q=\{c_1,c_2,c_3,...,c_q\}$  yields a factor of 
\beq
G(c_1)\,G(c_2)\,G(c_3)...G(c_q)=\prod_{i=1}^{q} \,G(c_i)
\eeq{Gci} 
which is associated with the operator \reff{QObject} acting on $Z[\vec{J}]$ (with $\vec{J}$ set to zero afterwards).  When the $q$ objects in \reff{QObject} are distributed among the $N\m 1$ boxes, the remaining $N\m 1\m q$ boxes yield a factor of $I[0]^{N\m 1\m q}=Z[0]\,I[0]^{-q}$ where $I[0]$ and $Z[0]$ are given by \reff{I0} and \reff{Z0} respectively. 

In \reff{Am}, we have an $m$-dimensional sum of $m$ two-derivative objects. We broke this sum up into cases involving $q$-dimensional sums over $q$ objects each with $2\,c_i$ derivatives where $i$ runs from $1$ to $q$. Each case corresponds to an integer partition of $m$ given by $Q=\{c_1,c_2,c_3,...,c_q\}$. There is a numerical factor associated with how many ways the original sum can be reduced to a case with a given partition $Q$. This is given by  
\begin{align}
F =\dfrac{m!}{c_1!c_2!...c_q!} \dfrac{\text{P}[Q]}{q!}
\label{factor}
\end{align}
where $\text{P}[Q]$ is the number of permutations of the partition $Q$. $F$ stems from a product of binomial coefficients with the extra factor $\tfrac{\text{P}[Q]}{q!}$ required if the partition has two or more members that are equal to each other. As an example, consider the integer partition $\{3,1,1\}$ of $m=5$. Then $\text{P}[\{3,1,1\}]=3$, $q=3$ and $c_1=3$, $c_2=1$ and $c_3=1$. The numerical factor associated with the partition $\{3,1,1\}$ is then $F=\frac{5!}{3!\, 1!\, 1!}\frac{3}{3!}=10$. 

We are now in a position to state the steps for obtaining the $A^{m}$ contribution for any non-integer $m$. The steps are:
\begin{itemize} 
\item Obtain the integer partitions of $m$. In Mathematica, they are generated by ``IntegerPartitions[m]''. 
\item Each partition $Q=\{c_1,c_2,c_3,...,c_q\}$ makes a contribution of
\beq
Z[0]\,I[0]^{-q}\,\prod_{i=1}^{q}(N-i) \,\prod_{\ell=1}^{q}\,G(c_{\ell})\,F
\eeq{Am1}
where $F$ is the numerical factor given by \reff{factor}. Substituting \reff{Gk} for the G's and \reff{I0} for $I[0]$ this simplifies to
\beq
 Z[0]\,\lambda^{-\frac{m}{2}}\, \prod_{i=1}^{q}(N-i)\, \prod_{\ell=1}^{q}\Gamma\big(\frac{2\,c_{\ell}+1}{4}\big)\,
\,\Big(4\, \Gamma\big(\frac{5}{4}\big)\Big)^{-q}\,\Big( \frac{\hbar}{\epsilon}\Big)^{\frac{m}{2}}\,F
 \eeq{AAm}
\item Sum the contributions from each partition of $m$. Multiply the result by $(2 + \omega^2\,\epsilon^2)^m$  to obtain the  total contribution from $A^{m}$.  
\end{itemize}
The result is an exact analytical expression containing functions of $N$ multiplied by Gamma functions. This is a very significant improvement over calculating numerically an $N$ dimensional integral with limits running from $-\infty$ to $\infty$ especially when $N$ is large. In the analytical expression, we simply substitute the value of $N$. 

\subsubsection{Procedure for $\hat{B}^k$ contribution}\label{Bkk}
Here we outline the procedure for obtaining the analytical expression given  by $\hat{B}^k \, Z[\vec{J}]\,\Bigr\rvert_{\vec{J}=0}$ where $\hat{B}$ is given by \reff{BB2} and $k$ is an even integer. This is somewhat more involved than the previous $\hat{A}^m$ case. The good news is that once this is completed, using the $\hat{A}^m$ and $\hat{B}^k$ parts to finally obtain $\hat{A}^m\,\hat{B}^k$ in the next section is less involved.  We begin here by writing out explicitly the operator $\hat{B}^k$:
\begin{align} 
\hat{B}^k&=(-2)^k\, \sum_{i_1,i_2,...,i_k=2}^{N-1}\frac{\delta}{\delta J_{i_1}}\,\frac{\delta}{\delta J_{i_1-1}}\,\frac{\delta}{\delta J_{i_2}}\,\frac{\delta}{\delta J_{i_2-1}} ...\frac{\delta}{\delta J_{i_k}}\,\frac{\delta}{\delta J_{i_k-1}}\nonumber\\
&=(-2)^k\, \sum_{i_1,i_2,...,i_k=2}^{N-1}\frac{\delta}{\delta J_{i_1}}\,\frac{\delta}{\delta J_{i_2}}\,...\frac{\delta}{\delta J_{i_k}}\,\frac{\delta}{\delta J_{i_1-1}}\,\frac{\delta}{\delta J_{i_2-1}}...\frac{\delta}{\delta J_{i_k-1}}
\label{Bk}
\end{align}
where we simply reordered the derivatives in the last line so that all the $i\,'$s are grouped together followed by all the $i-1\,'$s. Recall that odd derivatives with respect to $J_{i}$ for a given $i$ yield zero. This means that derivatives with a given label $i$ must come in even powers. Note that there is in total $2\,k$ derivatives in \reff{Bk}. \textit{However, there are only $k$ labels to specify since $i_{\ell}$ determines the value of $i_{\ell}-1$}. 

Consider the case $k=6$ where there is $12$ derivatives in total. One possibility is to combine them into two six-derivative objects. This occurs when $i_1=i_2=i_3=i_4=i_5=i_6$ ( label as $i_1$). This yields  
\beq
\Big(\frac{\delta}{\delta J_{i_1}}\Big)^6\, \Big(\frac{\delta}{\delta J_{i_1-1}}\Big)^6
\eeq{J6}
which we label as $P=(6,6)$. $P$ tells us how the $2\,k$ derivatives are divided up. $P=(6,6)$ tells us that $12$ derivatives are divided into two objects of 6 derivatives each. There is only one sum to perform (over $i_1$ from $2$ to $N-1$). When acting on $Z[\vec{J}]$, which can be thought of as a string of $N\m 1$ boxes, the number of ways to place the two consecutive objects (one pair) in the $N-1$ boxes is simply $N-2$. Using \reff{Central}, each six-derivative object when placed in a box contributes a factor of $G(3)$ for a total factor of $G(3)^2$ where $G(k)$ is defined in \reff{Gk}. The remaining $N\m 3$ boxes contribute $I[0]^{N-3}=Z[0]\,I[0]^{-2}$ where $I[0]$ and $Z[0]$ are given by \reff{I0} and \reff{Z0} respectively. The contribution from $(6,6)$ is then 
\beq
(6,6) = (N\m 2)\,G(3)^2\,Z[0]\,I[0]^{-2}\,.
\eeq{B66}
Another possibility is $i_1=i_2=i_3=i_4$  with  $i_5=i_6$. This yields 
\beq
\Big(\frac{\delta}{\delta J_{i_1}}\Big)^4\, \Big(\frac{\delta}{\delta J_{i_1-1}}\Big)^4\Big(\frac{\delta}{\delta J_{i_2}}\Big)^2\, \Big(\frac{\delta}{\delta J_{i_2-1}}\Big)^2\,.
\eeq{J42}
 The derivative with $i_1$ label is to the power of $4$ because we equated four $i\,'\,$s together. However, there is another way to obtain $i_1$ to the power of $4$ without equating $4$ $i\,$s. You can equate two $i\,'\,$s to $i_1$ (two powers of $i_1$) and then equate two other $i\,$s to $i_1+1$. The $i-1$ partners of those two latter $i\,'\,$s will then be equal to $i_1$ yielding two extra powers of $i_1$ for a total of $4$ powers. We can therefore have $i_1=i_2$, $i_3=i_4=i_1+1$ and $i_5=i_6=i_1+2$. This yields  
\beq
 \Big(\frac{\delta}{\delta J_{i_1-1}}\Big)^2\,\Big(\frac{\delta}{\delta J_{i_1}}\Big)^4\, \Big(\frac{\delta}{\delta J_{i_1+1}}\Big)^4\,\Big(\frac{\delta}{\delta J_{i_1+2}}\Big)^2\,.
 \eeq{J222}
In both cases, \reff{J42} and \reff{J222}, we have $P=(2,2,4,4)$ i.e. the $12$ derivatives are divided among $4$ objects, one with $2$ derivatives, another with $2$, another with $4$ and another with $4$ (in writing $P=(2,2,4,4)$ we are not concerned with order so that $(2,4,2,4)$ would have been just as valid). When they are placed in the $N\m 1$ boxes, $2$ derivatives contributes $G(1)$ and $4$ derivatives contributes $G(2)$ for a total factor of $G(2)^2 \,G(1)^2$. In both cases, the remaining number of boxes is $N-5$ which contributes $I[0]^{N\m 5}=Z[0] \,I[0]^{-4}$. In the case \reff{J42}, there are two labels $i_1$ and $i_2$. There are four objects but the first two are consecutive (one pair) and the last two are also consecutive (a second pair). So you have two distinguishable pairs. What is the number of ways to distribute those two pairs among the $N\m 1$ boxes? This is equivalent to distributing two distinguishable objects among $N \m 3$ boxes (not $N \m 1$ since the two pairs take up two extra boxes). The answer is then $(N\m 3)(N\m 4)$. We also have a numerical factor associated with how many ways we can choose $4 \,i\,'\,$s from $6\,i\,'\,$s i.e. the set $(i_1,i_2,i_3,i_4,i_5,i_6)$. The answer is $\binom{6}{4}=15$. In  the case \reff{J222}, we have four consecutive objects and clearly there is $N\m 4$ ways to distribute them among $N \m 1$ boxes.  We also have a numerical factor associated with how many ways we can equate $2\,i\,'\,$s to $i_1$, $2\,i\,'\,$s to $i_1+1$ and $2\,i\,'\,$s to $i_1+2$ among $6\,i\,'\,$s. This factor is $\tfrac{6!}{2!\,2!\,2!}=90$. Summing the contribution to $P=(2,2,4,4)$ from both cases we obtain 
\beq
(2,2,4,4)=\Big(90\,(N\m 4) +15\, (N\m 3)\,(N\m 4)\Big)\,G(2)^2\,G(1)^2\,Z[0]\,I[0]^{-4}\,.
\eeq{B24}  
The other contributions to $\hat{B}^6$ can be shown to be 
\begin{align}
(4,6,2) &= 30\,(N\m 3) \,G(3)\,G(2)\,G(1) Z[0]\,I[0]^{-3}\nonumber\\
(2,2,2,2,4)& =90\,(N\m 4)\,(N\m 5)\, G(2)\,G(1)^4\,Z[0]\,I[0]^{-5}\nonumber\\
(2,2,2,2,2,2)&=15\,(N\m 4)\,(N\m 5)\,(N\m 6) \,G(1)^6\,Z[0]\,I[0]^{-6}\,.
\label{Cases}
\end{align}
We are now in a position to provide the steps to automate this process for any even value of $k$ i.e. construct a program for a symbolic software package that generates the analytical expressions.   The steps are:
\begin{itemize}
\item Multiply by two the integer partitions of $k/2$. Then include the permutations for each partition. In Mathematica, this can be achieved using the built-in functions ``IntegerPartitions'' and ``Permutations''. For $k=6$, this yields the partitions 
$\{6\},\!\{4,2\},\{2,4\},\!\{2,2,2\}\!.$ A partition is labeled  $\{a_1,a_2,...,a_t\}$ where $t$ can take on integer values from $1$ to $k/2$.

\item Associate $i_1$, $i_2$, $i_3$, etc. with  $f(1,0)$, $f(2,0)$, $f(3,0)$, respectively. Associate $i_1+1$, $i_2+1$, $i_3+1$, etc. with  $f(1,1)$, $f(2,1)$, $f(3,1)$, respectively. In general, associate $f(m,n)$ with $i_m+n$. Generate a table with rows and columns of $f\,'$s (which we refer to as an ``f-table'') so that a given partition with $d$ members can be equated with any row that has $d$ columns. Since the minimum and maximum length (number of elements) of the partitions are $1$ and $k/2$ respectively, the f-table consists of rows that range from $1$ to $k/2$ columns inclusively. Each row starts with $f(1,0)$. $f(m,n)$ must be preceded by $f(m,n-1)$ (for $n >0$ ) and preceded by $f(m-1,n)$ (for $m>1$).  A row with one column is therefore $f(1,0)$.  A row with two columns can be $f(1,0)\, f(2,0)$ or $f(1,0)\,f(1,1)$. Those are the only possibilities with two columns. You \textit{cannot} for example have $f(1,0)\,f(1,2)$ because $f(1,2)$ must be preceded by $f(1,1)$.  A row with three columns can be either $f(1,0)\, f(2,0)\,f(3,0)\,$, $\,f(1,0)\,f(1,1)\,f(2,0)\,$ or $f(1,0)\,f(1,1)\,f(1,2)$. Those are the only possibilities for three columns.  You \textit{cannot} have for example $f(1,0)\,f(2,0)\,f(2,1)$ because $f(2,1)$ must be preceded by $f(1,1)$ or  $f(1,0)\,f(2,1)\,f(2,2)$ because $f(2,1)$ must be preceded by $f(2,0)$. 

There is a \textit{unique} f-table for a given value of $k$.  The f-table for $k=6$ is 
\begin{align}
&f[1,0]\nonumber\\
&f[1,0] \,f[1,1]\nonumber\\
&f[1,0]\, f[2,0]\nonumber\\
&f[1,0] \,f[1,1] \,f[1,2]\nonumber\\
&f[1,0] \,f[1,1] \,f[2,0]\nonumber\\
&f[1,0] \,f[2,0] \,f[3,0]
\label{ftable}
\end{align}
We created a program that generates the rows of an f-table for any even $k$.

\item Equate each partition with $m$ elements to a row of the f-table with $m$ columns . For example, in the case of $k=6$, the partition $\{4,2\}$ has two elements and can be equated with either row $(f[1,0],f[1,1])$ or row $(f[1,0], f[2,0])$ in the f-table \reff{ftable}. In the first case one has $f(1,0)=4$ and $f(1,1)=2$; this means that $4\,i\,'$s  are equal with label $i_1$ and $2\,i\,'$s are equal with label $i_1+1$ respectively. Here the $i-1$ partners are  $4\,i\,'$s equal to $i_1-1$ and $2\,i\,'$s  equal to $(i_1+1) -1=i_1$. Note that we have 4+2=6 derivatives with $i_1$ (it is not 4 but 6 due to the 2 extra $i_1$ stemming from an $i-1$ partner where $i=i_1+1$ so that $(i_1+1) -1=i_1$). We also have 4 derivatives with $i_1-1$ and 2 derivatives with $i_1+1$. This yields 12 derivatives in total as expected. This case was therefore a contributor to $P=(4,6,2)$ in \reff{Cases}. The second case has $f(1,0)=4$ and $f(2,0)=2$. This means $4\,i\,'$s  are equal with label $i_1$ and  $2\,i\,'$s are equal with label $i_2$. The $i-1$ counterparts yield $4\,i\,'$s equal to $i_1-1$ and $2\,i\,'$s  equal to $(i_2-1)$. So there are 4 derivatives with $i_1$, 4 derivatives with $i_1-1$,  2 derivatives with $i_2$ and 2 derivatives with $i_2-1$. This case was therefore a contributor to $P=(2,2,4,4)$ in \reff{B24}. 

\item The rules for obtaining $P$ from a row in the f-table are the following: a) each $f$ has two entries. Any $f$ whose second entry is zero becomes an element of $P$. b) Neighbouring $f\,'$s  that have the same first entry are added together and become an element of $P$. c) In a sequence of $f\,'$s with the same first entry, the last one becomes an element of $P$. If there is only one $f$ in the sequence (so that its second entry is zero), this still holds and must be included as a member of $P$. Note that together with rule a), an $f$ whose second entry is zero can appear twice in $P$.  

Applying these rules, the $P$ associated with $(f[1,0],\!f[1,1])$ is $\!(f[1,0],\! f[1,0] \p f[1,1], \!f[1,1])\!.$ In our above example, $f(1,0)\!=\!4$ and $f(1,1)\!=\!2$ and this yields $P=(4,6,2)$. The $P$ associated with $(f[1,0], f[2,0])$ is $(f[1,0], f[1,0], f[2,0], f[2,0])$. In our above example, $f(1,0)=4$ and $f(2,0)=2$ and this yields $P=(4,4,2,2)$. Let us look at a more complicated example. The $P$ associated with a row $f(1,0)\,f(1,1)\,f(1,2)\,f(2,0)\,f(2,1)\,f[3,0]\\$ is $(f[1,0], f[1,0] \p f[1,1], f[1,1]\p f[1,2], f[1,2], f[2,0],\! f[2,0]\p f[2,1], f[2,1], f[3,0],f[3,0]).$   

\item We saw above that each partition when equated with a row in the f-table yields a $P=(b_1,b_2,...,b_q)$ where $b_i$ is the number of derivatives of object $i$ and $q$ is the total number of objects.  We extract from $P$ the contribution of the product of $G$ factors (for a sample, see \reff{B24} and \reff{Cases}). This is given by
\beq
\prod_{i=1}^q \, G(b_i/2)
\eeq{GF}
where $G(y)$ is given by \reff{Gk}. We also know that after $q$ objects are placed in $N\m 1$ boxes there remains $N\m 1\m q$ boxes. This yields a factor of $Z[0]\, I[0]^{-q}$. Let $u$ be the highest value of the first entry of the $f\,'$s in a given row. For example, in the row $f[1,0] f[1,1] f[2,0]$ we would have $u=2$ whereas in the row $f[1,0] f[2,0] f[3,0]$ we would have $u=3$. We define $B=u-1$. This yields a  polynomial in $N$ of degree $u=B+1$ given by 
\beq
\prod_{h=q-B}^{q} (N-h)\,.
\eeq{Prods}
Such products of $N\,'$s can be viewed in  examples like \reff{Cases} and \reff{B24}. The only thing left to determine is the numerical factor. There is a numerical factor associated with how many ways one can group and equate the different $k$ $\,i\,'$s that yield a given partition  $ \{a_1,a_2,...,a_t\}$ with a given $P$. The numerical factor is given by
\beq
F=\frac{k!}{a_1!a_2!...a_t!}\frac{1}{S}
\eeq{FB}
where $S$ is a symmetry factor: if a row in the f-table is invariant under the exchange of a set of $z$ integers in the first entry of $f$ then $S=z!$. The set of $z$ integers are chosen among $1, 2,3,..,k/2$. For example, in the row $f[1,0] f[2,0]$ if we exchange $1$ and $2$ in the first entry of each $f$, we obtain $f[2,0] f[1,0]$ which is counted as the same row (even though the $f\,'$s appear in a different order now, we assume that we re-order them into their regular order. After re-ordering, they yield the the same row.)  Here $z=2$ and this implies $S=2!$. For a row like  $f[1,0] f[1,1] f[2,0]$, exchanging $1$ and $2$ in the first entry yields a different row. So there is no invariance and $S$ is simply unity.  A row like $f[1,0] f[2,0] f[3,0]$ is invariant under the exchanges of all three integers $1,\,2$ and $3$. So $z=3$ and $S=3!$ in that case. 

\item Putting all the above results together, the contribution of a partition  $ \{a_1,a_2,...,a_t\}$ that is equated with a row in the f-table whose highest first entry is $u$ and yields $P=(b_1,b_2,...,b_q)$ is given by
\beq
Z[0]\, I[0]^{-q}\,F\,\prod_{h=q-B}^{q} (N-h)\,\prod_{i=1}^q \, G(b_i/2)\,
\eeq{BCont}
where $B=u-1$, $F$ is the numerical factor given by \reff{FB}, $q$ is the number of entries in $P$ and $G(y)$ is given by \reff{Gk}.   
\item  Sum the contributions \reff{BCont} from all partitions and multiply the total by $(-2)^k$ (see \reff{Bk}). This yields $\hat{B}^k \, Z[\vec{J}]\,\Bigr\rvert_{\vec{J}=0}$.
\end{itemize}

\subsubsection{Obtaining the $\hat{A}^m\,\hat{B}^k$ contribution} \label{AmmBkk}
Our final goal is to outline the procedure for obtaining the analytical expression given  by $\hat{A}^m \,\hat{B}^k \, Z[\vec{J}]\,\Bigr\rvert_{\vec{J}=0}$ where $m$ is a non-negative integer and $k$ is an even integer.  This term belongs to the order $n=m+k$ in the series expansion \reff{HypSeries}. We begin by writing out explicitly the operator $\hat{A}^m\,\hat{B}^k$:
\begin{align}
\hat{A}^m \,\hat{B}^k&=(-2)^k\,(2 + \omega^2\,\epsilon^2)^m\!\! \nonumber\\ \sum_{i_1,...,i_m=1}^{N-1} \sum_{i_{m+1},...,i_{m+k}=2}^{N-1}&\big(\frac{\delta}{\delta J_{i_1}}\big)^2 \big(\frac{\delta}{\delta J_{i_2}}\big)^2 ...\big(\frac{\delta}{\delta J_{i_m}}\big)^2\,
\frac{\delta}{\delta J_{i_{m+1}}}\,\frac{\delta}{\delta J_{i_{m+1}-1}}\nonumber\\&\qquad\frac{\delta}{\delta J_{i_{m+2}}}\,\frac{\delta}{\delta J_{i_{m+2}-1}}...\frac{\delta}{\delta J_{i_{m+k}}}\,\frac{\delta}{\delta J_{i_{m+k}-1}}
\label{AmBk}
\end{align}
To determine the terms in the analytical expression for $\hat{A}^m\,\hat{B}^k$ require us to find the the same three quantities we previously found for $\hat{B}^k$ in \reff{BCont} and for  $\hat{A}^m$ in \reff{Am1}. Those three quantities are  $P=(b_1,b_2,...,b_q)$, which yield the product of $q$ $G\,'$s,  $B$ (which together with $q$ determines the polynomial in $N$) and a numerical factor $F$. So once we determine a given $P$, $B$ and $ F$ for $\hat{A}^m\,\hat{B}^k$ this yields a term given  by
\beq
Z[0]\, I[0]^{-q}\,F\,\prod_{h=q-B}^{q} (N-h)\,\prod_{i=1}^q \, G(b_i/2)\,.
\eeq{TermAmBk} 
Note that $q$ is the number of elements in $P$. So the goal is to generate all the possible $P\,'$s, $B\,'$s and $F\,'$s and add all their contributions to obtain the final analytical expression.  The $\hat{A}^m$ part by itself already has its own  $P_A$, $B_A$, $F_A$ (and $q_A$) and the $\hat{B}^k$ part by itself has its own $P_B$, $B_B$, $F_B$ (and $q_B$). These have already been determined in sections \reff{Amm} and \reff{Bkk} respectively.  Note that  $P_A= 2 Q=\{2\,c_1,2\,c_2,2\,c_3,...,2\,c_{q_A}\}$ where $Q$ was introduced in \reff{QQ} as an integer partition of $m$ and we have relabeled $q$ in \reff{QQ} by $q_A$. $B_A$ is simply $q_A-1$ and $F_A$ is given by \reff{factor}. $P_B$,  $B_B$ and $F_B$ are exactly the same as the $P$, $B$ and $F$ respectively obtained in section \reff{Bkk} except that we now add the subscript $B$. The quantities $q_A$ and $q_B$ are the numbers of elements in $P_A$ and $P_B$ respectively. 

The $P$, $B$ and $F$ for $\hat{A}^m\,\hat{B}^k$  can be obtained by combining $P_A$ and $P_B$, $B_A$ and $B_B$ and $F_A$ and $F_B$ respectively. Let $I_{min}$ be the minimum of $m$ and $k$.  There are $I_{min}+1$ different cases to consider. Each case combines the $\hat{A}$ and $\hat{B}$ parts differently.   

The first case is associated with keeping the $\hat{A}$ and $\hat{B}$ parts separate by not making any $i$ associated with $\hat{A}$ equal to an $i$ assocaited with $\hat{B}$. The answer is then straightforward: $P$ is the union of the original $P_A$ and $P_B$ (hence it automatically follows that $q=q_A+ q_B$), $F=F_A\,F_B$, $B=B_A+B_B+1$. Note that $B$ is \textit{not} equal  to $B_A+B_B$. Recall that $B$ is the number of independent $i$ labels minus 1. For example, if you have two independent labels associated with $\hat{A}^m$ and two independent labels associated with $\hat{B}^k$, this yields  $B_A=2-1=1$ and  $B_B=2-1=1$ respectively. Combined, there are four independent labels in total and this implies  $B=4-1=3$. This is equal to $B_A+B_B+1$.  

The second case is obtained by transferring one element of $P_A$ to one element of $P_B$. We add the transferred element from $P_A$ to the element of $P_B$ to obtain the new $P_B$. The new $P_B$ has the same number of elements as before but the value of one of its elements changes. The new $P_A$ is obtained by removing the transferred element. For example, say $P_A=(4,2)$ and $P_B=(2,4,6)$. Then $P_{B_{new}}=\{(6,4,6), (2,8,6), (2,4,10),(4,4,6), (2,6,6), (2,4,8)\}$ and $P_{A_{new}}=\{(2), (2), (2), (4), (4), (4)\}$. Their union yields 
\beq
P=\{(2,6,4,6), (2,2,8,6), (2,2,4,10),(4,4,4,6), (4,2,6,6), (4,2,4,8)\}\,.
\eeq{PPP}
There are $6$ $P\,'$s in total each with $q=4$. For each $P$ there is an associated $B$ where  $B=(B_A+B_B+1)-1=B_A+B_B$.  Note that we subtract $1$ compared to the previous case because we transferred one element of $P_A$ and in the process we lose one independent $i$. The numerical factor of all $6$ $P\,'$s is always the same: $F=F_A\,F_B$ because there are no new permutations. Each of the $6$ $P\,'$s (together with their associated $B\,'$s) contributes a term \reff{TermAmBk}.

The third case is obtained by transferring two separate elements of $P_A$ to two separate elements of $P_B$. We add each transferred element from $P_A$ to the respective element of $P_B$ to obtain the new $P_B$. Again, the new $P_B$ has the same number of elements as before but the values of two of its members will change. The new $P_A$ is obtained by removing the two transferred elements. For example, say $P_A=(4,2,3)$ and $P_B=(2,4,6)$. Since we are transferring two elements, we find all possible permutations of $P_A$ that have two elements. This yields 
\beq
\{\,(4,2)\,,\,(2,4)\,,\,(4,3)\,,\,(3,4)\,,\, (2,3)\,,\,(3,2)\,\}
\eeq{PA2}

Note that we include both $(4,2)$ and $(2,4)$ because they both make a contribution. We will illustrate here only the case of adding $(4,2)$ to $P_B$ (the procedure is the same for all the other members of \reff{PA2}). We will label this case new1.  We add $(4,2)$ to two elements of $P_B$. This means adding $4$ to one element at a given position and $2$ to an element at a different position to the right.  There are three possible groupings of two ordered elements of $P_B$: $(2,4,_)\,,\, (2,_,6)$ and $(_,4,6)$ where the lower dash means this element is left alone when we add $(4,2)$. This yields  $P_{B_{new1}}=(6,6,6), (6,4,8), (2,8,8)$ and $P_{A_{new1}}=\{(3), (3), (3)\}$. Their union yields $P=\{(3,6,6,6), (3,6,4,8), (3,2,8,8)\}$. So there are three $P\,'$s each with $q=4$, They each have an associated $B=(B_A+B_B+1)-2=B_A+B_B-1$ (we subtract $2$ from the original (``no transfer'') case because we transferred two elements of $P_A$ to $P_B$ so that we lost two independent $i\,'s$). The numerical factor is always the same: $F=F_A\,F_B$ because all permutations of $P_A$ are taken into account in the list \reff{PA2}. So each of the three $P\,'$s we found makes a separate contribution given by \reff{TermAmBk}. One simply repeats the procedure for all the different permutations \reff{PA2} and add their respective contributions. 

This process continues until we can no longer transfer elements from $P_A$ to $P_B$. The maximum number of elements that can be transferred is $I_{min}$. For example if $m=2$ and $k=4$, then $I_{min}=2$.  If $m=6$ and $k=4$, then $I_{min}=4$. Together with the ``no transfer'' case, there are  $I_{min}+1$ cases to consider.  We add the contributions \reff{TermAmBk} from all $I_{min}+1$ cases and multiply the total by the factor $(-2)^k\,(2 + \omega^2\,\epsilon^2)^m$ that appears in \reff{AmBk}. We have completed our task of outlining the procedure for obtaining the analytical expression given  by $\hat{A}^m \,\hat{B}^k \, Z[\vec{J}]\,\Bigr\rvert_{\vec{J}=0}$ where $m$ is a non-negative integer and $k$ is an even integer. In Appendix A, we write down the fifth order ($n=5$) contribution to the series \reff{HypSeries} generated by our program. 

We have already encountered in section \reff{ThirdO} an example of an $\hat{A}^m\,\hat{B}^k$ contribution: the case $\hat{A}\,\hat{B}^2$ which occurs at third order ($n=3$). This is given by the sum of \reff{B31} and \reff{B32} multiplied by $4\,(2 + \omega^2\,\epsilon^2)$:
\beq 
4\,(2 + \omega^2\,\epsilon^2) \,Z[0]\,
\Big(\frac{\Gamma\big(\frac{3}{4}\big)}{\Gamma\big(\frac{5}{4}\big)}\Big)\,\Big(\frac{\hbar}{\epsilon}\Big)^{3/2}\,\dfrac{1}{\lambda^{3/2}}\,\,\Big[\frac{1}{64} \,(N\m 2)\,(N\m 3)\,
\Big(\frac{\Gamma\big(\frac{3}{4}\big)}{\Gamma\big(\frac{5}{4}\big)}\Big)^2
+\frac{1}{8} \,(N\m 2)\,\Big]\,.
\eeq{A1B2}
Note that the $\frac{1}{\lambda^{3/2}}$ dependence is what is expected from a third order contribution to the series \reff{SeriesForm2}.

\subsection{Numerical results for the second series for different values of $\lambda$}

The second series expansion for the Euclidean path integral $K_E$ is given by \reff{Series2}. Upon discretization we divide the time interval $\mathcal{T}$ into $N$ equal intervals $\epsilon= \mathcal{T}/N$ and the path integral contains $N-1$ integrals. This yields the series \reff{SeriesDis} where $K_E$ is now a function of $N$. We showed that this series is equivalent to \reff{HypSeries} which is expressed in terms of functional derivatives of a generating functional $Z[\vec{J}]$ given by \reff{ZJ} and composed of products of generalized hypergeometric functions. We evaluate this series at strong coupling where the first series failed i.e. at 
$\lambda=1$ and $\lambda=10$ \footnote{The series converges also at the small values of $\lambda=0.01$ and $\lambda=0.1$ but its convergence is slow. It is also less important since we saw that the first series is reliable in that weak-coupling regime.}. We use the same values for the parameters as in the first series: $\omega=\mathcal{T}=\hbar=m=1$. The exact value from direct numerical integration is obtained for a given $N$. The continuum corresponds formally to the limit as $N\to \infty$ but in practice is reached at a finite $N$ to a given accuracy. For $\lambda=1$ and $\lambda=10$, direct numerical integration yields a continuum value at three decimal accuracy of $0.342$ (for any $N \ge 7$) and $0.237$ (for any $N \ge 9$) respectively. For a given $N$, we evaluate the discretized series expansion \reff{HypSeries} up to an order $n$ that converges to a value within a certain accuracy.  This value can then be compared to the exact value from direct numerical integration. Below we tabulate results up to $N=5$ for $\lambda=1$ and up to $N=7$ for $\lambda=10$. For a given $N$, there is overall a very good agreement between the series and the exact value. This confirms that the second series works well at strong coupling. At larger $N$, the matching is not as exact but this is entirely a numerical issue and not one of principle. One limitation, is that as $N$ increases, the series requires more terms to converge and this in turn requires more numerical precision and hence more computational resources. It is therefore of interest to see if one can formulate a second series for the QM path integral that converges faster. We discuss this further in the conclusion.   


\begin{table}[t]
\includegraphics[scale=0.8]{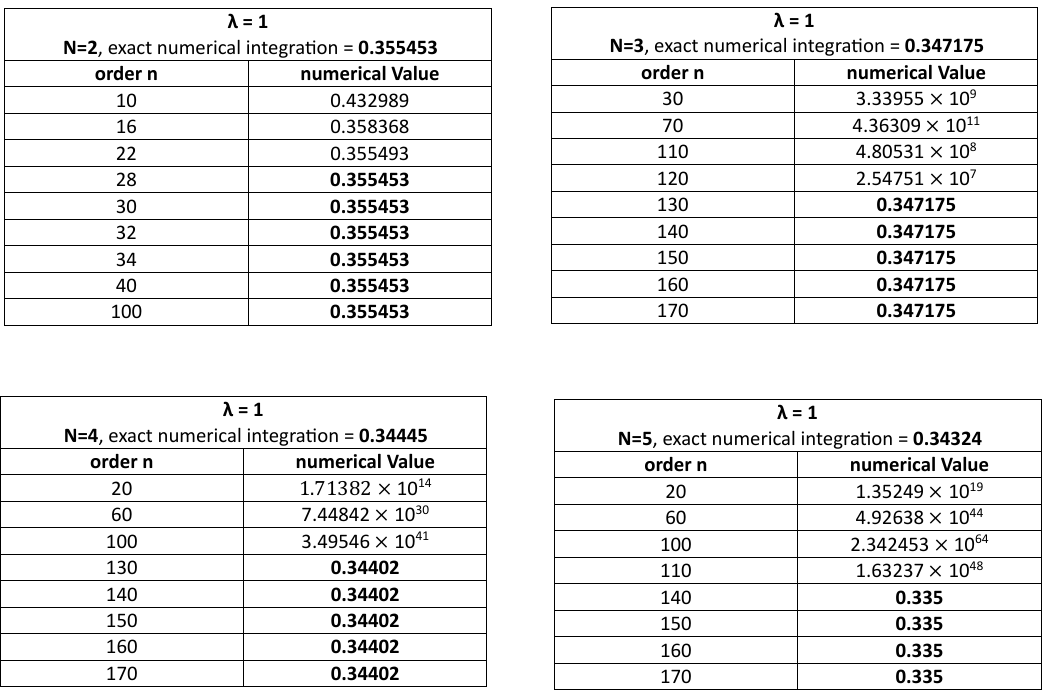}
\caption{The results of the series for $\lambda=1$ are tabulated for $N=2$ up to $N=5$. For each $N$ the exact value obtained from direct numerical integration is quoted to six decimal places. The series and exact value match to six decimal places for $N=2$ and $N=3$, to three decimal places for $N=4$ and two decimal places at $N=5$.  As $N$ increases, the series remains correct but converges at higher orders which requires more numerical precision and hence more computational resources. Direct numerical integration converges to three decimal places to the value of $0.342$ for any $N\ge 7$. This can be taken as the continuum value (formally, the $N\to \infty$ limit).}
\label{Table_Lambda1}
\end{table}   

\begin{table}[t]
		\includegraphics[scale=0.7]{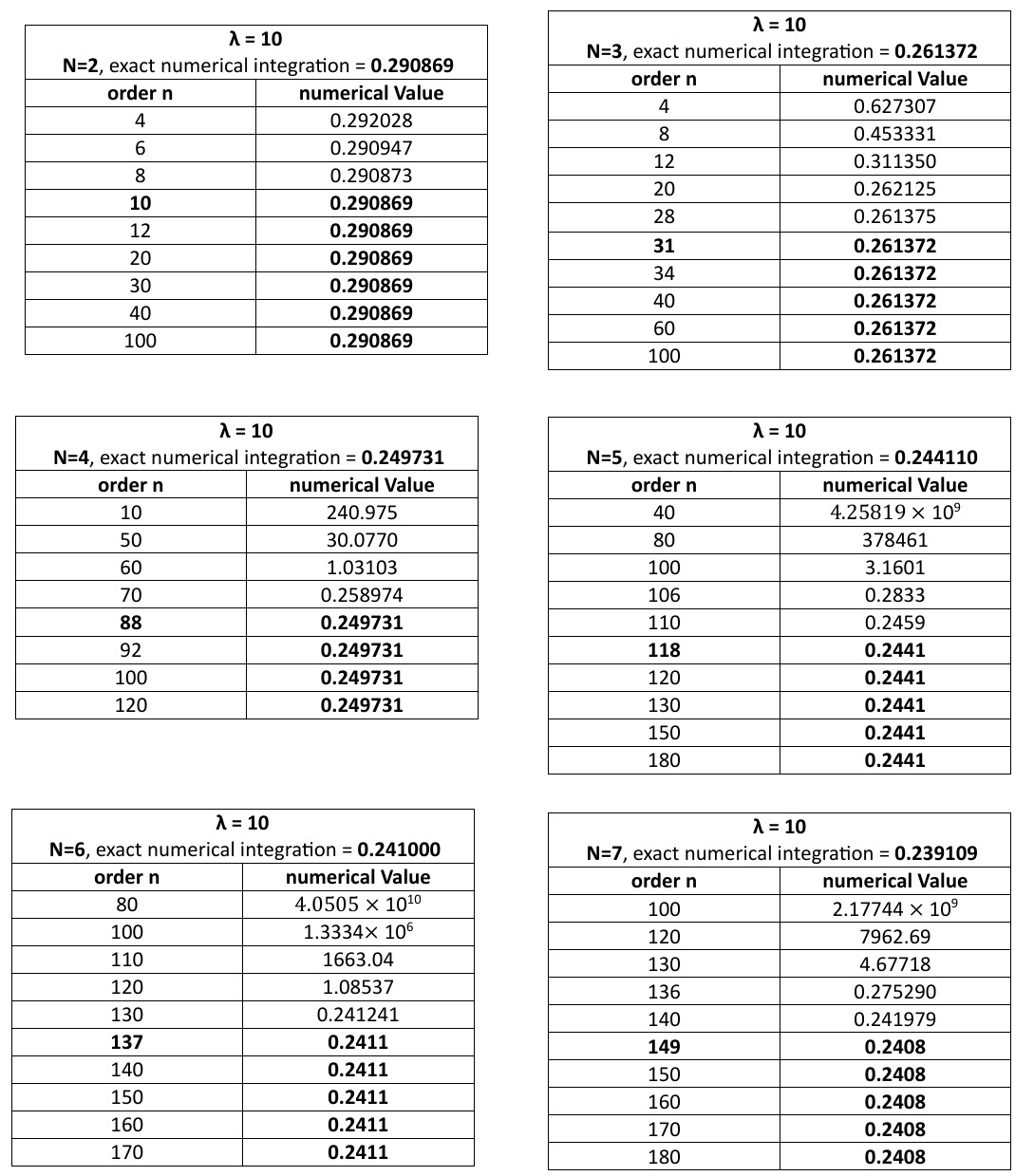}      
\caption{The results of the series for $\lambda=10$ are tabulated for $N=2$ up to $N=7$. The exact result from direct numerical integration is quoted to six decimal places for each $N$. For $N=2$, $N=3$ and $N=4$ the series matches the exact result fully (i.e. to six decimal places). For $N=5$ and $N=6$ the series matches the exact result to four decimal places and at $N=7$, effectively at three decimal places. Again, as $N$ increases, the series converges at higher orders which requires more numerical precision and hence more computational resources. Direct numerical integration converges to three decimal places to the value of $0.237$ for any $N\ge 9$. This can be taken as the continuum value (formally, the $N\to \infty$ limit).}
\label{Table_Lambda10}
\end{table}

\section{Conclusion}

In this work we presented two types of series expansions that are valid at strong coupling and applied it to a basic one-dimensional integral and path integral in quantum mechanics (QM) containing quadratic terms as well as a quartic interaction with coupling $\lambda$. We began with the basic integral. The first series $F_1(n)$ was  obtained by expanding the quartic term in powers of the coupling and was an asymptotic series. We plotted this series and this showed, in a concrete and transparent fashion how the asymptotic series actually behaves as a function of the coupling $\lambda$ and the order $n$.  At weak coupling, the series after a few orders reaches the correct value (to eight digit accuracy) and plateaus at that value over many orders before diverging. That is analogous to perturbative calculations using Feynman diagrams yielding accurate results at weak coupling in QFT. At strong coupling the series fails completely: it never gets close to the correct value and diverges from it right from the start. In contrast to the first series, the second series, $F_2(n)$, obtained by expanding the quadratic part, was an absolutely convergent series in inverse powers of the coupling. We plotted this function at weak, intermediate and strong coupling and it converged to the correct value at all couplings. It converges must faster at strong coupling than at weak coupling and hence is more suited to the strong coupling regime. 

We then resolved a paradox: why does the first series diverge when it stems from an original integral which is finite? In the original integral the quartic part dominates over the quadratic part asymptotically (as $x ->\infty$). The series expansion of the quartic part however is valid for an arbitrary large but finite $x$ (i.e. if one takes the limit as $x \to \infty$ and then one sums the series, it diverges, regardless of how many orders one goes up to. For arbitrary large but finite $x$, summing the series converges). Therefore, to capture the asymptotic behaviour of the original integral, which is dominated by the quartic part,  the series expansion of the integrand must be integrated to finite instead of infinite limits (run from $x=-\beta$ to $x=\beta$ where $\beta$ is a finite real number).  The resulting series $S(n,\beta)$ in powers of the coupling is, remarkably, an absolutely convergent series for any finite $\beta$. In contrast to the second series, it converges faster at weak coupling than at strong coupling and we showed, using a series representation of the incomplete gamma function, that there is a weak-strong coupling duality between the two series. In practice, $\beta$ is not required to be that large to obtain the exact result corresponding to the $\beta \to \infty$ limit. For our parameters, we obtained the exact result to eight digit accuracy for weak and strong coupling without going beyond $\beta=4$ (see table of values in \reff{Incomplete}). This series circumvents Dyson's argument because the original basic integral at negative $\lambda$ is finite if the limits of integration are finite. In the QM case, there is an interesting physical explanation. Dyson's argument would rely on the fact that at negative $\lambda$ the potential yields tunneling and hence has an unstable vacuum. However, with finite integration limits, there are infinite walls at $x=\pm \beta$ which prevents tunneling from occurring. So Dyson's argument is no longer applicable. 

We also saw how the same reasoning could be used to explain why perturbative expansions in QFT yield asymptotic series. The interaction contains a product of more than two fields and therefore dominates the quadratic part asymptotically (as the fields tend to infinity). The series expansion of the interaction part is however valid only for finite field (can be arbitrarily large but is finite). Therefore, in order for the perturbative series to capture the asymptotic behaviour of the original integral, which is dominated by the interaction part, one should be integrating the fields to a finite value $\beta$ not infinity. The series expansion for any quantity of physical interest (e.g. correlation functions) should then be absolutely convergent for a given $\beta$ (just like the series $S(n,\beta)$ that we obtained for the basic integral was an absolutely convergent series for a given $\beta$). The correct physical result corresponds formally to the $\beta \to \infty$ limit. Practically, this is reached by increasing $\beta$ until one obtains a value that does not change to within a desired level of accuracy (as an example see table of values \reff{Incomplete} for $S(n,\beta)$). 

For the quantum mechanical path integral, we considered the quartic anharmonic oscillator system. We obtained the terms for the first and second series expansions. As in the basic integral, the first series was an asymptotic series in powers of the coupling and the second series was an absolutely convergent series in inverse powers of the coupling.  So this aspect is not affected when we pass from a basic integral to a QM path integral. For the second series, the generating functional $Z[J]$ is a product of generalized hypergeometric functions in contrast to the usual Gaussians. The terms in the series can be obtained by taking functional derivatives of $Z[J]$. We obtained exact analytical formulas for the $n$th order terms in the series. The expressions are functions of $N$ (here $N$ is the number of segments the time interval is divided into in the discretized version of the path integral). For a general order $n$, the procedure is mathematically complicated, so we first worked out the expressions for the first three orders explicitly by hand (at third order, the expressions are already long). The general mathematical procedure was implemented in a Mathematica program that generates the analytical formulas for the $n$th order terms and in Appendix A we showcased the formula generated for $n=5$.  We presented numerical results at strong coupling $\lambda=1$ and $\lambda=10$ for different values of $N$ and these matched the exact numerically integrated value for that $N$ (to a given level of accuracy depending on $N$). This confirms that the second series works very well at strong coupling in a physical context (quantum mechanics of quartic anharmonic oscillator). As $N$ increased, one had to sum more terms in the series which required more numerical precision and computational resources (see comments in next paragraph on how this might be improved). For the first series, we generated the terms via functional derivatives of the forced harmonic oscillator (the generating functional, as it includes the source term $J(\tau) \,x$). We presented numerical results for weak, intermediate and strong coupling. For weak and intermediate coupling the results matched the exact numerically integrated value to within three or four decimal places. At strong coupling, the series departs significantly from the exact result and fails completely as expected from an asymptotic series.      

For future work, it would be of significant interest to work out the QM analog of the absolutely convergent series $S(n,\beta)$ in powers of the coupling that we obtained for the basic integral. This would be obtained by simply replacing the path integration limits in the first series from infinity to finite $\beta$. In this paper, we dedicated considerable time to developing the second series for the QM case because this is the series that is most suited to strong coupling; it would converge much faster at strong coupling compared to the QM analog of $S(n,\beta)$. The second series for the QM case was based on expanding the entire quadratic part which includes the kinetic term and any other quadratic terms (e.g. harmonic oscillator). As $N$ increased, more orders were needed to reach convergence and as we already mentioned, this requires more numerical precision and hence more computational resources. One possible way to remedy this situation and obtain a faster convergence is to not expand the entire quadratic part. As a simple illustration, consider the expression $e^{-\lambda(x_1^4 +x_2^4) -a (x_1^2 +x_2^2 -2 \,x_1 x_2)}$ which contains quartic and quadratic terms ($\lambda$ and $a$ are positive constants). If you integrate this expression over all $x_1$ and $x_2$ one does not of course obtain an analytical result (otherwise no series expansion would be needed). However, you do obtain an analytical result (modified Bessel functions) if you expand only $2\,a x_1 x_2$ but keep $-a (x_1^2 +x_2^2)$ in the exponential. In other words, one is not required to expand the entire quadratic part of the expression. Convergence would be significantly faster since $-a( x_1^2+x_2^2)$ is negative and would make the exponential decrease even faster compared to having only the quartic part (note that the discretized kinetic term in the Eulclidean action would have a term proportional to $-(x_2-x_1)^2= -(x_1^2+x_2^2)+ 2 \,x_1x_2$ so that $x_1^2 +x_2^2$ has a negative sign in front). Moreover, we would no longer need to expand $x_1^2+x_2^2$ leading to greater simplicity and convergence. The only drawback is that one cannot now add a source term in the exponential because with the quadratic part 
$-a( x_1^2+x_2^2)$ present, one no longer obtains an analytical result. So one would need to use an alternative to functional derivatives with respect to a source $J$. One possibility is to use a combination of derivatives with respect to $\lambda$ and $a$ (note that expanding $x_1\,x_2$ only requires even powers to be brought down since any odd powers yield zero). This is suggestive and there may be other clever techniques. In any case, this is worth exploring as this could lead to a considerable improvement in the speed of convergence of the second series in QM and QFT cases. Note that absolute convergence of the second series is ensured as long as the interaction part remains in the exponential. 

An important future goal is to apply our series expansions to realistic four-dimensional quantum field theories like QED and QCD.  In particular, one would like to use the series to calculate various correlation functions at strong coupling in these theories. It has been known analytically for a while that for $U(1)$ and $SU(3)$ four-dimensional lattice gauge theories, the expectation value of Wilson loops at strong coupling obey an area law corresponding to a confinement phase \cite{KW, Seiler,Gat,Drouf, Schwartz,ChengLi}. It has also been proven analytically that there is a non-confining (Coulomb) phase in four-dimensional $U(1)$ lattice gauge theory \cite{Guth}. This has been confirmed by Monte Carlo simulations in four-dimensional compact (discrete) QED \cite{Laut}. In other words, at zero temperature, there is a \textit{phase transition} as the coupling decreases from strong to weak. In contrast, lattice simulations \cite{LQCD} suggest that for four-dimensional QCD no such phase transition at zero temperature occurs when the coupling decreases from strong to weak, so that it is always in the confined phase. That has yet to be proven analytically using any technique. It would therefore be of great interest to see if the series expansions presented here can shed some light on this long-standing problem.     

\begin{appendices}
\numberwithin{equation}{section}
\setcounter{equation}{0}
\section{Analytical expression for fifth order contribution to series generated by program}
We worked out by hand the series \reff{HypSeries} to first, second and third order in sections \reff{FirstO}, \reff{SecondO} and \reff{ThirdO} respectively. Already at third order, expressions are quite lengthy and complicated. As the order increases,  the complexity increases significantly so that the task becomes increasingly laborious and one is more prone to make some error. In sections \reff{Amm},\reff{Bkk} and \reff{AmmBkk} we outlined the steps required to automate this process (writing a program to generate the expressions using symbolic software like Mathematica). In this appendix we will write down the explicit analytical expression for the fifth order ($n=5$) term in the series \reff{HypSeries}. The $n$th order term is given by \reff{Ordern}. The fifth order term is therefore
\begin{align}
C\, \frac{\hat{Q}^5}{5!} \, Z[\vec{J}]\,\Bigr\rvert_{\vec{J}=0}
 \label{Order5}
 \end{align} 
where $C$ is given by 
\beq
C=\Big( \dfrac{m}{2\,\pi\,\epsilon\, \hbar}\Big)^{N/2}
\eeq{CC2}
and
\begin{align}
\hat{Q}^5&=\Bigg[- \frac{m}{2\,\epsilon\,\hbar}\Big(\hat{A}+\hat{B}\Big)\Bigg]^5\nonumber\\
&=- \frac{m^5}{2^5\,\epsilon^5\,\hbar^5}\sum_{j=0}^{2}\binom{5}{2\,j}\,\hat{A}^{\,5-2 j}\,\hat{B}^{\,2\,j}\nonumber \\
&=- \frac{m^5}{2^5\,\epsilon^5\,\hbar^5}\Big(\hat{A}^{5}+ 10 \,\hat{A}^{3}\,\hat{B}^{2}+ 5\,\hat{A}\,\hat{B}^{4}\Big)\,.
\label{Q5}
\end{align}
The operators $\hat{A}$ and $\hat{B}$ are given by \reff{AA} and \reff{BB} respectively. We wrote a program that generates the analytical expression for the contribution  $\hat{A}^{m}\,\hat{B}^{k}$ i.e.  $\hat{A}^{m}\,\hat{B}^{k} \,Z[\vec{J}]\,\Bigr\rvert_{\vec{J}=0}$ where $m$ is any non-negative integer and $k$ is any positive even integer (including zero). We write down below the expressions generated by our program for $\hat{A}^{5}$, $\hat{A}^{3}\,\hat{B}^{2}$ and $\hat{A}\,\hat{B}^{4}$. 
\begin{align}
\hat{A}^5&=(2 + \omega^2 \,\epsilon^2)^5  \frac{\text{Z[0]}}{\left(\frac{\lambda  \epsilon }{\hbar }\right)^{5/2}}\Bigg[\frac{(N-5) (N-4) (N-3) (N-2) (N-1) \,\Gamma \left(\frac{3}{4}\right)^5}{1024 \,\Gamma \left(\frac{5}{4}\right)^5}
\nonumber\\&\quad+\frac{5 (N-4) (N-3) (N-2) (N-1) \,\Gamma \left(\frac{3}{4}\right)^3}{128\, \Gamma \left(\frac{5}{4}\right)^3}+\frac{5 (N-3) (N-2) (N-1)\, \Gamma \left(\frac{7}{4}\right) \Gamma \left(\frac{3}{4}\right)^2}{32 \Gamma \left(\frac{5}{4}\right)^3}\nonumber\\&\qquad+\frac{5 (N-2) (N-1) \Gamma \left(\frac{9}{4}\right) \Gamma \left(\frac{3}{4}\right)}{16\, \Gamma \left(\frac{5}{4}\right)^2}+\frac{15 (N-3) (N-2) (N-1) \Gamma \left(\frac{3}{4}\right)}{64 \Gamma \left(\frac{5}{4}\right)}\nonumber\\&\qquad\quad+\frac{5 (N-2) (N-1) \Gamma \left(\frac{7}{4}\right)}{8 \, \Gamma \left(\frac{5}{4}\right)}+\frac{(N-1) \Gamma \left(\frac{11}{4}\right)}{4 \,\Gamma \left(\frac{5}{4}\right)}\Bigg]\,.
\label{A5}
\end{align}
\begin{align}
\hat{A}^{3}\,\hat{B}^{2}&=4\,(2 + \omega^2 \,\epsilon^2)^3\, \frac{\text{Z[0]}} {\left(\frac{\lambda  \epsilon }{\hbar }\right)^{5/2}}
\Bigg[\frac{(N-5) (N-4) (N-3) (N-2) \Gamma \left(\frac{3}{4}\right)^5}{1024 \Gamma \left(\frac{5}{4}\right)^5}\nonumber\\&\qquad+\frac{9 (N-4) (N-3) (N-2) \Gamma \left(\frac{3}{4}\right)^3}{256 \Gamma \left(\frac{5}{4}\right)^3}+\frac{7 (N-3) (N-2) \Gamma \left(\frac{7}{4}\right) \Gamma \left(\frac{3}{4}\right)^2}{64 \Gamma \left(\frac{5}{4}\right)^3}\nonumber\\&\qquad\qquad+\frac{(N-2) \Gamma \left(\frac{9}{4}\right) \Gamma \left(\frac{3}{4}\right)}{8 \Gamma \left(\frac{5}{4}\right)^2}+\frac{3 (N-3) (N-2) \Gamma \left(\frac{3}{4}\right)}{16 \Gamma \left(\frac{5}{4}\right)}+\frac{3 (N-2) \Gamma \left(\frac{7}{4}\right)}{8 \Gamma \left(\frac{5}{4}\right)}\Bigg]\,.
\label{A3B2}
\end{align}
\begin{align}
\hat{A}\,\hat{B}^{4}&=16\,(2 + \omega^2 \,\epsilon^2)\, \frac{\text{Z[0]}} {\left(\frac{\lambda  \epsilon }{\hbar }\right)^{5/2}}
\Bigg[\frac{3 (N-5) (N-4) (N-3) \Gamma \left(\frac{3}{4}\right)^5}{1024 \Gamma \left(\frac{5}{4}\right)^5}+\frac{9 (N-4) (N-3) \Gamma \left(\frac{3}{4}\right)^3}{128 \Gamma \left(\frac{5}{4}\right)^3}\nonumber\\&+\frac{3 (N-3) \Gamma \left(\frac{7}{4}\right) \Gamma \left(\frac{3}{4}\right)^2}{32 \Gamma \left(\frac{5}{4}\right)^3}+\frac{3 (N-3) \Gamma \left(\frac{3}{4}\right)}{16 \Gamma \left(\frac{5}{4}\right)}+\frac{(N-3) (N-2) \Gamma \left(\frac{3}{4}\right)}{64 \Gamma \left(\frac{5}{4}\right)}+\frac{(N-2) \Gamma \left(\frac{7}{4}\right)}{8 \Gamma \left(\frac{5}{4}\right)}\Bigg]\,.
\label{A1B4}
\end{align}
The fifth order term in the series \reff{HypSeries} is therefore given by
\begin{align}
K_{E_{(n=5)}}&=-\Big( \dfrac{m}{2\,\pi\,\epsilon\, \hbar}\Big)^{N/2}\, \frac{m^5}{2^5\,\epsilon^5\,\hbar^5}\,\Big(\hat{A}^{5}+ 10 \,\hat{A}^{3}\,\hat{B}^{2}+ 5\,\hat{A}\,\hat{B}^{4}\,\Big)\nonumber\\\\
&=-\Big( \dfrac{m}{2\,\pi\,\epsilon\, \hbar}\Big)^{N/2}\, \frac{m^5}{2^5\,\epsilon^5\,\hbar^5}\,\frac{\text{Z[0]}} {\left(\frac{\lambda  \epsilon }{\hbar }\right)^{5/2}}\nonumber\\
&\Bigg[(2 + \omega^2 \,\epsilon^2)^5 \bigg(\frac{(N-5) (N-4) (N-3) (N-2) (N-1) \Gamma \left(\frac{3}{4}\right)^5}{1024\, \Gamma \left(\frac{5}{4}\right)^5}
\nonumber\\&\quad+\frac{5 (N-4) (N-3) (N-2) (N-1) \Gamma \left(\frac{3}{4}\right)^3}{128\, \Gamma \left(\frac{5}{4}\right)^3}+\frac{5 (N-3) (N-2) (N-1) \Gamma \left(\frac{7}{4}\right) \Gamma \left(\frac{3}{4}\right)^2}{32 \Gamma \left(\frac{5}{4}\right)^3}\nonumber\\&\qquad+\frac{5 (N-2) (N-1) \Gamma \left(\frac{9}{4}\right) \Gamma \left(\frac{3}{4}\right)}{16 \,\Gamma \left(\frac{5}{4}\right)^2}+\frac{15 (N-3) (N-2) (N-1) \Gamma \left(\frac{3}{4}\right)}{64 \,\Gamma \left(\frac{5}{4}\right)}\nonumber\\&\qquad\qquad+\frac{5 (N-2) (N-1) \Gamma \left(\frac{7}{4}\right)}{8\, \Gamma \left(\frac{5}{4}\right)}+\frac{(N-1) \Gamma \left(\frac{11}{4}\right)}{4\, \Gamma \left(\frac{5}{4}\right)}\bigg)\nonumber\\
&+40\,(2 + \omega^2 \,\epsilon^2)^3\, \bigg(\frac{(N-5) (N-4) (N-3) (N-2) \Gamma \left(\frac{3}{4}\right)^5}{1024 \,\Gamma \left(\frac{5}{4}\right)^5}\nonumber\\&\qquad+\frac{9 (N-4) (N-3) (N-2) \Gamma \left(\frac{3}{4}\right)^3}{256\, \Gamma \left(\frac{5}{4}\right)^3}+\frac{7 (N-3) (N-2) \Gamma \left(\frac{7}{4}\right) \Gamma \left(\frac{3}{4}\right)^2}{64 \,\Gamma \left(\frac{5}{4}\right)^3}\nonumber\\&\qquad\qquad+\frac{(N-2) \Gamma \left(\frac{9}{4}\right) \Gamma \left(\frac{3}{4}\right)}{8\, \Gamma \left(\frac{5}{4}\right)^2}+\frac{3 (N-3) (N-2) \Gamma \left(\frac{3}{4}\right)}{16\, \Gamma \left(\frac{5}{4}\right)}+\frac{3 (N-2) \Gamma \left(\frac{7}{4}\right)}{8 \,\Gamma \left(\frac{5}{4}\right)}\bigg)\nonumber\\
&+80\,(2 + \omega^2 \,\epsilon^2)\, \bigg(\frac{3 (N-5) (N-4) (N-3) \Gamma \left(\frac{3}{4}\right)^5}{1024 \Gamma \left(\frac{5}{4}\right)^5}+\frac{9 (N-4) (N-3) \Gamma \left(\frac{3}{4}\right)^3}{128 \,\Gamma \left(\frac{5}{4}\right)^3}\nonumber\\&\qquad+\frac{3 (N-3) \Gamma \left(\frac{7}{4}\right) \Gamma \left(\frac{3}{4}\right)^2}{32 \Gamma \left(\frac{5}{4}\right)^3}+\frac{3 (N-3) \Gamma \left(\frac{3}{4}\right)}{16\, \Gamma \left(\frac{5}{4}\right)}+\frac{(N-3) (N-2) \Gamma \left(\frac{3}{4}\right)}{64 \Gamma \left(\frac{5}{4}\right)}\nonumber\\&\qquad\qquad+\frac{(N-2) \Gamma \left(\frac{7}{4}\right)}{8\, \Gamma \left(\frac{5}{4}\right)}\bigg)\Bigg]\,.
\label{kl}
\end{align}

The above fifth order term is a function of $N$ and proportional to $Z[0]$ times the inverse of $\lambda^{5/2}$. A general $n$th order term would be proportional to $Z[0]$ times the inverse of $\lambda^{n/2}$. Recall that $Z[0]$ contains inverse powers of the coupling $\lambda$ and is given by:
\begin{align*}
Z[0]= I[0]^{N-1}=\Big[2\,\Gamma\big(\frac{5}{4}\big) \,\Big(\frac{\hbar}{\epsilon\,\lambda}\Big)^{1/4}\Big]^{N-1}\,.
\end{align*}
\end{appendices}

\section*{Acknowledgments}
A.E. acknowledges support from a discovery grant of the National Science and Engineering Research Council of Canada (NSERC).

\end{document}